%% file: Draft2022.tex
\documentclass[11pt,a4paper]{scrartcl}

\usepackage{ILD}
\usepackage{arydshln}
\usepackage[symbol]{footmisc}
\usepackage{feynmf}
\usepackage{textpos}
\usepackage{enumitem}
\usepackage{verbatim}
\usepackage{upgreek}
\usepackage{caption}
\usepackage{subcaption}
\usepackage{color}
\def\Gammaqq{\ensuremath{\Gamma_{q\bar{q}}}\xspace}
\def\Gammahad{\ensuremath{\Gamma_{hadrons}}\xspace}

\def\kaonness{\ensuremath{\Delta_{\dEdx-K}}\xspace}

\def\epsilonb{\ensuremath{\epsilon_{b}}\xspace}
\def\epsilonc{\ensuremath{\epsilon_{c}}\xspace}
\def\epsilonuds{\ensuremath{\epsilon_{uds}}\xspace}
\def\epsilonb2{\ensuremath{\epsilon^{2}_{b}}\xspace}
\def\epsilonc2{\ensuremath{\epsilon^{2}_{c}}\xspace}
\def\epsilonuds2{\ensuremath{\epsilon^{2}_{uds}}\xspace}

\def\costheta{\ensuremath{\cos \theta}\xspace}

\def\costhetaq{\ensuremath{\cos \theta_{q}}\xspace}

\def\fb{fb\ensuremath{^{-1}}\xspace}

\def\bbbar{\ensuremath{b}\ensuremath{\overline{b}}\xspace}
\def\qqbar{\ensuremath{q}\ensuremath{\overline{q}}\xspace}
\def\ccbar{\ensuremath{c}\ensuremath{\overline{c}}\xspace}
\def\eebb{\ensuremath{e^{-}e^{+}\rightarrow b\bar{b}}\xspace}%\ensuremath{e^{\mbox{\scriptsize +}}}\ensuremath{\rightarrow}\ensuremath{b}\ensuremath{\overline{b}}\xspace}
\def\eecc{\ensuremath{e^{-}e^{+}\rightarrow c\bar{c}}\xspace}
\def\eeqq{\ensuremath{e^{-}e^{+}\rightarrow q\bar{q}}\xspace}
\def\ee{\ensuremath{e^{-}e^{+}}\xspace}

\def\cme{\ensuremath{c.m.e.}\xspace}
\def\eLpR{\ensuremath{e_L^{-}e_R^{+}}\xspace}
\def\eRpL{\ensuremath{e_R^{-}e_L^{+}}\xspace}
\def\eLpRqq{\ensuremath{e_L^{-}e_R^{+}\rightarrow q\bar{q}}\xspace}
\def\eRpLqq{\ensuremath{e_R^{-}e_L^{+}}\rightarrow q\bar{q}\xspace}

\def\dEdx{\ensuremath{dE/\/dx}\xspace}
\def\Afbb{\ensuremath{A^{b\bar{b}}_{FB}}\xspace}
\def\AFBb{\ensuremath{A^{b\bar{b}}_{FB}}\xspace}
\def\Afb{\ensuremath{A_{FB}}\xspace}
\def\AFB{\ensuremath{A_{FB}}\xspace}

\def\Rb{\ensuremath{R_{b}}\xspace}
\def\Rc{\ensuremath{R_{c}}\xspace}
\def\Rq{\ensuremath{R_{q}}\xspace}
\def\Rqp{\ensuremath{R_{q\prime}}\xspace}

\def\Afbq{\ensuremath{A^{q\bar{q}}_{FB}}\xspace}
\def\AFBc{\ensuremath{A^{c\bar{c}}_{FB}}\xspace}

\def\bquark{\ensuremath{b}-quark\xspace}

\def\cquark{\ensuremath{c}-quark\xspace}

\def\Zpole{\ensuremath{Z}-pole\xspace}

\def\Bc{\ensuremath{Vtx}-method\xspace}
\def\Kc{\ensuremath{K}-method\xspace}

\def\Pb{\ensuremath{P_{chg.}}\xspace}
\def\Qb{\ensuremath{Q_{chg.}}\xspace}

\def\PbB{\ensuremath{P_{chg.,M_{1}}}\xspace}
\def\PbK{\ensuremath{P_{chg.,M_{2}}}\xspace}
\def\QbB{\ensuremath{Q_{chg.,M_{1}}}\xspace}
\def\QbK{\ensuremath{Q_{chg.,M_{2}}}\xspace}

\title{Experimental methods and prospects on the measurement of electroweak $b$ and $c$-quark observables at the ILC operating at 250 GeV}

\ildpublic{PHYS}{2023}{001}% Public note 
\date{\formatdate{19}{6}{2023}}

\addauthor{A. Irles}{\institute{1} \footnote{First author \href{mailto:adrian.irles@ific.uv.es}{adrian.irles@ific.uv.es}}}
\addauthor{R. P\"oschl}{\institute{2}}
\addauthor{F. Richard}{\institute{2}}

\addinstitute{1}{IFIC, Universitat de Val\`encia and CSIC, C./ Catedr\'atico Jos\'e Beltr\'an 2, E-46980 Paterna, Spain}
\addinstitute{2}{Universit{\'e} Paris-Saclay, CNRS/IN2P3, IJCLab, 91405 Orsay, France}

\abstract{
  This paper describes a comprehensive experimental study on viability and prospects for the measurement of electroweak observables in \eebb and \eecc processes at the International Linear Collider (ILC) operating at 250 GeV of centre of mass energy. The ILC
  will produce electron and positron beams with different degrees of longitudinal polarisation (up to 80$\%$ for electrons and $30\%$ for positrons). The studies are based on a detailed simulation of the International Large Detector (ILD) concept. 
  This will allow to inspect in detail the four independent chirality combinations of the electroweak couplings 
  to electrons and other fermions and also perform background free analysis.
  The ILD design is based on the particle flow approach and the excellent vertexing and tracking capabilities, \mbox{including} charged hadron identification thanks to the \dEdx.
  We evaluate the main sources of experimental systematic uncertainties and identify the key design aspects of the accelerator and detector that are crucial to achieve the required per mil level accuracy that matches the expected statistical accuracy.

}

\titlecomment{This work was carried out in the framework of the ILD concept group}

\graphicspath{ {./logos/}{./figures/} }

\begin{document}

% generates the title page
\titlepage

%\newpage

\tableofcontents
%\newpage

% include source for sections
%\let\clearpage\relax
%\include{selection}

\input{sections/1_intro}

\input{sections/2_observable} %formalism
\input{sections/3_ILCILD} %description of ILC and ILD.
\input{sections/4_analysis_simulation} 
\input{sections/5_analysis_preselection}

\input{sections/6_analysis_R}

\input{sections/7_analysis_AFB}

\input{sections/8_systematics}

\input{sections/9_results}

\section*{Acknowledgements}

We are grateful to the ILD Publication and Speakers Bureau and, in particular, K. Kawagoe, M. Berggren and K. Fuji 
for their work and support as the editorial board team. We also acknowledge R. Settles and U. Einhaus for reading the draft and providing valuable comments.
Finally, we would like to acknowledge S. Bilokin for the studies during his PhD that motivated part of this note.

We would like to thank the LCC generator working group and the ILD software working group for providing the simulation and reconstruction tools and producing the Monte Carlo samples used in this study.
This work has benefited from computing services provided by the ILC Virtual Organization, supported by the national resource providers of the EGI Federation and the Open Science GRID.
A. Irles is funded by projects PID2021-122134NB-C2 (PEICTI 2021-2023) and by the Generalitat Valenciana (Spain) under the grant number CIDEGENT/2020/21.
A. Irles also acknowledges the financial support from the MCIN with funding from the European Union NextGenerationEU and Generalitat Valenciana in the call Programa de Planes Complementarios de I+D+i (PRTR 2022) Project \textit{Si4HiggsFactories}, reference ASFAE$/2022/015$.

\newpage%
%\appendix

\clearpage
% add references
\printbibliography[title=References]
%\section*{References}
%\bibliographystyle{JHEP}
%\bibliography{references} 

\end{document}

%% file: sections/1_intro.tex
\section{Introduction}
\label{sec:intro}

Despite its success, the Standard Model (SM) does not explain the striking mass hierarchy in the fermion sector. 
Models of new physics featuring extra-dimensions \cite{Yoon:2018xud,Funatsu:2017nfm,Funatsu:2020haj}
may explain this mass hierarchy. 
Furthermore, the LEP/SLC anomaly in \eebb is still unexplained \cite{Djouadi:2006rk}. The effects of new physics may differ for different fermion chiralities and additional terms associated with various mediators (SM $Z$ and $\gamma$ or beyond SM $Z^{\prime}$ or mixing of these). This motivates the study of quark pair production in high energy $e^{-}e^{+}$ collisions at past lepton colliders \cite{ALEPH:2005ab} and at future ones \cite{Bilokin:2017bor,Bilokin:2017lco,bilokin:tel-01826535,Irles:2019xny,Irles-Quiles:2019mdp,Okugawa:2019ycm,Irles:2020gjh,Okugawa:2022zmt}.   

The present paper intends to document the main experimental challenges for the precise measurements of the \bquark and \cquark observables
in the International Large Detector (ILD) of the International Linear Collider (ILC) \cite{Behnke:2013lya,ILD:2020qve}
colliding polarised beams of electrons and positrons at 250 GeV centre of mass energy.
These studies assume an integrated luminosity of 2000 \fb for the 250 GeV program (ILC250).

The document is organised as follows: Section \ref{sec:observable} introduces  the definition of the observables and the main processes
involved. In Section \ref{sec:ILCILD}, the main design aspects of the ILC and the ILD are described, emphasising the beam polarisation aspects, the particle flow approach and the expected excellent tracking and vertexing capabilities of the ILD.
In Section \ref{sec:simulation}, the framework for the event generation,
the simulation tools (full simulation including a high level of detector realism)
and the reconstruction algorithms are discussed. The expected particle identification and flavour tagging performances at ILD are also addressed in Section \ref{sec:simulation}.
Section \ref{sec:analysis_preselection} focuses on the event reconstruction, signal selection, and separation from the background.
Sections \ref{sec:analysis_R} and \ref{sec:analysis_AFB} detail the experimental methods for the measurement of the 
observables defined in Section \ref{sec:observable} ($R$ and \AFB). In Section \ref{sec:systematics} 
a comprehensive assessment of systematic uncertainties is given. 
Finally, in Section \ref{sec:results}, the expected results at the ILC250 are discussed.

%% file: sections/2_observable.tex
%\tableofcontents
\section{Definition of the experimental observables}
\label{sec:observable}

At leading order (LO) and with $\sqrt{s}\gg m_{q}$ ($\beta\approx1$) the differential cross section for
\eeqq with 100\% polarised beams at the centre of mass energy of $\sqrt{s}$ can be written as:

\begin{align}
    \frac{d\sigma_{\eLpR\rightarrow\qqbar}}{d\costhetaq}= \frac{s}{32\pi} [ (1+\costhetaq)^{2} |Q_{e_{L}q_{L}}|^{2} +  (1-\costhetaq)^{2} |Q_{e_{L}q_{R}}|^{2} ] \\
    \frac{d\sigma_{\eRpL\rightarrow\qqbar}}{d\costhetaq} = \frac{s}{32\pi} [ (1+\costhetaq)^{2} |Q_{e_{R}q_{R}}|^{2} +  (1-\costhetaq)^{2} |Q_{e_{R}q_{L}}|^{2} ] 
    \label{eq:sigma}
\end{align}
following the same notation as in \cite{Funatsu:2020haj} were $\theta_{q}$ is the angle between quark and incoming electron. The following notation is used: \eLpR for the cases in which the electron beam has 100\% left polarisation
and the positron beam has 100\% right polarisation (and vice versa for \eRpL).

The helicity amplitudes $Q_{e_{X}q_{Y}}$ are given by
\begin{align}
        Q_{e_{X}q_{Y}} \equiv \sum_{i} \frac{g^{X}_{V_{i}e}g^{Y}_{V_{i}q}}{(s-m_{V_{i}}^{2})+im_{V_{i}}\Gamma_{V_{i}}}
    \label{eq:helicity}
\end{align}
where $g^{X}_{V_{i}f}$ are the couplings of the $X$-handed and $Y$-handed fermions $f$ to the vector boson $V_{i}$, 
and $m_{V_{i}}$ and $\Gamma_{V_{i}}$ are the mass and total decay width of $V_{i}$. 
In the absence of new resonances, only contributions from the photon and Z boson are considered:

\begin{align}
        Q^{SM}_{e_{X}q_{Y}} = \frac{e^{2}}{s} + \frac{g^{X}_{Ze}g^{Y}_{Zq}}{(s-m_{Z}^{2})+im_{Z}\Gamma_{Z}}
    \label{eq:helicity_SM}
\end{align}

As described in Section \ref{sec:ILCILD}, the ILC will operate with partially polarised electron and positron beams. 
The ILC250 physics program foresees a total integrated luminosity of 2000 \fb
shared in four different data sets with different beam longitudinal polarisation values: $45\%$ of the luminosity with the configuration $P_{\ee}=(-0.8,+0.3)$; $45\%$ with $P_{\ee}=(+0.8,-0.3)$; $5\%$ with $P_{\ee}=(-0.8,-0.3)$; and $5\%$ with $P_{\ee}=(+0.8,+0.3)$. This notation uses the negative sign for the left-handed polarisation and the positive for the right-handed polarisation fraction.
The differential cross section for \eeqq with partially polarised beams is given by:

\begin{align}
    \frac{d\sigma_{\eeqq}}{d\costhetaq}(P_{e^{-}},P_{e^{+}},\costhetaq) = (1-P_{e^{-}}P_{e^{+}}) \frac{1}{4} \left[
    {(1-P_{eff}) \frac{d\sigma_{\eLpRqq}}{d\costhetaq} + (1+P_{eff}) \frac{d\sigma_{\eRpLqq}}{d\costhetaq}} \right]
    \label{eq:sigma}
\end{align}
where $P_{eff}$ is defined as
\begin{align}
    P_{eff}\equiv \frac{P_{e^{-}}-P_{e^{+}}}{1-P_{e^{-}}\cdot P_{e^{+}}}
\end{align}

Initial state photon radiation (ISR) alters the cross section w.r.t.\ the nominal centre-of-mass energy.  
Therefore it is necessary to only allow for restricted photon energy. 
The QCD and photon final state radiation (FSR) also impact the definition of the observables. 
For instance, when a hard gluon is emitted by any of the quarks produced in the hard scattering, 
the momentum of the final state quark pair is affected, and the topology of the event may be modified, leading to a potential breaking of the expected two jets, back-to-back kinematics.
The cross section is therefore redefined as a function of the invariant mass of the outcoming quark-pair at parton level invariant mass and acollinearity. 
A simplified definition of acollinearity is used:

\begin{align}
    \sin{\Psi_{acol}}=\frac{|\vec{p_{q}} \times \vec{p}_{\bar{q}}|}{|\vec{p_{q}}|\cdot|\vec{p}_{\bar{q}}|}
    \label{eq:acol}
\end{align} with the extra requirement of $\cos{\Psi_{acol}}<0$. A cut on this variable can be used
to avoid infrared divergences in the calculations due to soft and/or collinear ISR and FSR. The system of reference used for the definitions is not the laboratory one but the event system of reference, defined by the thrust-axis\footnote{The thrust-axis is defined as the $\hat{n}$ that maximises the following expression $T=\frac{\sum\limits_{i=q,\bar{q}} \vec{p}_{i}\cdot \hat{n}}{\sum\limits_{i=q,\bar{q}} \vec{p}_{i}}$.\label{footnote:T}} of the \qqbar system.
A detailed study and optimisation of this cut should be done using dedicated NLO-generated events,
which are not yet available in the ILC software framework (\textrm{ i.e.}\
within the chain of the full detector simulation effects and beam features implementation).
However, it is essential to remark that, as shown in \cite{AlcarazMaestre:2020fmp}, 
the impact of the QCD radiation on electroweak observables is expected to be at the order of the per mil level if a reasonable cut on the acollinearity is applied.
In the following, our signal will be defined by the acollinearity smaller than $0.3$ and the quark-pair invariant mass larger than 140 GeV.
The Equation \ref{eq:sigma} is therefore changed to:
\begin{align}
    \frac{d\sigma_{\eeqq}}{d\costhetaq}(\costhetaq) \rightarrow  \frac{d\sigma^{cont.}_{\eeqq}}{d\costhetaq}(\sin{\Psi_{acol}}<0.3,m_{q\bar{q}}>140\text{ GeV},\costhetaq).
    \label{eq:sigma3}
\end{align}
For simplicity, in the following, these two cuts will always be implicit to signal processes. Therefore, when the notation \eeqq appears from now on,
it should be read as an abbreviation of $\eeqq(\sin{(Psi_{acol}}<0.3,m_{q\bar{q}}>140\text{ GeV})$.
The complementary definition is defined as \qqbar background.
The most significant contribution to this background is the \textrm{Radiative Return} background 
(due to the return to the \Zpole mass due to energy losses of the beam during because of ISR). 
For simplicity, in the following, we will use this denomination to refer to all the $\qqbar$ background events.

\begin{figure}[!ht]
  \centering
    \begin{tabular}{cc}
        \includegraphics[width=0.45\textwidth]{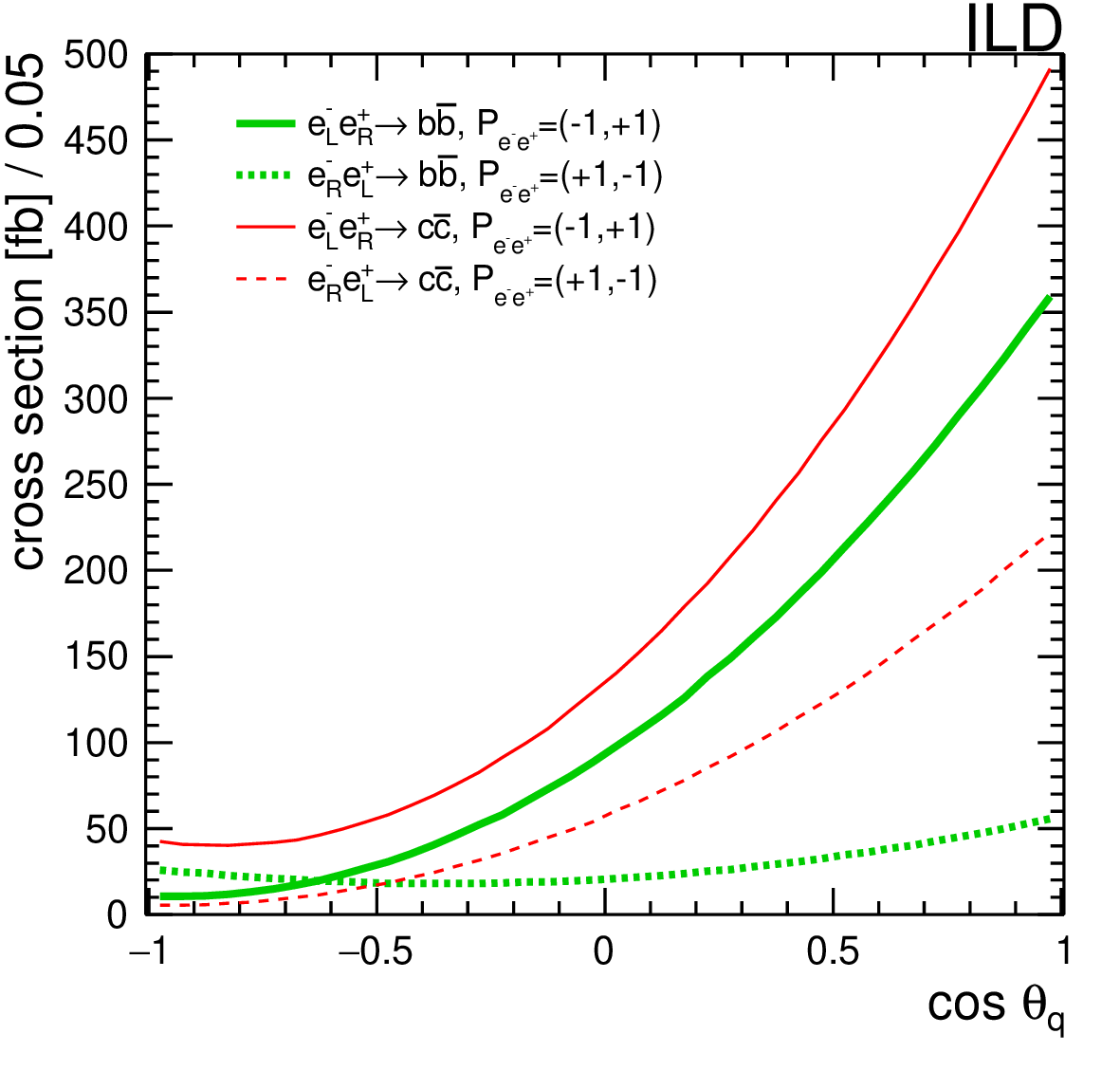} &
        \includegraphics[width=0.45\textwidth]{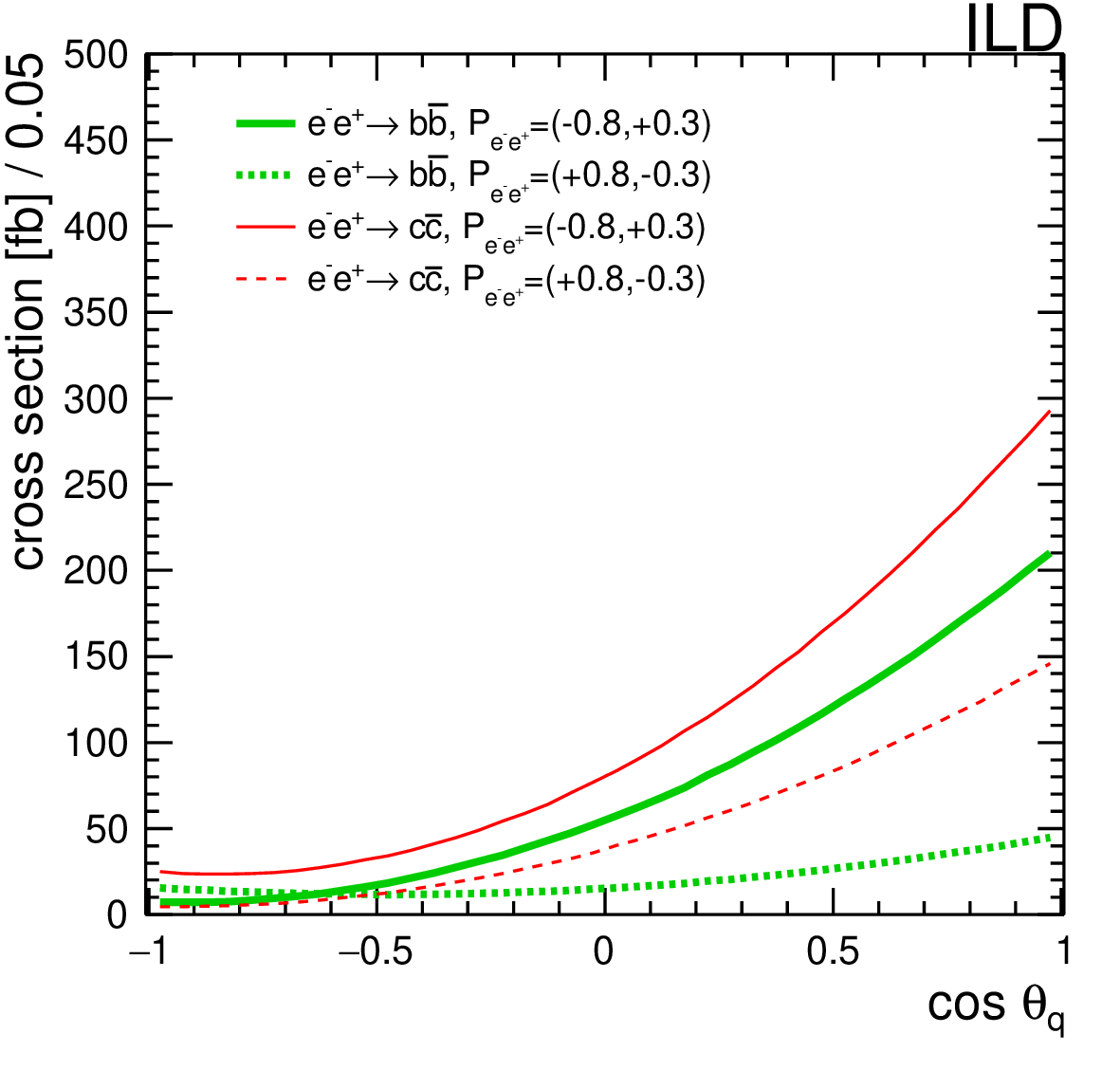} 
     \end{tabular}	
  \caption{\label{fig:diff_cross_kcut_acol} Differential cross section predictions at LO for $b$ and $c$-quark pair production assuming different beam polarisation scenarios (100$\%$ polarisation case in the left figure and ILC250 polarisation scenarion in the right figure).}
\end{figure}

Other observables besides the cross section (differential or inclusive) can be defined. One is the hadronic fraction \Rq, commonly used in past experiments. It is defined as  
$R_{q}^{m_{Z}} = \Gammaqq/\ \Gammahad$, with $\Gammaqq$ being the 
partial width for a Z decaying into $q\bar{q}$ pair of a given flavour and $\Gammahad$ the sum of the  
partial width of Z decaying to all quarks except the top quark. 
Experiments at LEP and SLC running at the Z-pole have largely studied this observable \cite{ALEPH:2005ab}.
In the continuum, as it is the ILC250 case, the observable needs to be redefined:
\begin{equation}
  \Rq= \frac{\sigma_{\eeqq}}{\sigma_{had.}}
\end{equation}
where $\sigma_{had.}$ is defined as $\sigma_{\eeqq}$ but integrated for all $q$ flavours except the top-quark. 
%Furthermore, this observable can also be defined as a deferentially as \Rqcostheta:
%\begin{equation}
%  \Rqcostheta= \frac{\sigma_{\eeqq}^{cont.}(|\costheta|)}{\sigma_{had.}^{cont.}(|\costheta|)}.
%\end{equation}
%%Note that \Rqcostheta=\Rq for all angles in the SM at leading order \todo{What about at higher orders?}.

Another observable traditionally used is the forward-backward asymmetry $\Afbq $\cite{ALEPH:2005ab}. 
It is defined as:
\begin{equation}
  \Afbq= \frac{\sigma^{F}_{\eeqq}-\sigma^{B}_{\eeqq}}{\sigma^{F}_{\eeqq}+\sigma^{B}_{\eeqq}}
\end{equation}
where $\sigma^{F/B}_{\eeqq}$ is the \eeqq cross section in the forward/backward
hemisphere as defined by the polar scattering angle $\theta_{q}$.

%The LO predictions for these two observables for heavy quark pair production at ILC250  are shown in Figure \ref{fig:cross_kcut_acol}.
The polar angle distributions are governed by the electroweak couplings of the electron beam and the quark under study and the interference between the photon and the $Z$ propagators. The strongest forward peaking is expected for the \bquark in case of a left-handed polarised electron beam due to the combination of the comparatively large coupling of the electron to the $Z$, the small electrical $b$-charge and the predominant left-handed coupling of the $b$ to the $Z$. 
The expected values of the differential cross section for the signal in different beam polarisation scenarios are shown in Figure \ref{fig:diff_cross_kcut_acol}.

%% file: sections/3_ILCILD.tex
%\tableofcontents
\section{The International Linear Collider and the International Large Detector}
\label{sec:ILCILD}

The International Linear Collider (ILC) \cite{Behnke:2013xla,Baer:2013cma,Adolphsen:2013jya,Adolphsen:2013kya,Behnke:2013lya}
is a linear electron-positron collider with polarised beams that will produce collisions at several energies.  
This document focuses on the study of the collisions at the centre of mass energy, \cme  of 250 GeV, called {\em ILC250} hereafter.
In the ILC250 case, the machine is expected to have an instantaneous luminosity between 
1.25 to $2.7 \times 10^{34}$cm$^{-2}$s$^{-1}$ \cite{ILD:2020qve} and 
beams polarised at the 80\% and 30\% for the electron and positron cases, respectively.
The ILC250 baseline physics foresees a total integrated luminosity of 2000\,\fb
shared in different beam polarisation settings: $900$ fb$^{-1}$ for $P_{\ee}=(-0.8,+0.3)$; 
$900$ fb$^{-1}$ for $P_{\ee}=(+0.8,-0.3)$;
$100$ fb$^{-1}$ for $P_{\ee}=(-0.8,-0.3)$, and 
$100$ fb$^{-1}$ for $P_{\ee}=(+0.8,+0.3)$ (with the $P_{\ee}$ notation explained in Section \ref{sec:intro}).
For this article, it is important to remind that the beam size at the interaction point is around $(\sigma_x,\,\sigma_y,\,\sigma_z) \approx (0.516, 0.008, 300)$\textmu m~\cite{Adolphsen:2013kya}.

The International Large Detector (ILD) \cite{Behnke:2013lya,ILD:2020qve} 
is one of the detectors proposed for collecting and exploiting the ILC data.
Its design has been optimised for the use of particle flow reconstruction algorithms \cite{Sefkow:2015hna} (PFA) to reconstruct and separate individual particles produced in the collisions
and to exploit the tracking capabilities of the
inner detectors maximally.
The ILD layout can be divided into four main sectors: the inner vertexing and tracking systems, the calorimetric systems,
the magnetic coil and the muon detection system.
For all the subsystems, several technological solutions are under discussion \cite{Behnke:2013lya,ILD:2020qve}.
It is out of the scope of this document to describe in detail the characteristics of the ILD and its subdetectors, but
since the methods described in this document heavily rely on the vertexing and the tracking and particle identification (PID) 
capabilities of the ILD, we briefly introduce them in the following paragraphs. 
For more technical details on these topics, the reader is referred to \cite{Behnke:2013lya,ILD:2020qve}.

The vertexing and tracking systems are divided into two groups: 
the innermost system based on silicon sensors and the central time projection chamber (TPC).

The vertex detector (VTX) is the closest detector to the beam pipe, with a minimum distance of 16  mm 
and a maximum distance of 60  mm. It is a pixel detector with pure barrel geometry composed
of three double layers. The first layer is twice shorter than the other two
to minimise the occupancy from beam background hits. 
The VTX is optimised to provide a resolution of the secondary vertices better
than 3 $\mu$m. Following the VTX detector are the silicon tracking systems: the silicon 
internal tracker (SIT) is placed in the barrel region between the VTX and the TPC, and the 
forward tracking detector (FTD) covers the region of shallow angles with respect to the beam.
The SIT also features barrel geometry, and it will offer tracking resolution parameters of 
5 $\mu$m with four 644 mm long layers placed between 155 and 301  mm distance of the beam pipe.
The FTD is composed of disks instead of a barrel geometry. Its acceptance starts at 4.8
degrees, and it complements the SIT coverage between 16 and 35 degrees. It will feature two sets of seven disks on each side of the VTX and SIT. The first two disks are 
installed close to the VTX and are equipped with highly granular pixel detectors to provide 
precise 3D points with a 3-5 $\mu$m resolution. 
The other five disks, featuring silicon strip sensors, 
have twice the outer radius compared to the pixel disks, extending out to the inner envelope of 
the TPC at a distance of 300 mm of the beam pipe. 

The TPC is a large volume time projection chamber allowing continuous 3D tracking and 
particle identification based on \dEdx. It is also barrel-shaped with an inner radius of 
329 mm and an outer radius larger than 1808 mm. It provides a single-point resolution of 
100 $\mu$m  over about 200 readout points and a \dEdx resolution of $\simeq 4.5\%$\footnote{In this document, only the TPC with pad-based electronics is considered. A version with pixel-electronics is expected to provide an improved \dEdx resolution of $\simeq 3.5\%$}.

%% file: sections/4_analysis_simulation.tex
\section{Event simulation and reconstruction}
\label{sec:simulation}

All results shown here are obtained using a detailed simulation of the 
ILD concept \cite{Behnke:2013lya}, in particular, the ILD-L model described in \cite{ILD:2020qve}.
The ILD detector geometry is implemented in the DD4HEP \cite{Frank:2014zya} 
the framework that provides the detector geometry, its material content and readout 
features interfaced to full simulation of the \texttt{Geant4} toolkit \cite{Agostinelli:2002hh,Allison:2006ve,Allison:2016lfl}. 
These frameworks and the different reconstruction algorithms are implemented in the \texttt{ILCSoft} toolkit. 
The events are generated with the
\texttt{WHIZARD} (v2.8.5) \cite{Kilian:2007gr} event generator. 
The matrix elements are implemented at leading order in electroweak theory and QED. 
However, QED ISR and FSR are also implemented in \texttt{WHIZARD}.  
Fragmentation (or hadronisation), including parton shower final state radiation, 
is provided by \texttt{Pythia} (v6.422) event generator \cite{Sj_strand_2006}. 
The beam energy spectrum and beam-beam interaction
producing incoherent \ee background pairs are generated with Guinea-Pig \cite{Schulte:1999tx}.
Other sources of background, such as $\gamma\gamma\rightarrow hadrons$ events, are generated separately
and overlaid to the simulated events. A description of the whole procedure to generate all SM processes is given in \cite{Berggren:2021sju}.

The signal cross sections for different polarisation scenarios are listed
in Table \ref{tab:crosssection}, together with the main background source: the radiative returns to the \Zpole events.
The cross sections of the other processes involving hadrons in the final state are listed in Table \ref{tab:crosssection_bkg}. 
Processes with leptons in the final state are ignored since these are expected to be easily 
identified.

\begin{table}[!ht]
  \centering
  \begin{tabular}{c|ccc|ccc}
    \hline
     & \multicolumn{3}{|c|}{ $\sigma_{\eeqq}$[fb]} & \multicolumn{3}{|c}{ Radiative Return BKG [fb]} \\
    \hline
    Polarisation & \bbbar & \ccbar & \qqbar ($q=uds$) & \bbbar & \ccbar & \qqbar ($q=uds$) \\
    \hline
    \eLpR & 4894.4 & 7068.1 & 16817.1  & 21087.0 & 18865.1 & 59227.7\\
    \eRpL & 1087.4 & 3006.9 & 5153.3 & 12872.4 & 11886.2 & 36410.3 \\
    \hline
  \end{tabular}
  \caption{Production cross section of the background originated by di-boson production. \label{tab:crosssection}}
\end{table}

\begin{table}[!ht]
 \centering
  \begin{tabular}{c|c|c|c}
    \hline
    & \multicolumn{3}{|c}{  $\sigma_{e^{-}e^{+}\rightarrow\,X}$ [fb]} \\
    \hline
	Polarisation &  $X= WW \rightarrow q_{1} \bar{q_{2}} q_{3} \bar{q_{4}}$ & $X= ZZ \rightarrow q_{1} \bar{q_{1}}q_{2} \bar{q_{2}}$ & $X= HZ \rightarrow q\bar{q}H$ \\
    \hline
    \eLpR & 14866.4 & 1405.1 & 343.0 \\
    \eRpL & 136.8 & 606.7 & 219.5 \\
      \hline
  \end{tabular}
  \caption{\label{tab:crosssection_bkg} Cross sections at 250 GeV for processes producing at least one pair of $q$-quarks using fully polarised beams. }
\end{table}

The size of the different analysed samples is the equivalent of between 2000-5000\fb for each process and each
beam polarisation scheme, assuming 100\% polarisation values.
Final results are scaled to the foreseen luminosity and beam polarisation schemes (see Section \ref{sec:ILCILD}).

\subsection{Tracking and particle flow} 
\label{sec:track}

The ILD track reconstruction is detailed in \cite{Gaede_2014,ILD:2020qve}. 
It is based on pattern recognition 
algorithms carried out independently in the different parts
of the ILD tracker system described in Section \ref{sec:ILCILD} (inner, forward and barrel regions). 
The pattern recognition step is followed by a combination of all the track 
candidates and segments for a final refit performed with a Kalman filter. 
This procedure relies on the detailed description of the detector material 
over all the surfaces that the particle has traversed, provided by the \texttt{DD4HEP}
framework. Dead material layers, such as cables, support structures or services, are accounted for in the simulation and reconstruction.
This is implemented in the \texttt{MarlinTrk} framework, part of the 
\texttt{ILCSoft} toolkit.

After the reconstruction of the charged particle tracks, the particle flow algorithm 
(PFA) is applied. The PFA applied is denominated \texttt{Pandora} \cite{Marshall:2015rfa}. The aim is to 
reconstruct every single particle generated in the 
collision and using the best information available in the
detector to determine its kinematics.
The reconstructed signals are clustered in single objects (associated with individual 
particles if the PFA works perfectly) which are denominated
particle flow objects (PFO).

\subsection{Vertex reconstruction} 
\label{sec:vtx}
After reconstructing all PFOs, a high-level reconstruction of vertices is carried out. This is done 
by the \texttt{LCFIPlus} package \cite{Suehara:2015ura}.
The primary vertex of the event is found in a tear-down procedure, starting with all 
tracks and gradually removing tracks with the largest $\chi^2$ -contribution 
up to a given $\chi^2$ -threshold.
In the second step, \texttt{LCFIPlus} tries identifying secondary vertices,
applying suitable requirements for invariant masses, momentum directions and $\chi^2$.
This is done before the jet reconstruction (described in Section \ref{sec:jet}),
using all tracks not associated with the primary vertex.
The procedure includes a rejection procedure for long-lived neutral particles decaying in two 
charged particles ($K_{S}^{0}\rightarrow \pi^{+}\pi^{-}$, photon conversion, etc.)
and a final step of refinement of the found vertices once the jet reconstruction 
is performed.
This refinement consists of two steps: reconstructing single tracks as pseudo 
vertices and recombining vertices within the same jet. For the first step, it 
is required that only one secondary vertex is found and one other track is found 
whose trajectory passes near the line connecting the primary and secondary vertex.
The track is tagged as a pseudo vertex if it 
satisfies some kinematic criteria further described in \cite{Suehara:2015ura}.
The second step is designed to recombine these pseudo vertices in the main vertex or 
recombine two vertices into one if it is kinematically compatible.
%once the loose filtering of neutral particles is 
%performed if the probability of the recombination is larger than the existence of the 
%two vertices (or one vertex plus one pseudo vertex).

While ILC has a 99\% % probability of reconstructing the relevant charged tracks, 
this probability decreases when the TPC segments of secondary tracks are
required to be connected to the VTX segments. This has been studied in detail in \cite{bilokin:tel-01826535}.
%While the absence of a significant offset is intrinsic and not recoverable, 
About half of the inefficiencies found in \cite{bilokin:tel-01826535} ($\sim5\%$) 
were due to bad associations of track segments reconstructed in different subdetectors of the ILD.
For tracks at low angles, $|\costheta|>0.9$, the VTX 
does not provide any information due to its limited angular coverage.
In this region, the FTD disks must fully perform the tracking reconstruction. The first FTD disks are placed at $\sim$20 cm from the 
interaction point. Therefore, due to multiple scattering in the material of the 
detector, the FTD alone gives an offset accuracy insufficient for average momenta. 
Also, mismatch of assignments of energy clusters  in transition regions (i.e. between endcap and barrel detectors)
contributed to the total inefficiency of track assignment to the reconstructed vertices.
A recovery procedure to match the tracks reconstructed only in the FTD but not associated with any reconstructed vertex was 
developed in \cite{bilokin:tel-01826535}. This study triggered the optimisation of the reconstruction tools 
in the newest release of the \texttt{ILCSoft} toolkit used in this document and for the latest benchmarking
studies of the ILD \cite{ILD:2020qve}. 
With the latest reconstruction tools, a value smaller than the $\sim5\%$ inefficiency 
is obtained, compared with the initial 10\% described in \cite{bilokin:tel-01826535}.
Further optimisations may be considered in the future and
including modifying the current 
geometry of the forward region and/or prolonging the first layer of 
the barrel.

\begin{figure}[!h]
  \centering
      \begin{tabular}{cc}
        \includegraphics[width=0.45\textwidth]{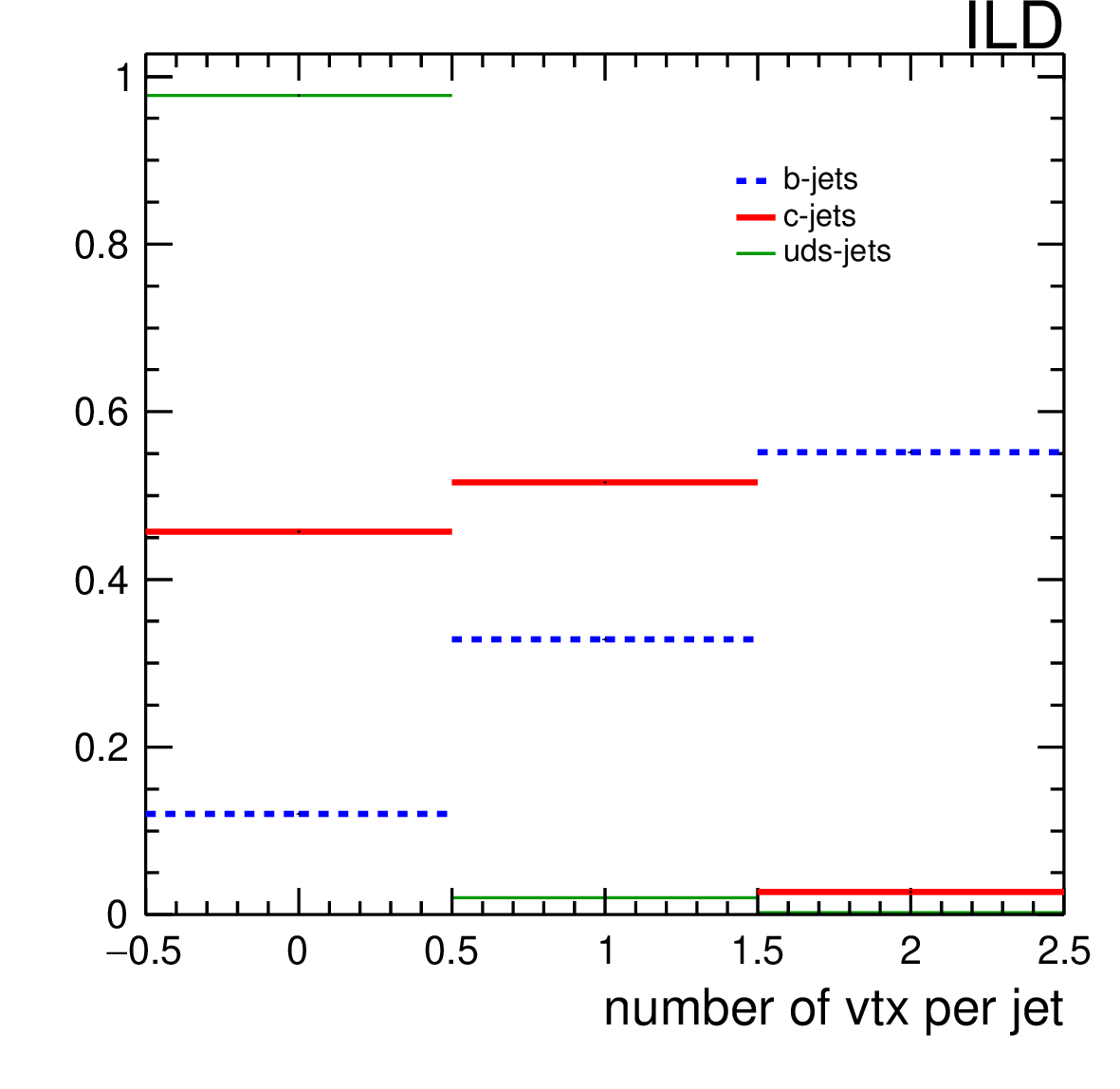} & \includegraphics[width=0.45\textwidth]{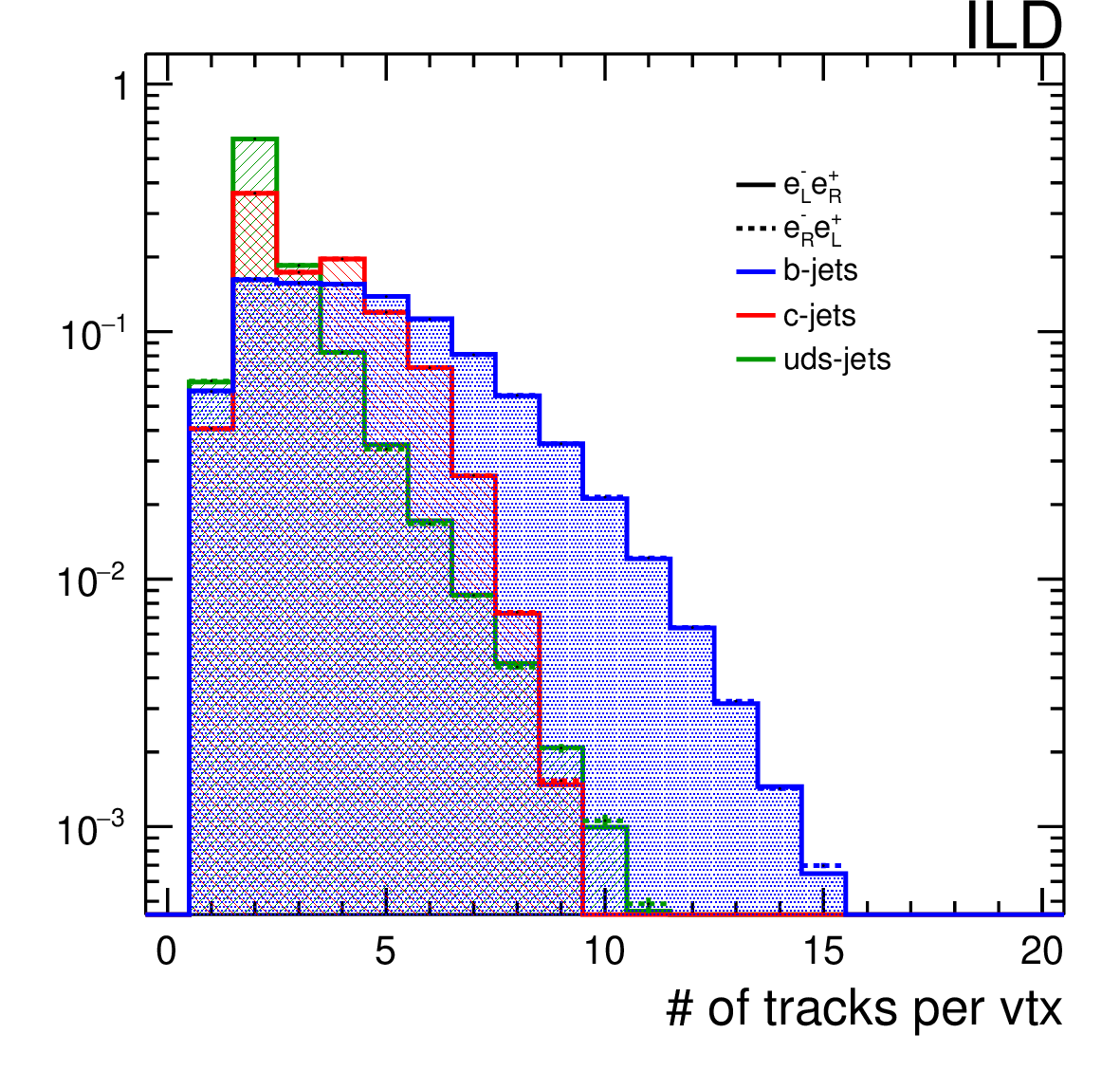} 
      \end{tabular}
      \caption{\label{fig:sectracks}
       Distribution of reconstructed secondary vertices per jet (left) and the total number of secondary tracks per reconstructed vertex (right) for different quark flavours. In both cases, the plot includes the pseudo-vertices made of only one track, as described in the text and in \cite{Suehara:2015ura}.
      }
\end{figure}

\begin{figure}[!h]
  \centering
      \begin{tabular}{cc}
        \includegraphics[width=0.4\textwidth]{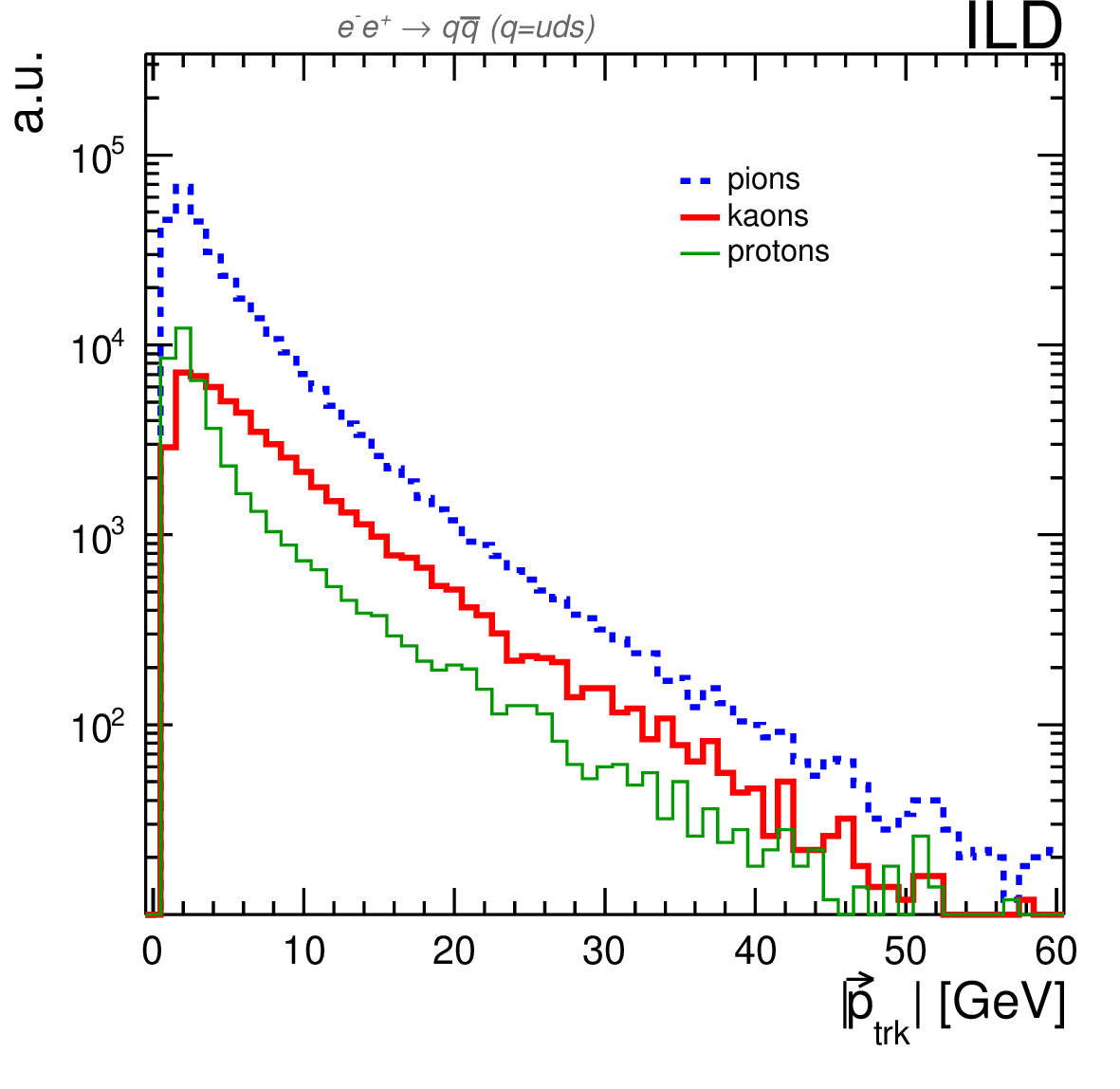}&
        \includegraphics[width=0.4\textwidth]{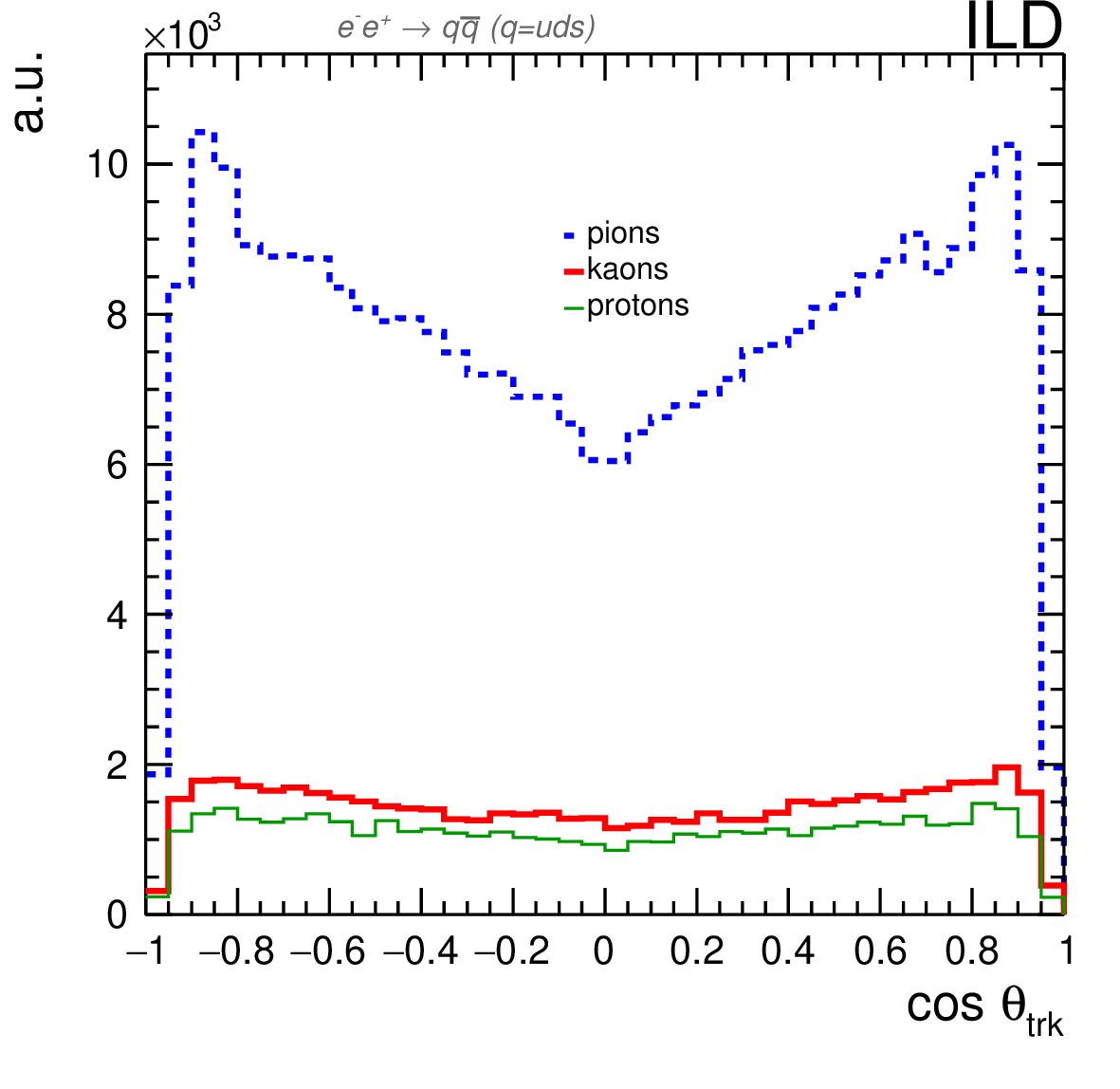} \\
        \includegraphics[width=0.4\textwidth]{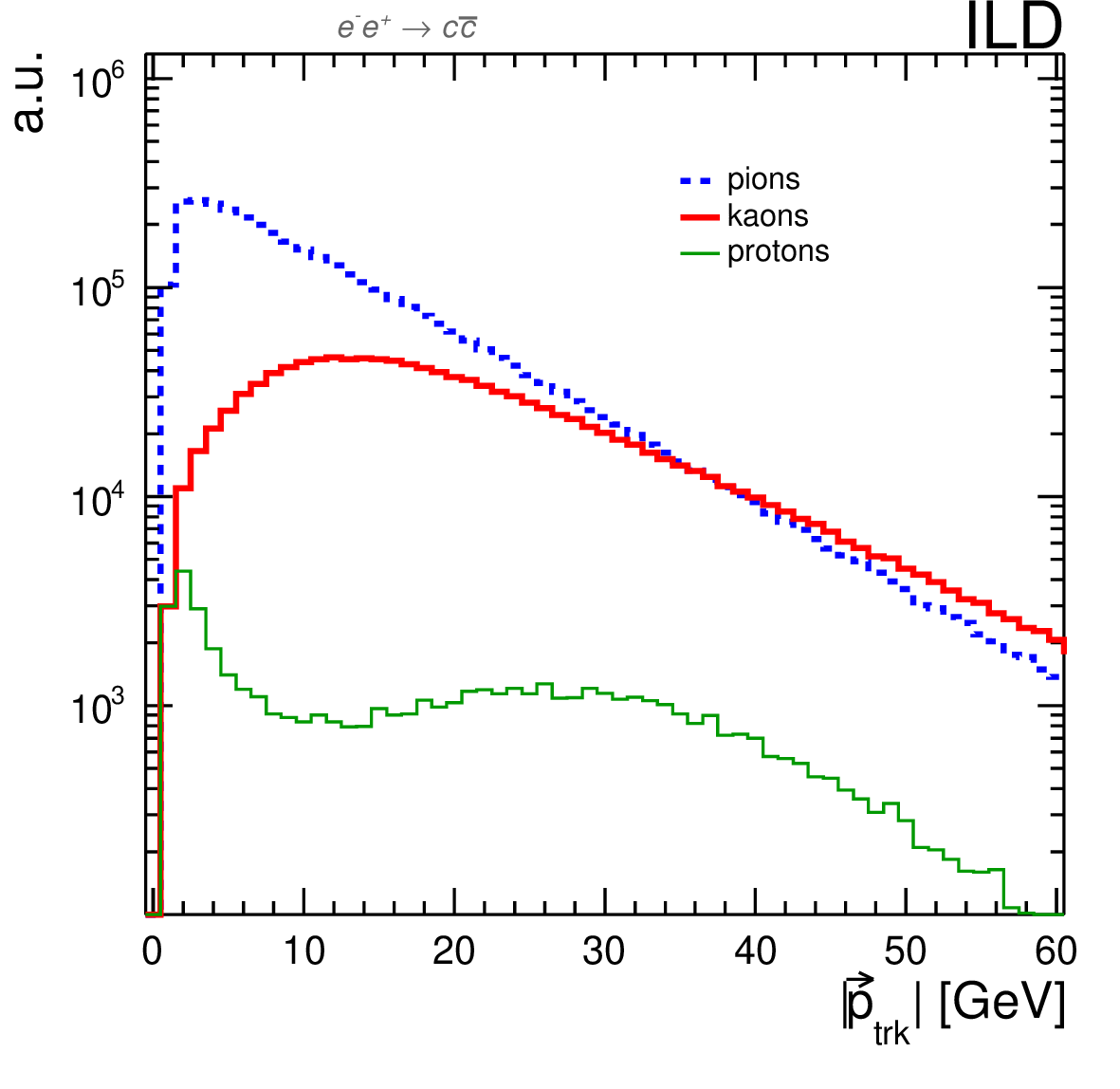} &
        \includegraphics[width=0.4\textwidth]{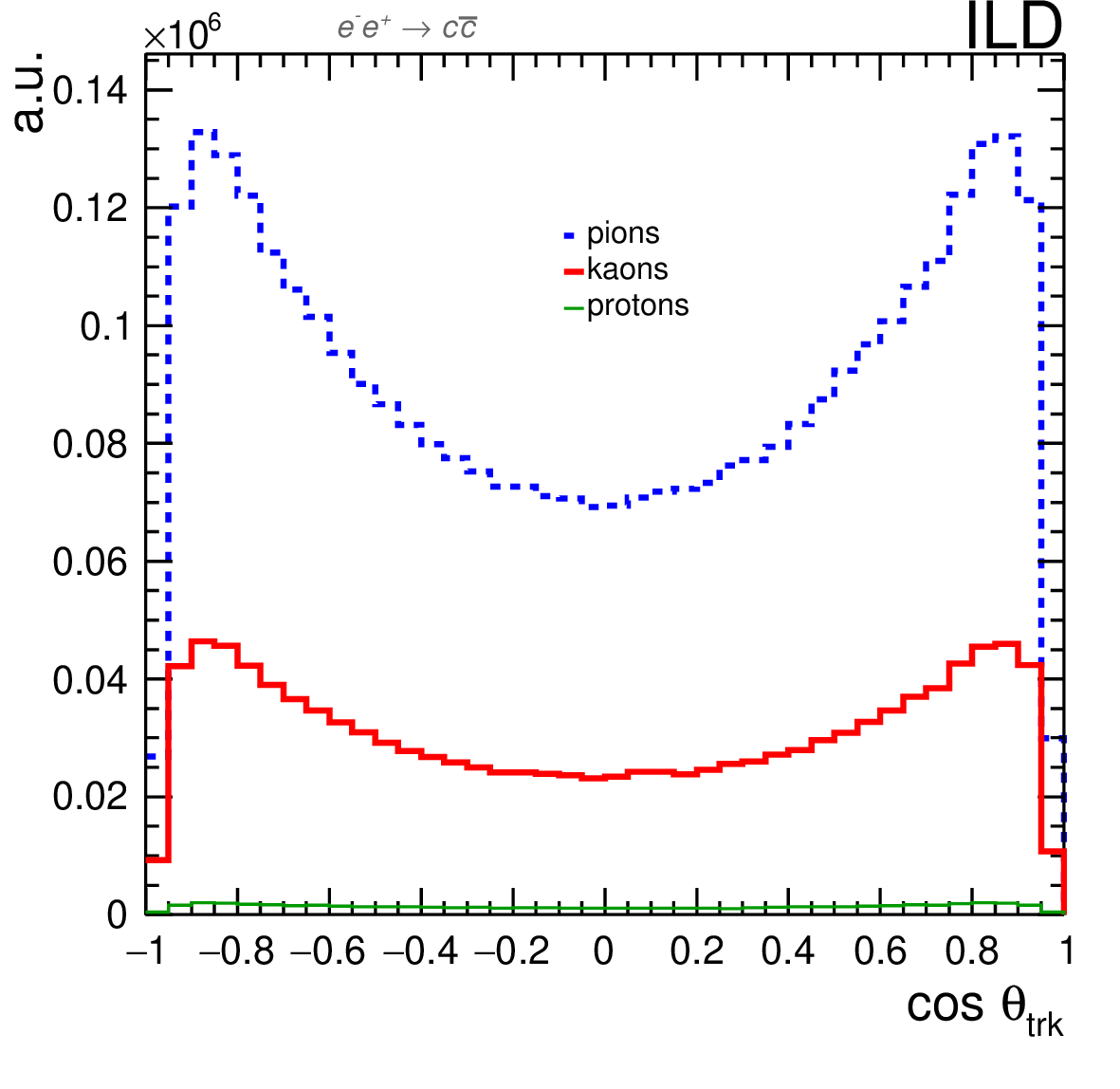} \\
        \includegraphics[width=0.4\textwidth]{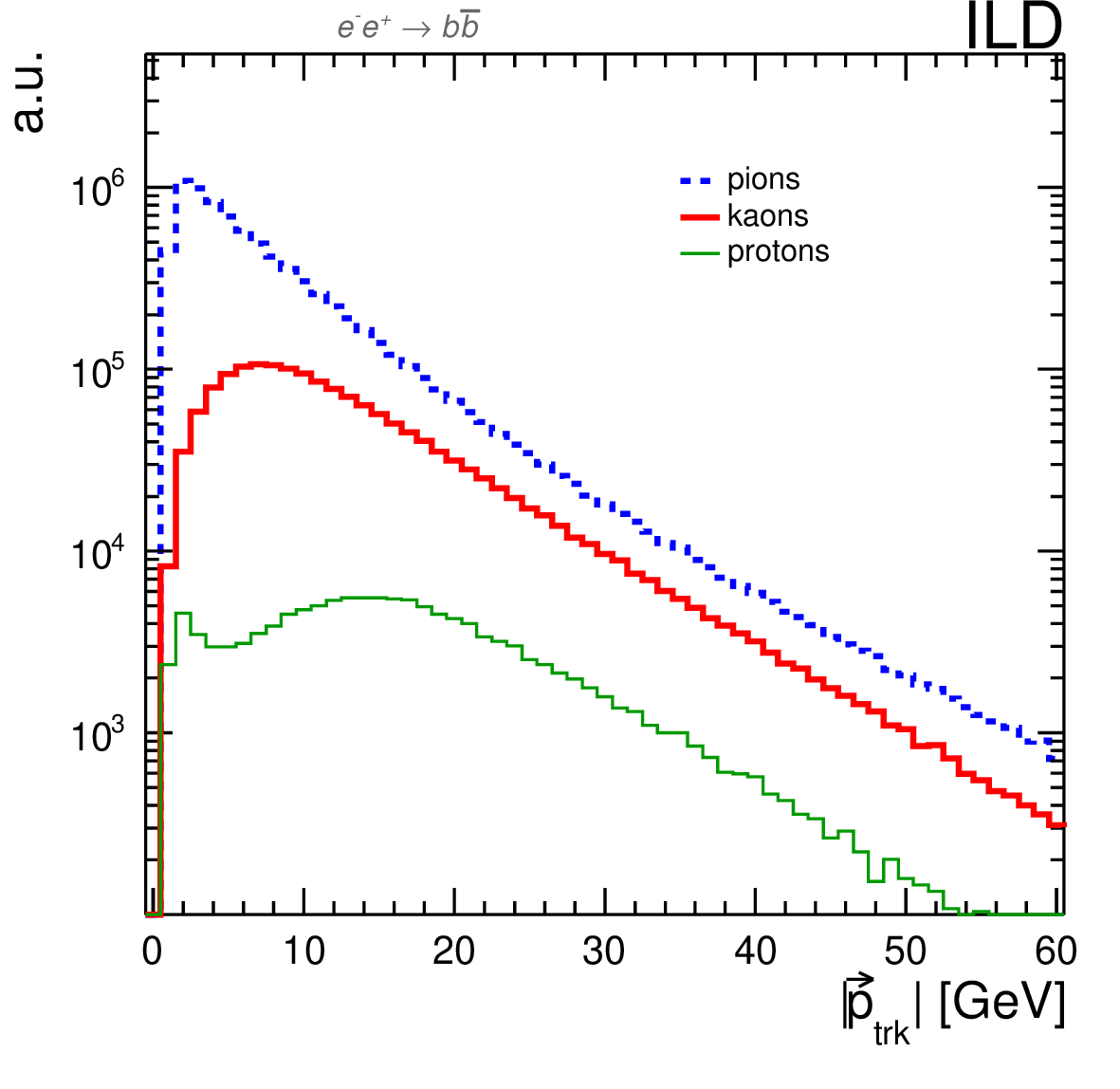} &
        \includegraphics[width=0.4\textwidth]{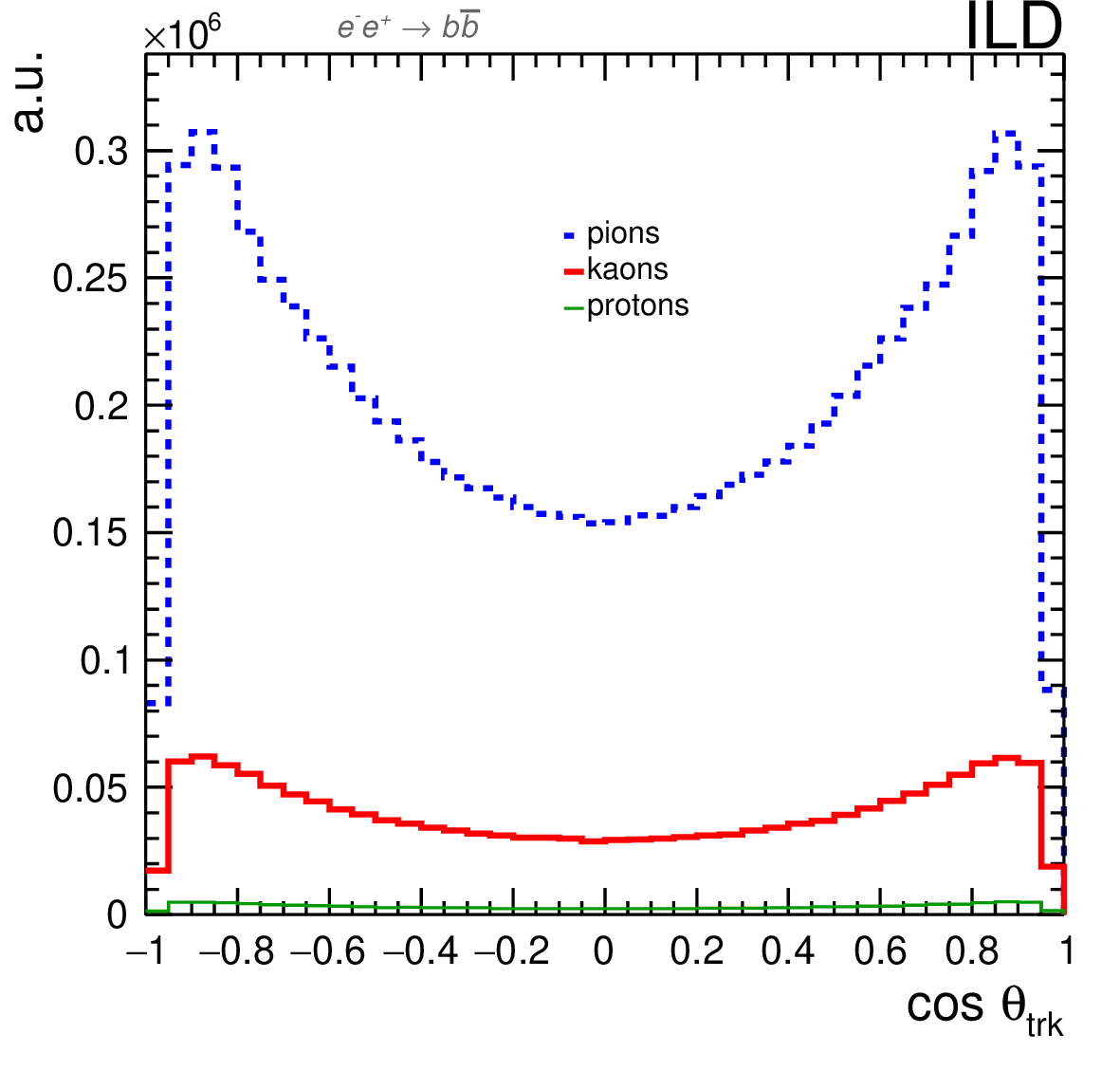} 
      \end{tabular}
      \caption{\label{fig:sectracks_kinemaitcs}
       Kinematic distributions for the momentum (left) and \costheta (right) for secondary tracks reconstructed in different \eeqq processes.
      }
\end{figure}

The analysis of the processes described in this note relies strongly on 
the precise reconstruction of primary and secondary vertices. 
It is, therefore, instructive to look at the number of secondary vertices and their related tracks.
The distributions of multiplicities of secondary vertices and secondary tracks per jet
reconstructed in different types of \eeqq 
events are shown in Figure \ref{fig:sectracks}. The plot on the left shows that, in more than $\sim95\%$ of the cases, no secondary vertices are reconstructed 
in events originating by light quark pairs.
In case of the \cquark, $\sim50\%$ of the time, a secondary vertex will
be reconstructed and only in $\sim5\%$ of the cases two vertices will be 
reconstructed.
In the case of the \bquark, up to 2 vertices per \bquark can be reconstructed with a 
probability of $\sim55\%$, and almost $90\%$ of the times at least one vertex will be 
reconstructed.
A set of kinematic distributions of the secondary tracks is shown in Figure \ref{fig:sectracks_kinemaitcs}.

\subsection{Jet reconstruction}
\label{sec:jet}

After the reconstruction of all the PFOs in the event and together with the vertex reconstruction, the jet clustering is performed. 
The  $VLC$ algorithm \cite{Boronat:2016tgd} for \ee colliders as implemented in \texttt{LCFIPlus} package is used.
This algorithm defines the following two distances, one between the PFO $i$ and $j$ and the other between the PFOs and the beam direction:
\begin{align}
    d_{ij} = 2min(E_{i}^{2\beta},E_{j}^{2\beta})\frac{1-\cos{\theta_{ij}}}{R^{2}} \\
    d_{iB} = E_{i}^{2\beta} \sin^{2\gamma}{\theta_{iB}}
\end{align}
with $\theta$ being the angle between the two PFOs. 
$R$, $\beta$ and $\gamma$ are free \textrm{ unitless} parameters of the algorithm. 
When $\beta=1$, $\gamma=0$ and $R=1$, the algorithm behaves in analogy with the 
Durham algorithm for clusters of PFOs far from 
the beam pipe but creates beam jets for clusters near the beam pipe. These beam jets are rejected, and 
their clusterisation depends on the value of the $R$ parameter.
In the following, $\beta=1$, $\gamma=0$ and $R=1$ and the algorithm in exclusive mode forcing it 
to always form two jets.

\subsection{Flavour tagging}
\label{sec:flavour_tagging}

\begin{figure}[!h]
  \centering
      \begin{tabular}{cc}
        \includegraphics[width=0.45\textwidth]{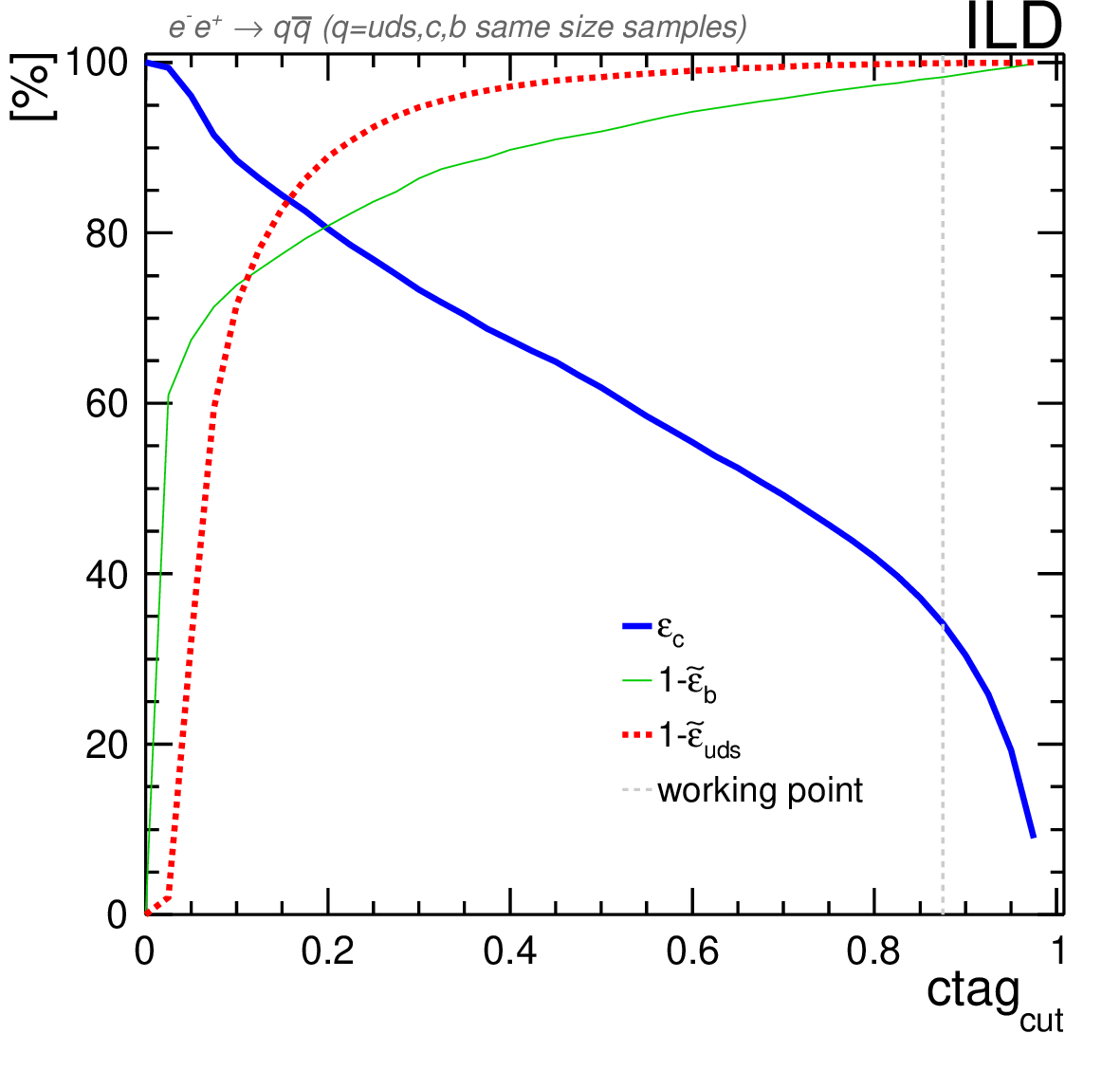}& 
        \includegraphics[width=0.45\textwidth]{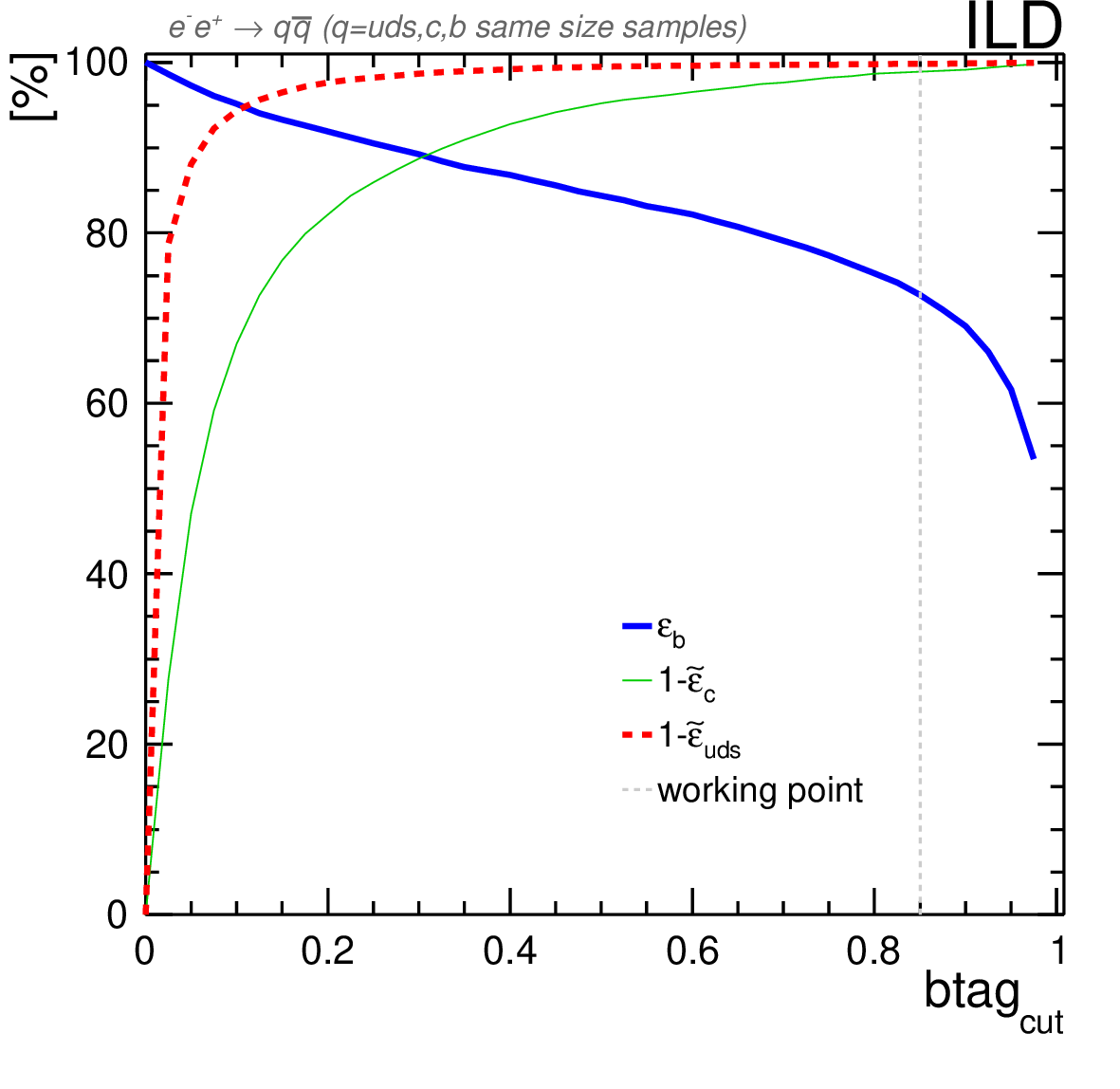} \\
    
      \end{tabular}
      \caption{\label{fig:btag} Left: efficiency of the correct \cquark tagging. Right: efficiency of the correct \bquark tagging. These numbers are estimated using Monte Carlo truth information in samples normalised to have the same number of events for each subset of $b$, $c$ and light ($uds$) quark flavours.
     }
\end{figure}

The \texttt{LCFIPlus} package also provides algorithms for jet flavour
tagging using boosted decision trees (BDTs) based on suitable
variables from tracks and vertices. 
The training of the BDTs is done with \eeqq events at $\cme=250$ GeV.
For each event, the jet flavour tagging algorithm gives a value of the \bquark 
(\cquark) likeness, $b_{tag}$ ($c_{tag}$), 
for the reconstructed jet. The efficiency of the $q$-quark tagging and its purity 
depends on the given
value of $q_{tag}^{cut}$ value that is required, such that $q_{tag}<q_{tag}^{cut}$.
This is shown in Figure \ref{fig:btag} for the \bquark and \cquark cases.
These numbers are calculated using samples with the same number of events for each 
subset of $b$, $c$ and $uds$ quark flavours.
The probability of correctly tagging a jet as originated by the assumed quark when 
using the \bquark tagging or \cquark tagging algorithms is shown in Figure \ref{fig:btag}. 
This probability is defined as $\varepsilon_{q}$. The $\tilde{\varepsilon}_{q\prime}$ 
accounts for the probability that when using a given
$q$ flavour tagging algorithm jets originated from other 
flavour $q\prime$ will be tagged as $q$.
As working points, the value of $b_{tag}>0.85$  and
the value of $c_{tag}>0.875$, are selected to keep the mistagging of other quarks below $\sim1\%$ while keeping maximal efficiency of correct tagging ($\sim 73\%$ for \bquark and $\sim34\%$ for \cquark).

\subsection{Particle Identification with \dEdx}
\label{sec:pid}

Both the \bquark and the \cquark frequently hadronise into charged kaons. Therefore, 
high performance of the charged kaon vs charged pion separation is crucial for the experimental study.

\begin{figure}[!ht]
\begin{center}
    \begin{tabular}{ccc}
      \includegraphics[width=0.3\textwidth]{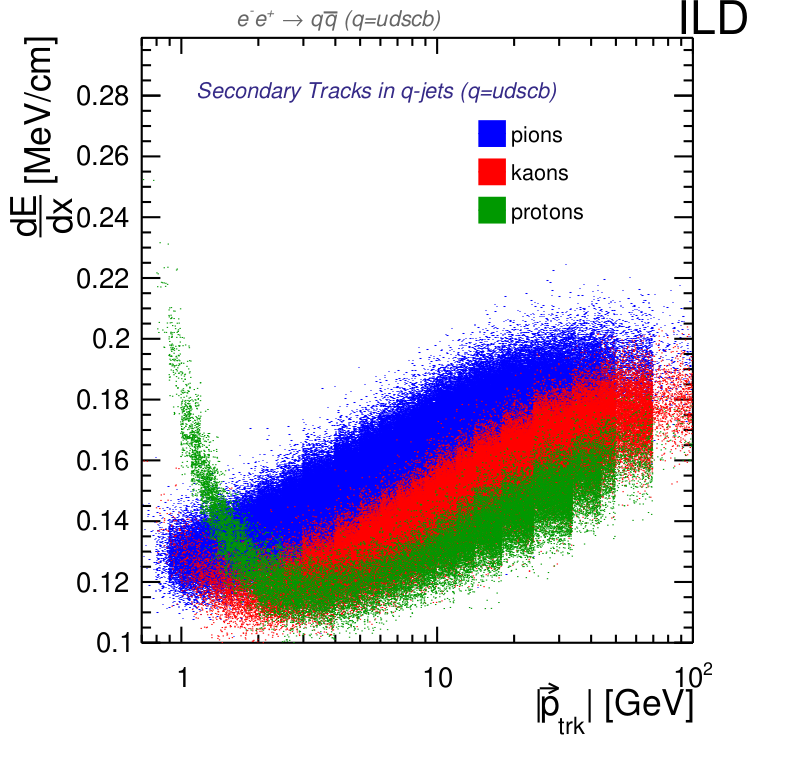} & %USE dEdx.eps for the final plot, but it has very large resolution!!
      \includegraphics[width=0.3\textwidth]{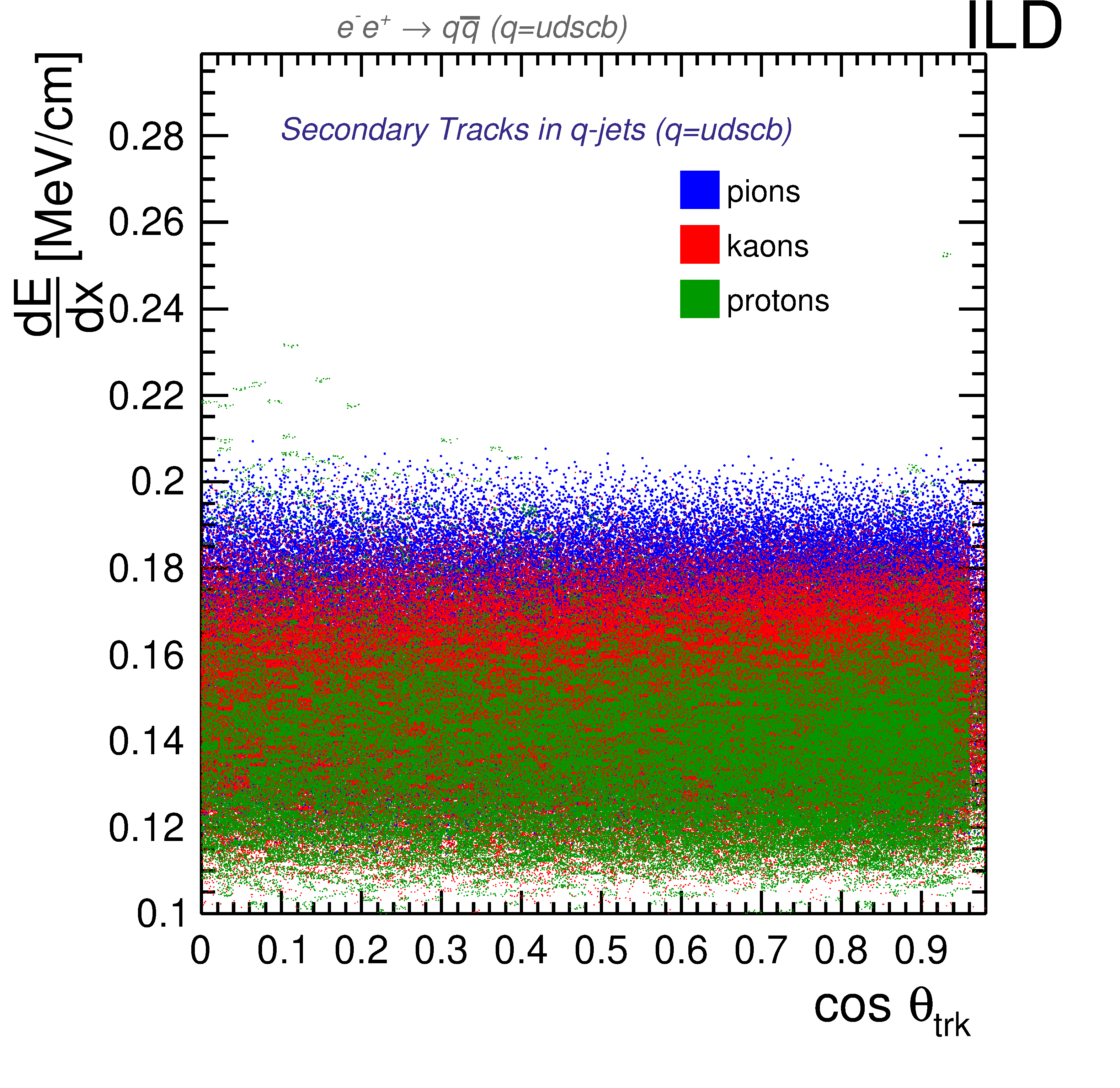} &
      \includegraphics[width=0.3\textwidth]{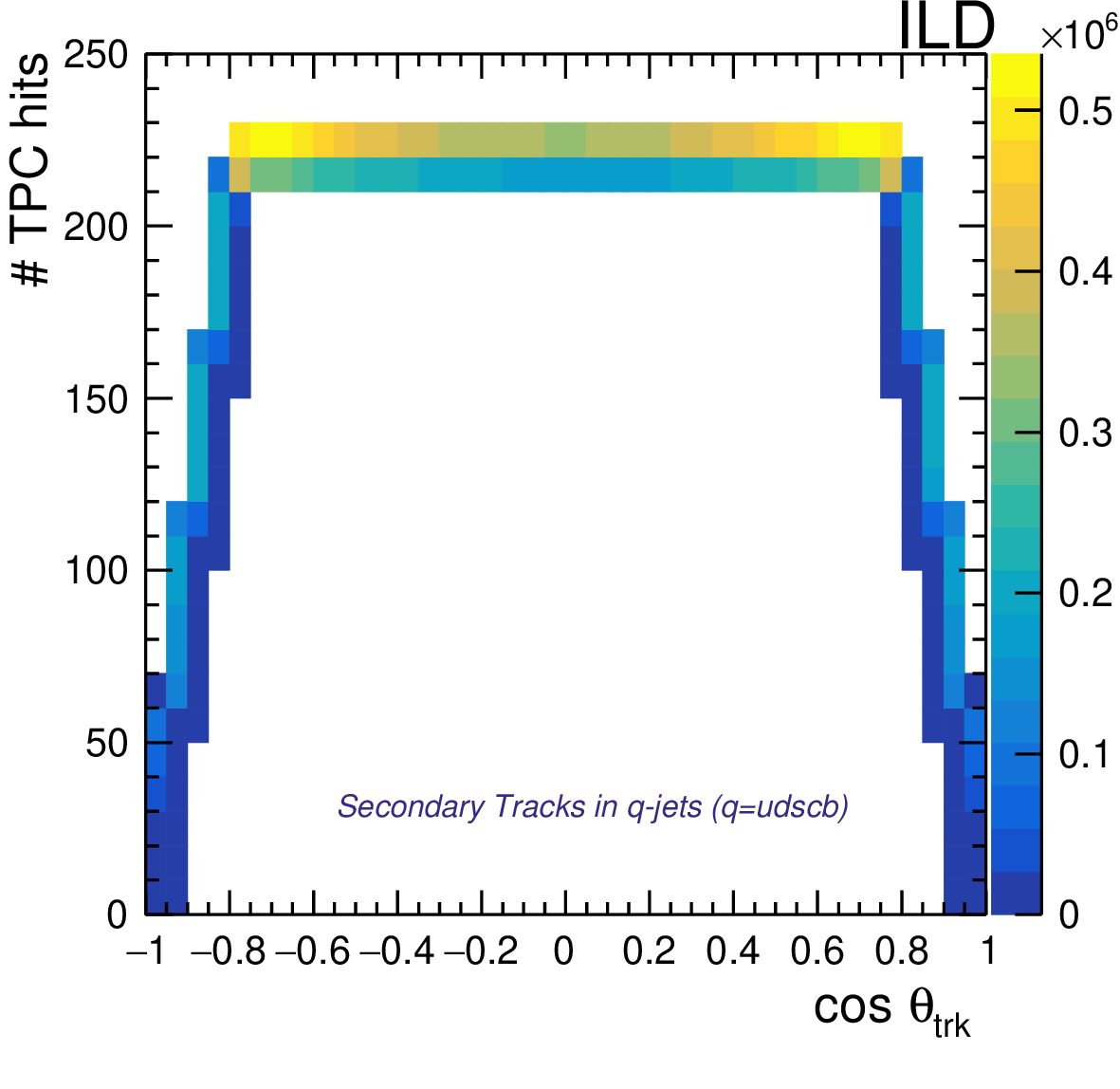} 
    \end{tabular}
\caption{Distribution of \dEdx as a function of the momentum of the secondary track (left plot) and the \costheta of the secondary track (middle plot) for different types of charged hadrons originated in secondary decays. The middle plot shows that the \dEdx mean value is stable over \costheta. The third figure shows the number of reconstructed TPC hits for tracks reconstructed at different angles. 
\label{fig:dEdx}}
\end{center}
\end{figure}

The left plot in Figure \ref{fig:dEdx} shows the \dEdx reconstructed from a truncated mean for charged
particle tracks in the TPC as a function of the particle momentum for different types of charged hadrons after correcting for an angular dependence of \dEdx 
\cite{bilokin:tel-01826535}. The relative resolution of \dEdx is about 4.5\%, adjusted to meet the measured resolution in 
beam test \cite{ILD:2020qve,LCTPC:2022pvp}.
These simulations, validated with beam test studies, show that a separation power between 
charged pions and kaons larger than 3
is possible for tracks with momentum larger than 3 GeV (see \cite{ILD:2020qve}, Figure 8.6). 
However, the TPC information only partially separates charged kaons and protons.
%ALEPH, with a larger TPC than DELPHI, has achieved similar resolution as ILD, but could not use it for $K^{\pm}/\pi^{\pm}$ separation because of its moderated readout granularity.  
%Other approaches, as the cluster counting approach, are under scrutiny at this moment by 
%the ILD and the Linear Collider TPC collaboration (LCTPC).
%A improvement of 30-50\% on the separation power of charged particles 
%seems feasible \cite{LCTPC:2022pvp}.
The middle plot in Figure \ref{fig:dEdx} shows the \dEdx as a function of the polar angle of the reconstructed track.
The third plot of the same figure shows the number of hits reconstructed in the TPC
compared with the angular distribution of the track. The
acceptance drop at large angles is due to the finite size of the barrel-shaped TPC. It could be noted here that due to the increased length of the track at $|\costheta|=0.8$, the particle identification capabilites have a maximum here which then quickly drops due to the acceptance losses at large angles. This is reflected in Figure \ref{fig:dEdx_eff_kineamtics}.

\begin{figure}[!ht]
\begin{center}
    \begin{tabular}{cc}
      \includegraphics[width=0.45\textwidth]{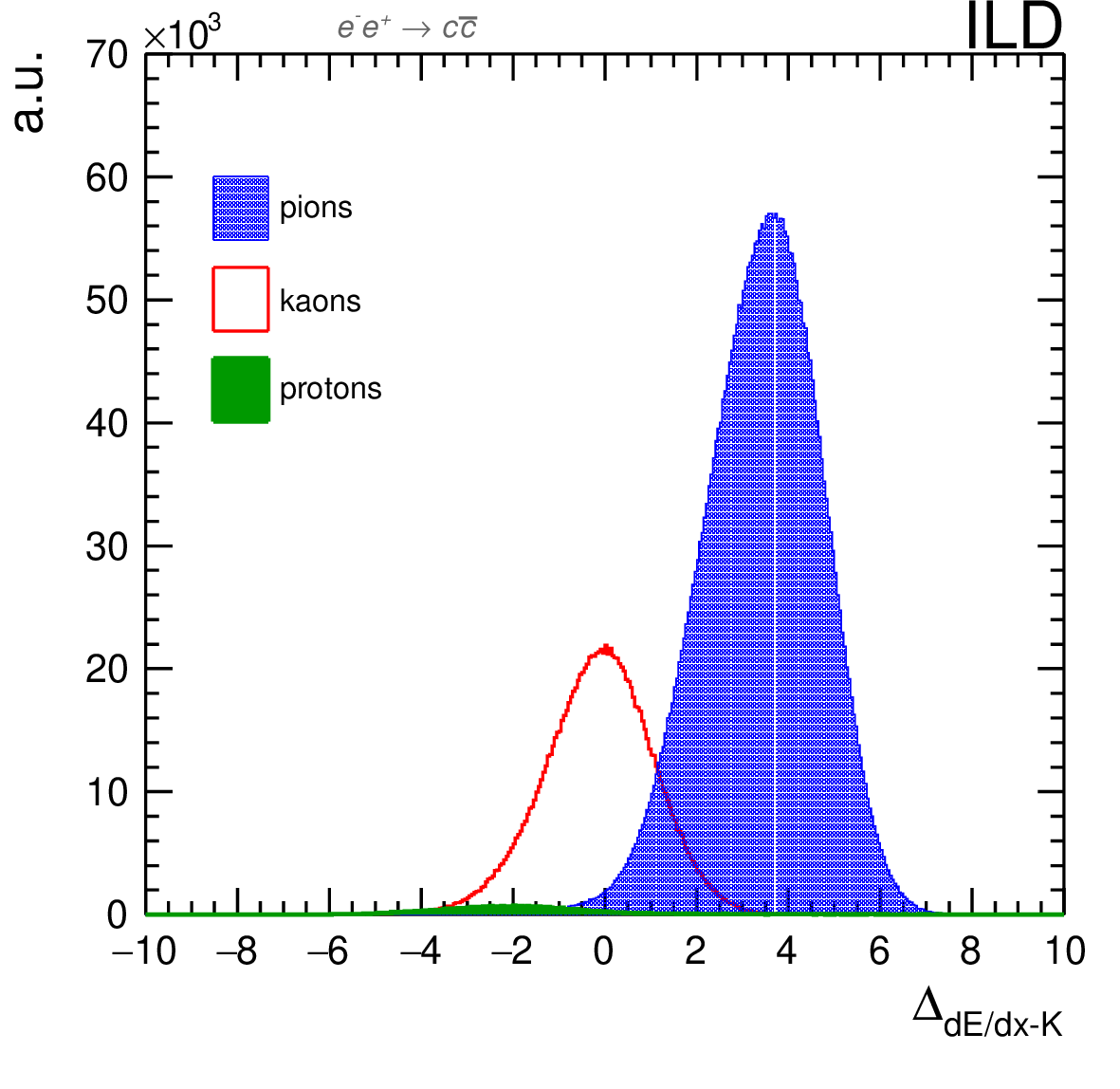} &
      \includegraphics[width=0.45\textwidth]{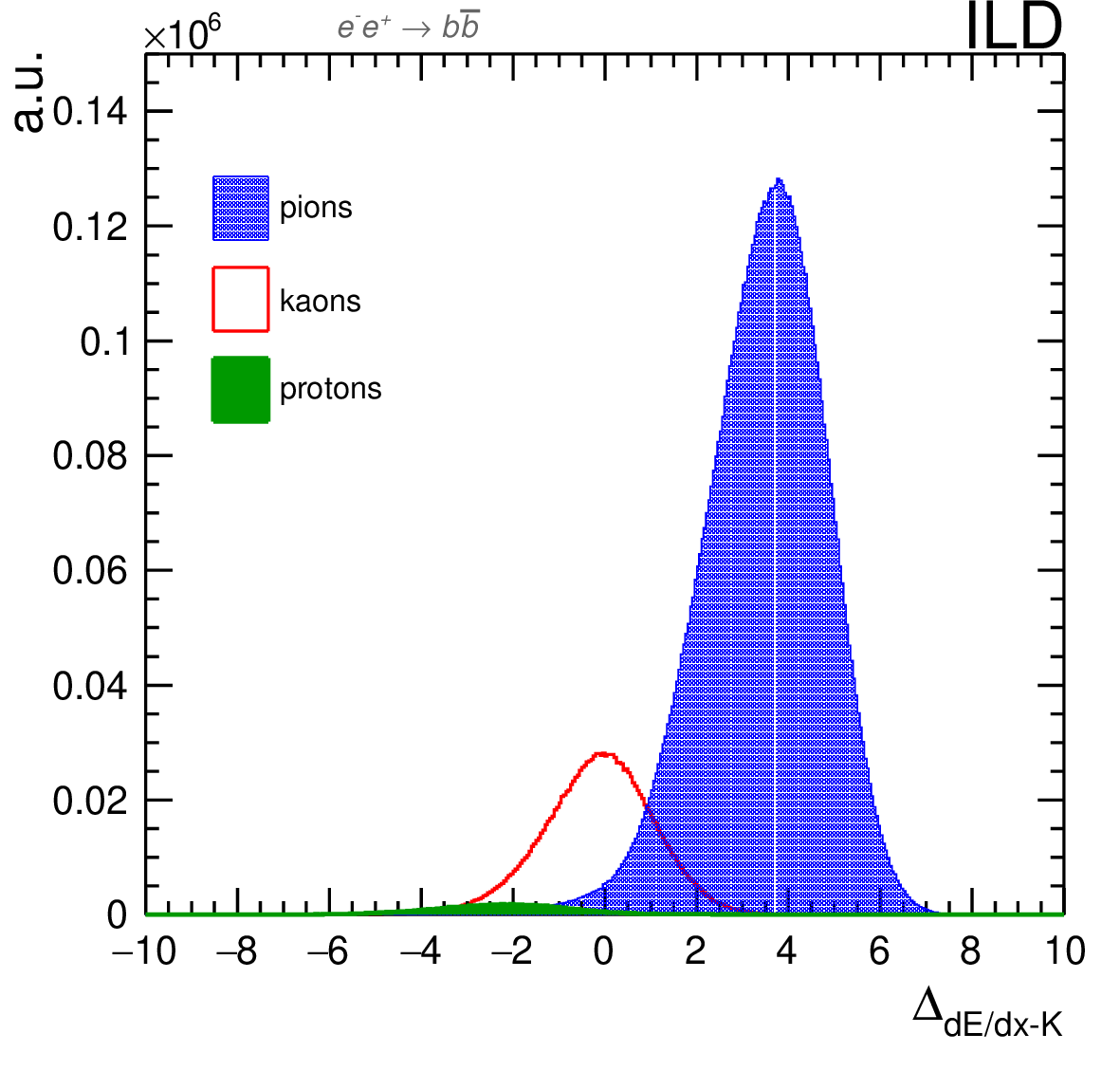} 
    \end{tabular}
\caption{Distribution of $\Delta_{\dEdx-K}$ for different processes (\eecc in the left and \eebb in the right). 
\label{fig:dEdx_dist}}
\end{center}
\end{figure}

\begin{figure}[!ht]
\begin{center}
    \begin{tabular}{cc}
      \includegraphics[width=0.45\textwidth]{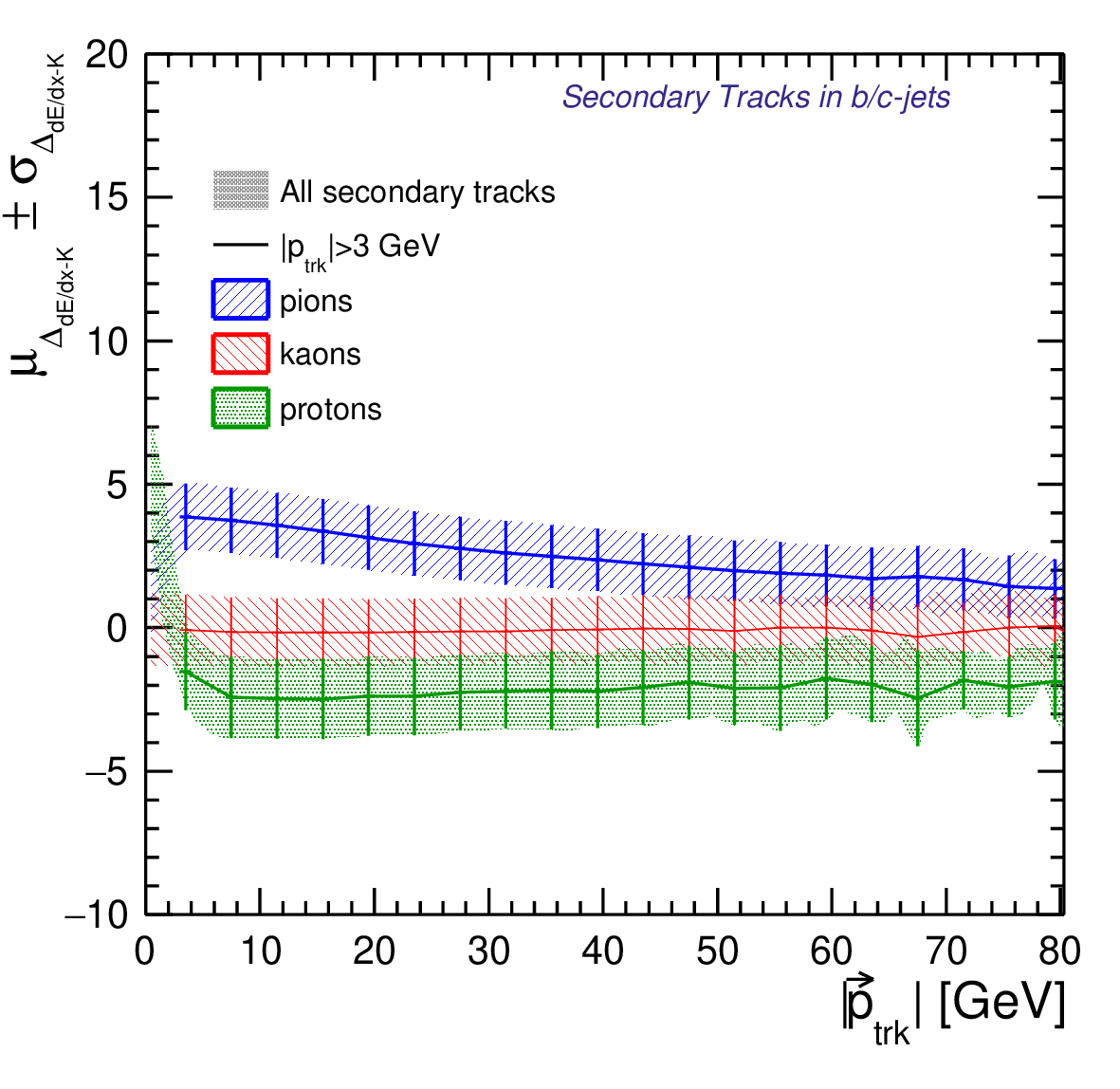}&
      \includegraphics[width=0.45\textwidth]{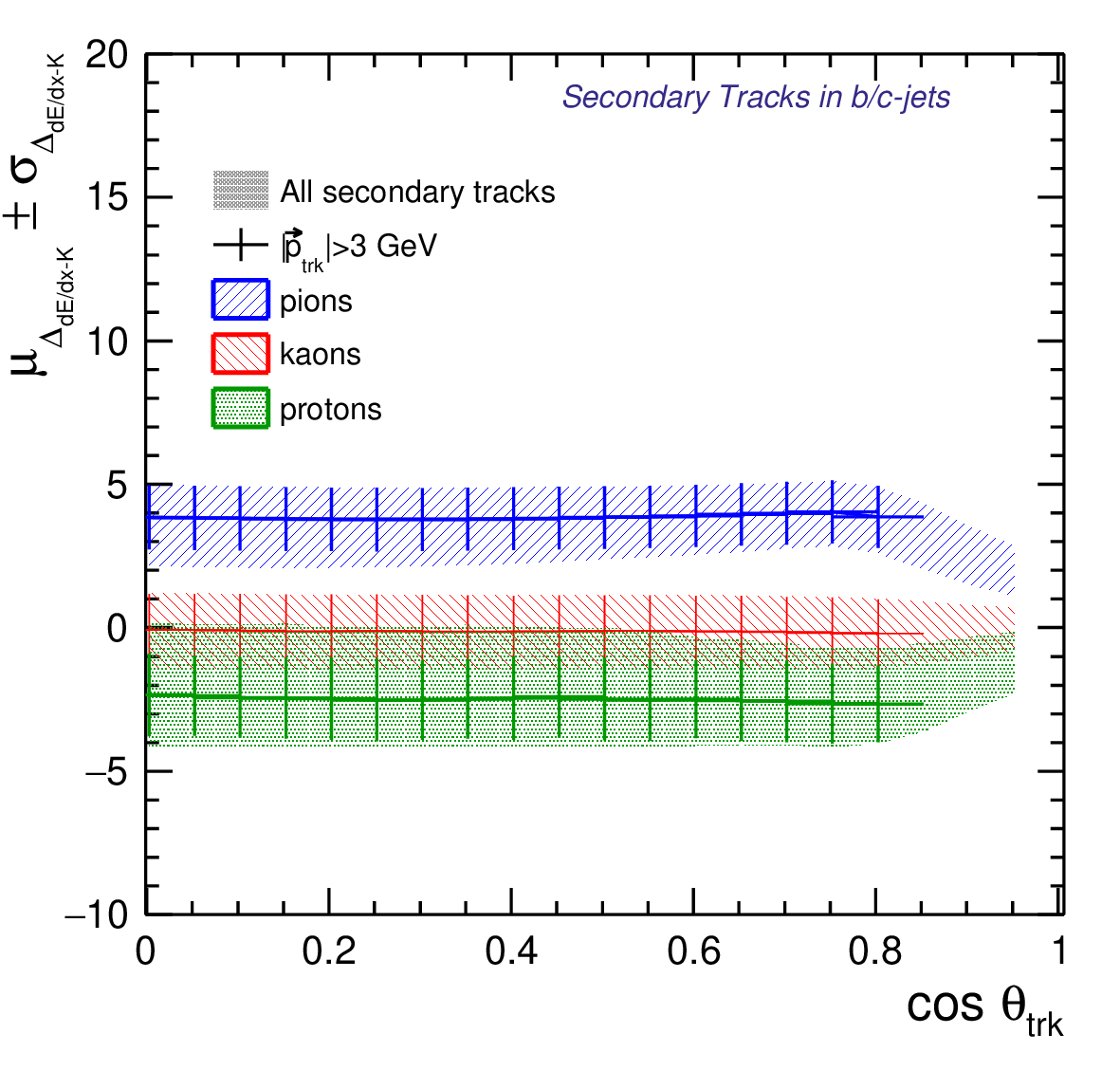}       
      \end{tabular}
\caption{Projection of the $\Delta_{\dEdx-K}$ distribution as a function of the momentum of the track (left column) or the \costheta of the track (right column). The different colour shaded areas correspond to different types of charged hadrons. The lines with error bars show the same information but for tracks with momentum larger than 3\,GeV to avoid the area in which the \dEdx distributions of the pions and kaons overlap.
\label{fig:dEdx_dist_projection} }
\end{center}
\end{figure}

\begin{figure}[!ht]
\begin{center}
    \begin{tabular}{cc}
      \includegraphics[width=0.35\textwidth]{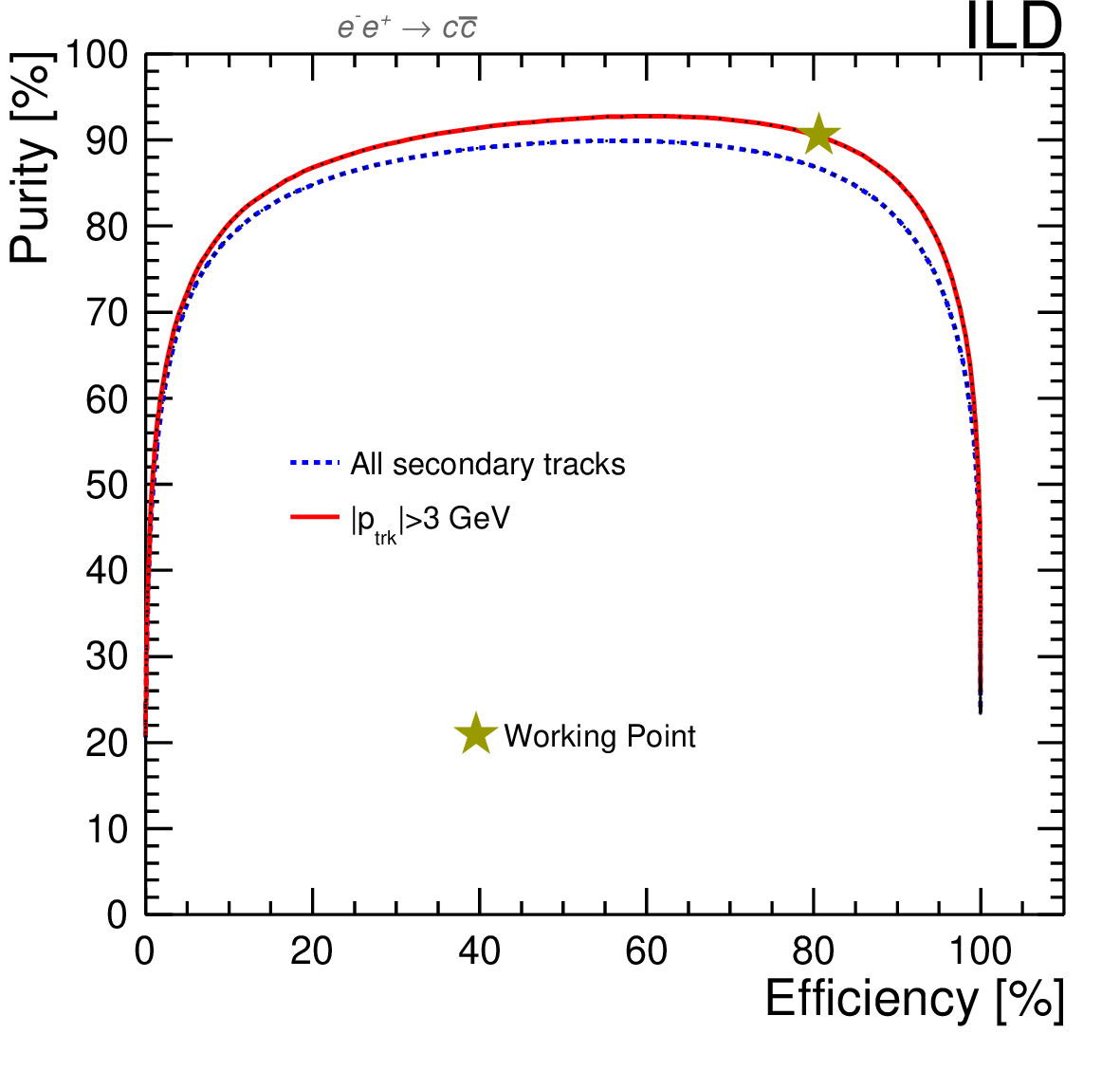} &
      \includegraphics[width=0.35\textwidth]{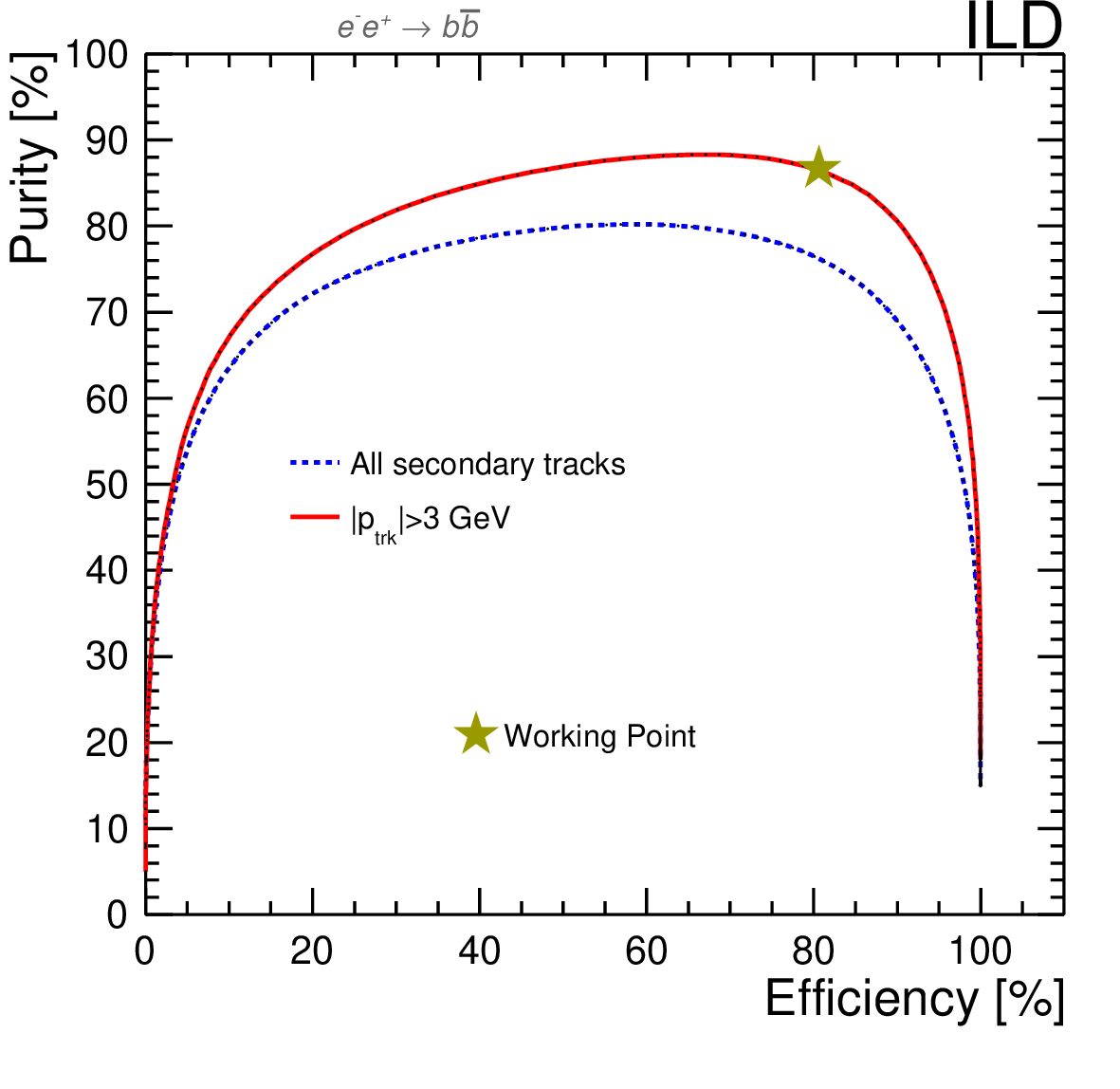} \\    
      \includegraphics[width=0.35\textwidth]{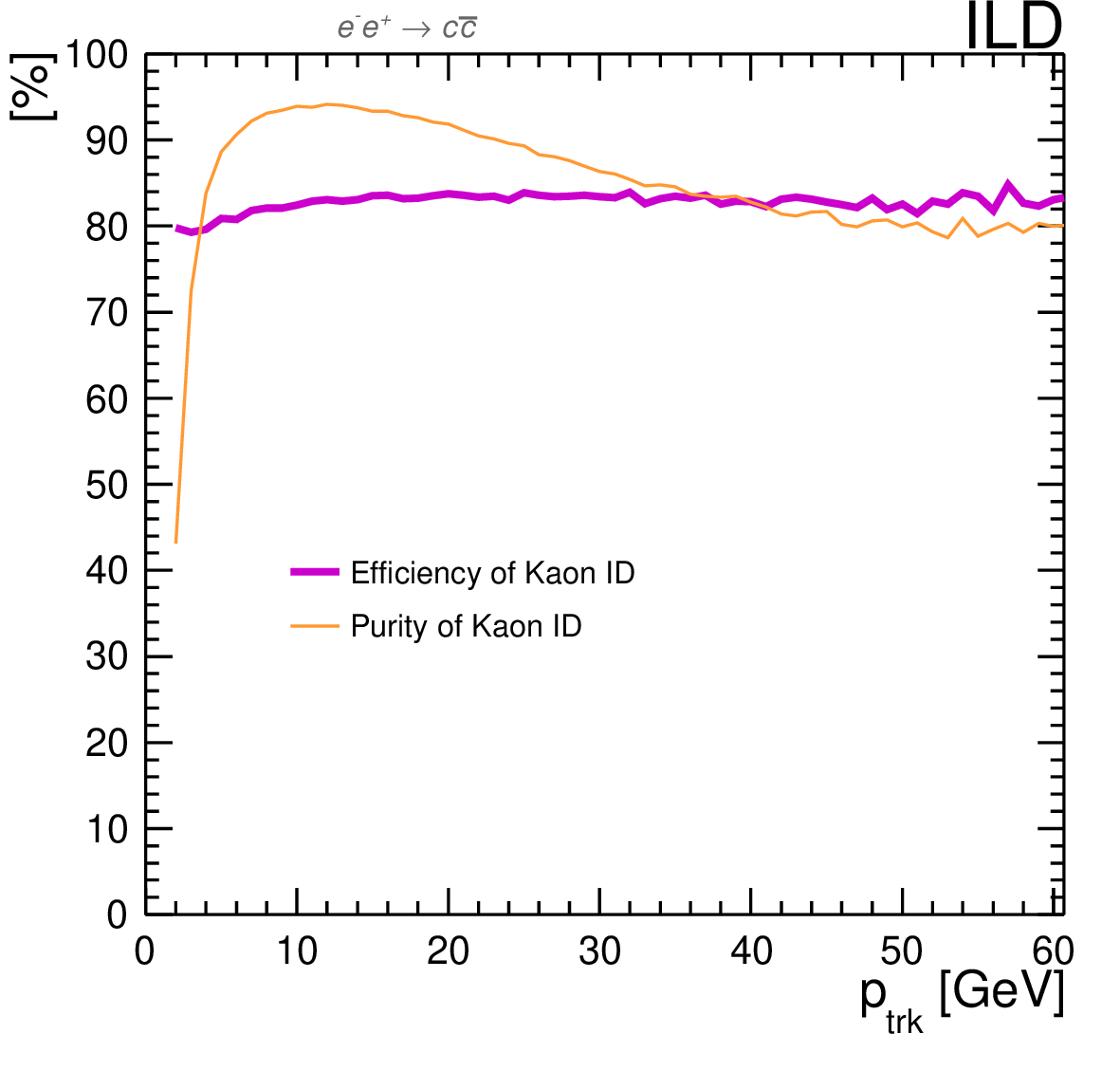} &
      \includegraphics[width=0.35\textwidth]{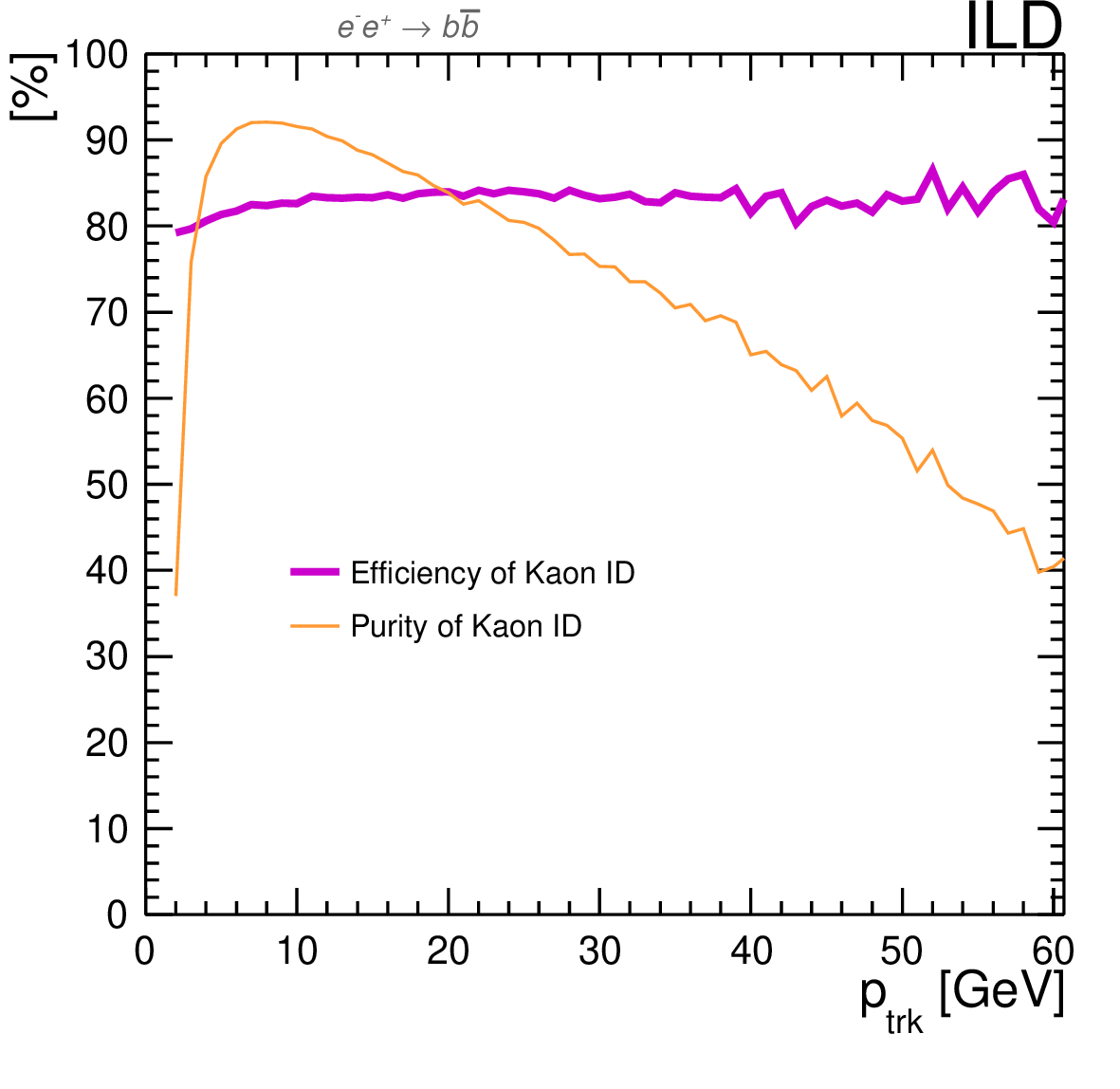} \\
      \includegraphics[width=0.35\textwidth]{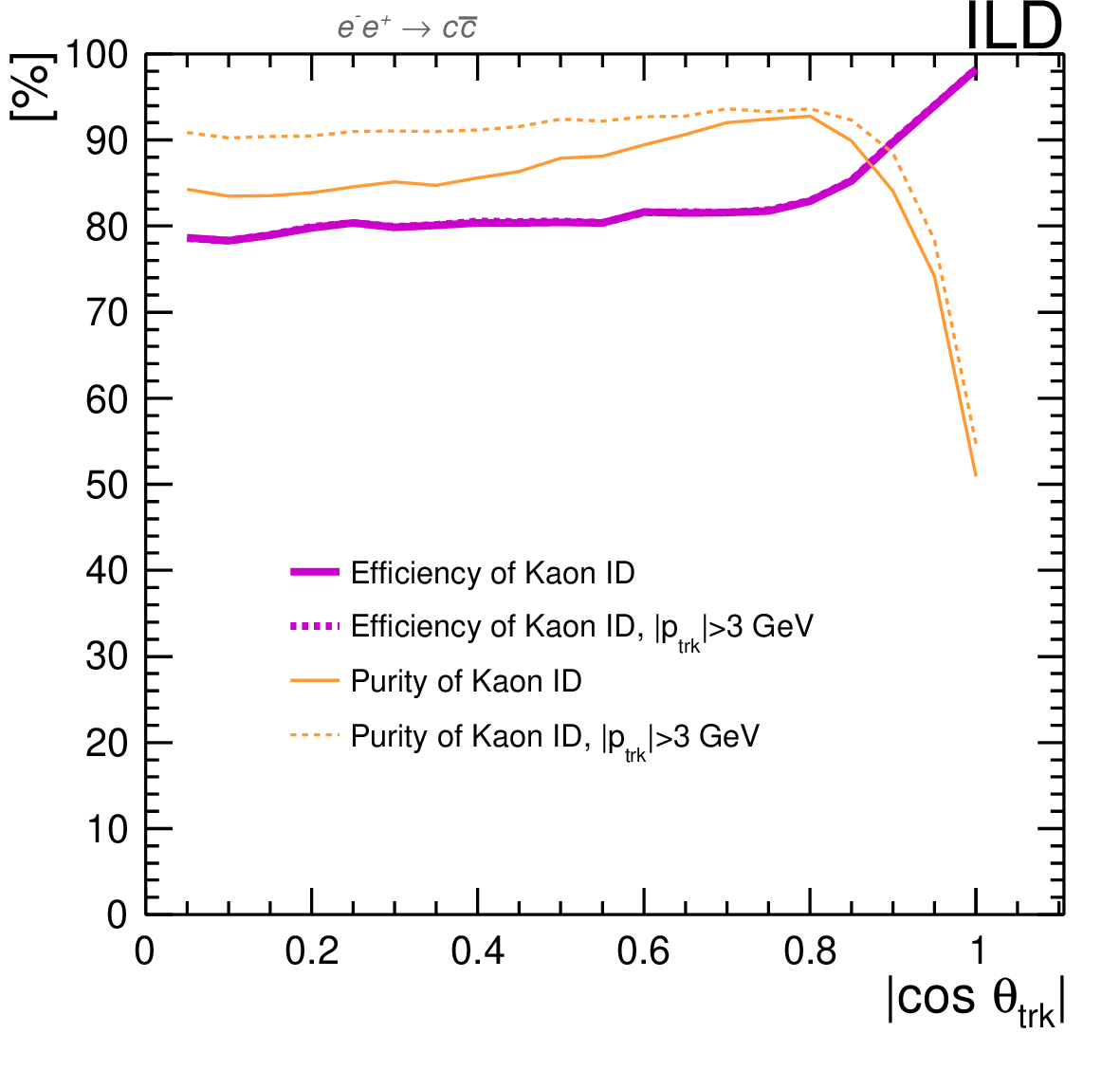} &
      \includegraphics[width=0.35\textwidth]{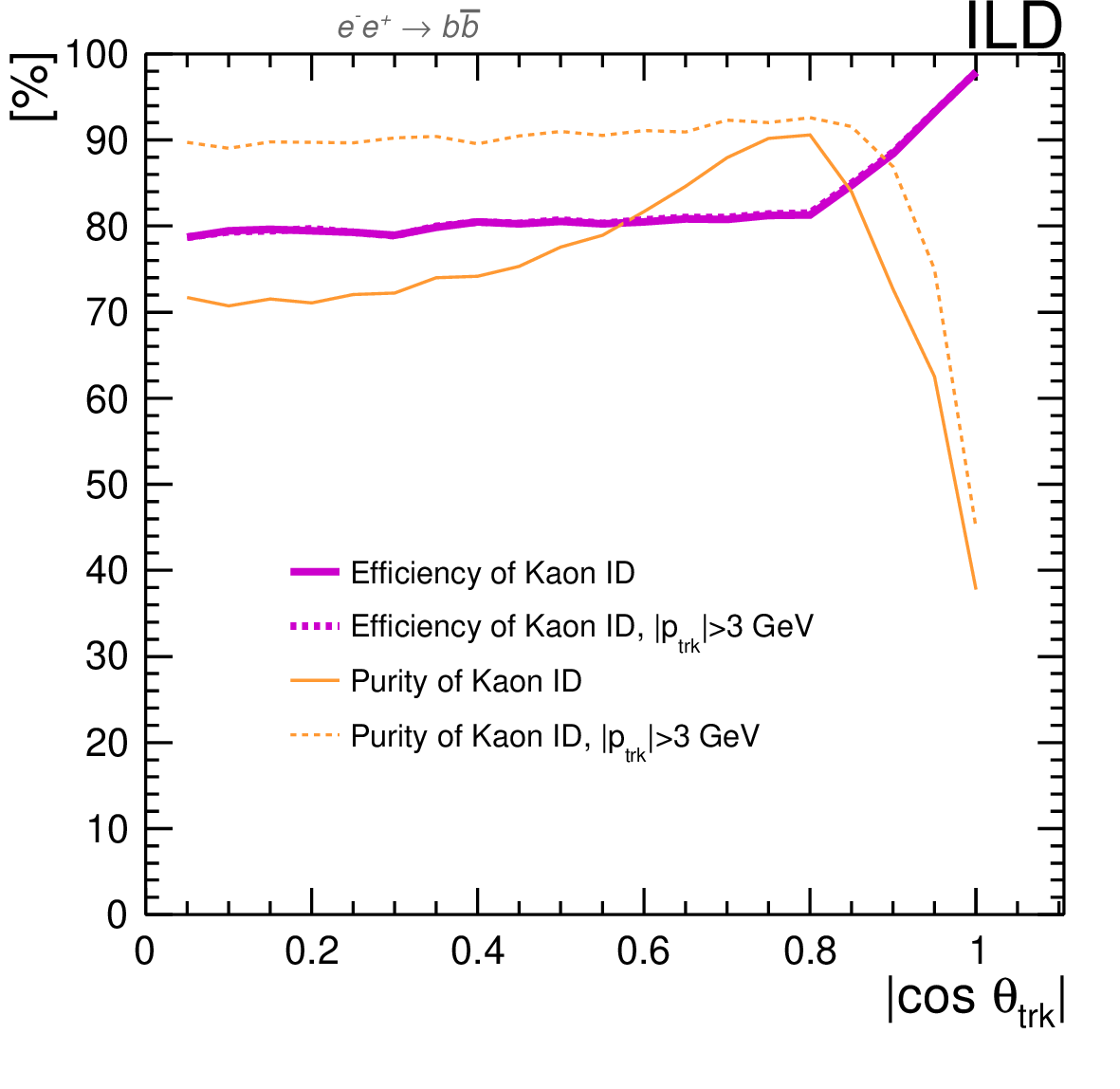} 
      \end{tabular}
\caption{Upper row: estimated efficiency and purity of charged kaon selection for \ccbar (left) and \bbbar (right) processes. The purity and efficiency is estimated from a 1-dimensional optimisation as described in the text. Middle row: same, as a function of the track's momentum. Lower row: same, as a function of the \costheta of the track. The differences in the purity vs momentum for $c$ and $b$-quarks can be explained by the difference in the momentum distribution of the different types of charged hadrons (see Fig.~\ref{fig:sectracks_kinemaitcs}). The drastic change in the purity vs polar angle for both types of quarks when requiring a momentum larger than 3\,GeV is due to the removal of the region on \dEdx vs momentum where the kaons and pions overlap.
\label{fig:dEdx_eff_kineamtics}}
\end{center}
\end{figure}

The charged Kaon identification is optimised 
using the \textit{kaonness} variable, $\Delta_{\dEdx-K}$, 
of a charged particle fully reconstructed in the TPC. It is defined as:
\begin{align}
   \Delta_{\dEdx-K}=\left(\frac{\dEdx_{exp} - \dEdx_{K,BB}}{\Delta\dEdx_{exp}}\right)
\end{align}
where $\dEdx_{exp}$ is the measured \dEdx for a given track, $\dEdx_{K,BB}$ is the expected \dEdx 
fitted by a Bethe-Bloch formula whose parameters are obtained from the simulation
and $\Delta\dEdx_{exp}$ is the expected experimental uncertainty for the \dEdx measurement, 
as obtained in beam test and simulation studies \cite{LCTPC:2022pvp}.
%These distributions, keeping the original sign of the term $(\dEdx_{exp} - \dEdx_{K,BB})$,
These distributions are shown in Figure \ref{fig:dEdx_dist} for two different assumptions
of the initial quark pair generated in the event. For both cases, only secondary tracks are considered. 
This variable is used to optimise the 
charged kaon and pion separation (the kaon proton separation is not a problem for the studies presented here 
due to the low proton multiplicity in secondary tracks originated by \cquark and \bquark fragmentation processes).
In Figure \ref{fig:dEdx_dist_projection} the mean value and \textit{rms} of the \kaonness 
The considered track's momentum and angle are shown as a function of the momentum and angle.
For completeness, this projection is shown for the cases of \eebb and \eecc although no difference is observed between them, as expected.
A minimum momentum of 3\,GeV is required to optimise the charged kaon-pion separation. 

The required \kaonness is thus optimised for the best identification of charged kaons reconstructed as secondary tracks.
Varying the requirement on $\kaonness<Y$, with arbitrary $Y$, the achievable efficiency and purity of charged kaon identification are estimated. 
The result of this study is shown 
in the first row of Figure \ref{fig:dEdx_eff_kineamtics}. 
It is found that with $-2.45<\kaonness<1.1$, charged kaons can be identified
with a global efficiency of 80\% and a global purity larger than 90\%.
Using this working point, efficiency and purity\footnote{Efficiency is defined as the ratio of the number of correctly identified kaons over the total number of kaons in the sample. Purity is defined as 1 minus the 
the ratio of wrongly identified as kaons over all particles identified a kaons.} 
are studied as a function of the measured momentum and angle of the tracks, finding that the efficiencies and purities remain constant 
in a broad range of the momentum and angle of the tracks.
This is shown in the last two rows of Figure \ref{fig:dEdx_eff_kineamtics}. 

%% file: sections/5_analysis_preselection.tex
\section{Event preselection}
\label{sec:analysis_preselection}

As described in the previous section, the event reconstruction and preselection start from a sample of 
fully reconstructed events of two jets.
The angular distribution of the jets is reconstructed in the reference frame of the thrust-axis of the event, calculated 
with all reconstructed particles as input for the formula from the footnote \ref{footnote:T} (page 4).
As suggested by Tables \ref{tab:crosssection} and \ref{tab:crosssection_bkg} the 
main contamination source 
is the radiative return events.
The second source of background contamination is pairs of SM heavy bosons
$HZ/ZZ/WW$ producing quarks in the final state.
Furthermore, the choice of cuts has been carefully designed to affect almost equally the five quark flavours to minimise modelling dependency, as explained in Section \ref{sec:analysis_doubletag}.

\subsection{Cuts against radiative return events}
\label{sec:analysis_preselection_radiative}

Most of the ISR photons will be collinear to the beam direction, and hence the reconstructed energy of the event will be much smaller than 250 GeV.
However, for the rest of the cases, the photon (or photons) will be reconstructed (fully or partially) 
in the detector volume.
In both cases, the two jets will no longer be back-to-back. 
In order to remove the ISR events,  a two-step selection procedure is performed: first, a veto on events
that have signals from the ISR photons in the 
detector volume; second, a veto on events that have no reconstructed photons but the topology of the two jets is not back-to-back.

\subsubsection*{ISR photon removal}%Identification of high energy ISR photons}
\label{sec:photon}

\begin{figure}[!pt]
  \centering
      \begin{tabular}{ccc}
        \includegraphics[width=0.3\textwidth]{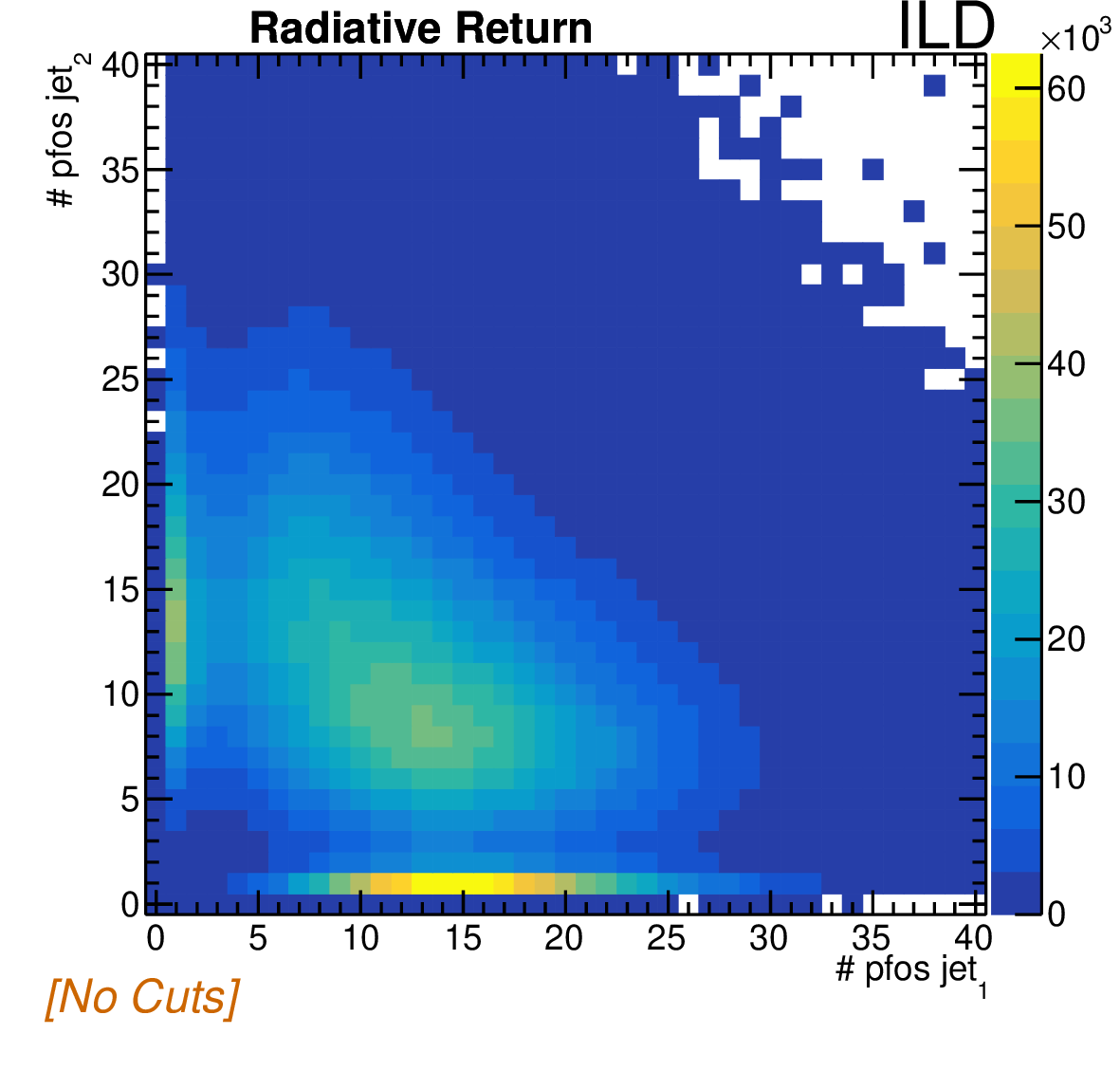} &
        \includegraphics[width=0.3\textwidth]{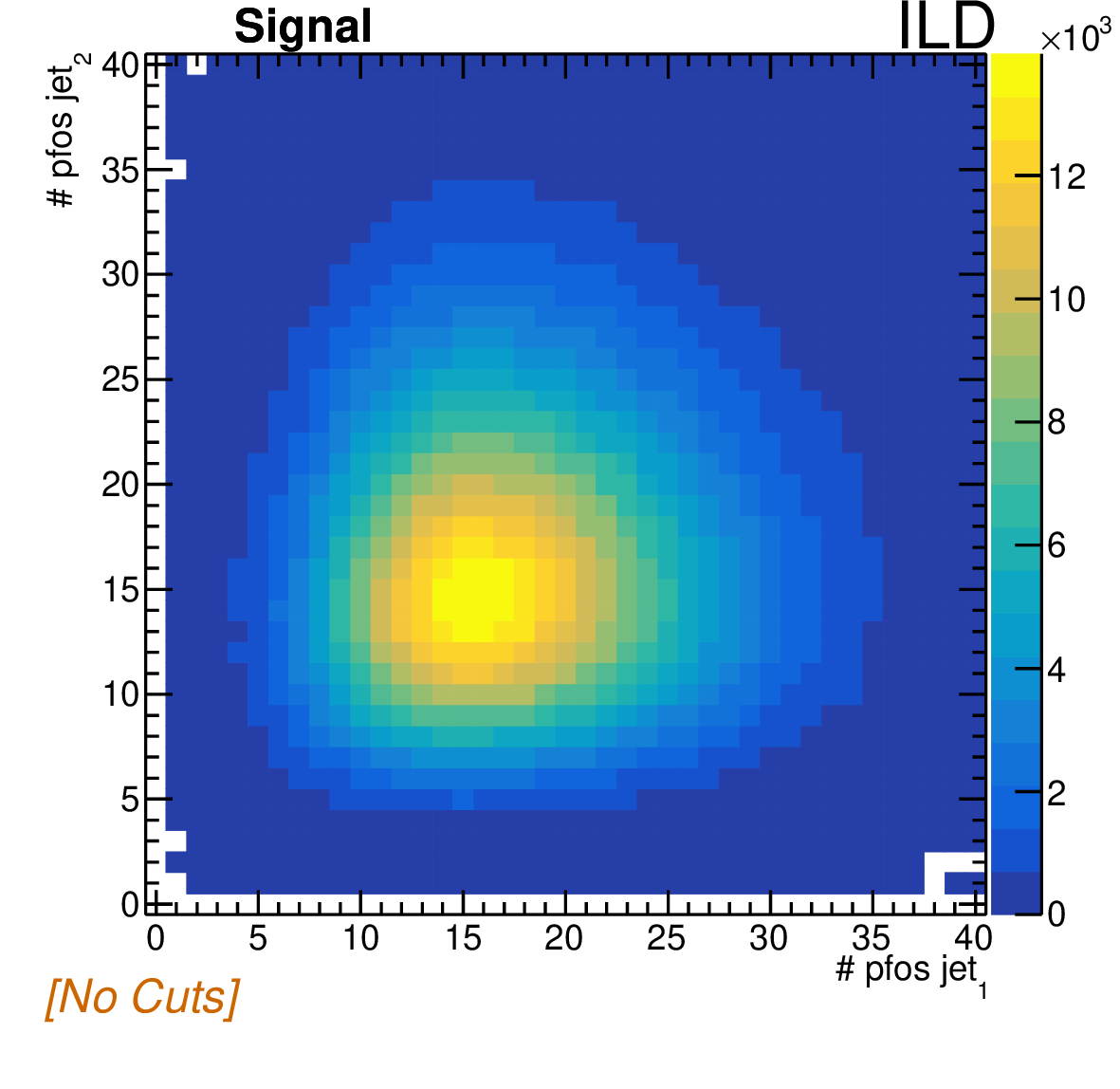} &
        \includegraphics[width=0.3\textwidth]{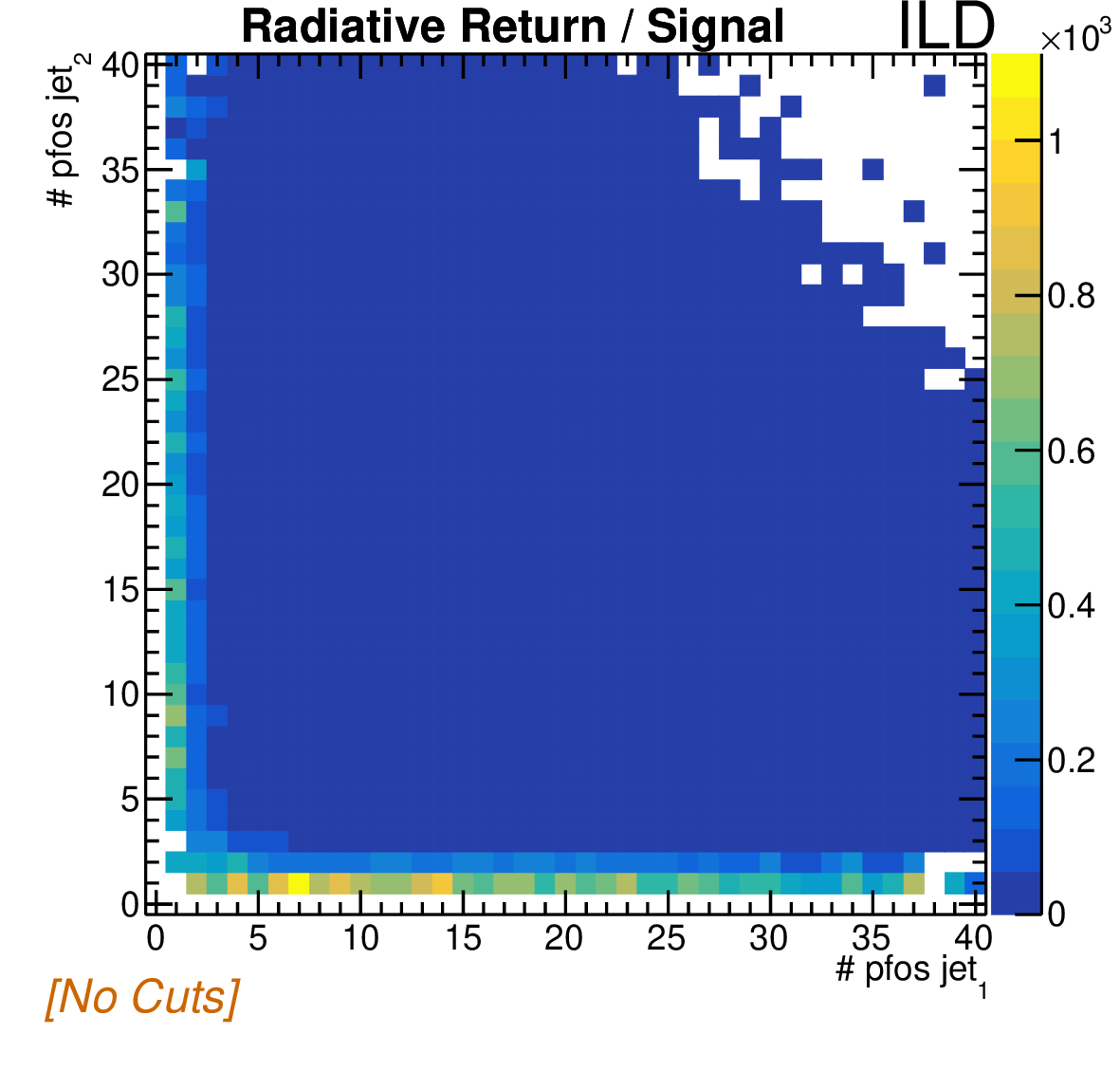}\\
        \includegraphics[width=0.3\textwidth]{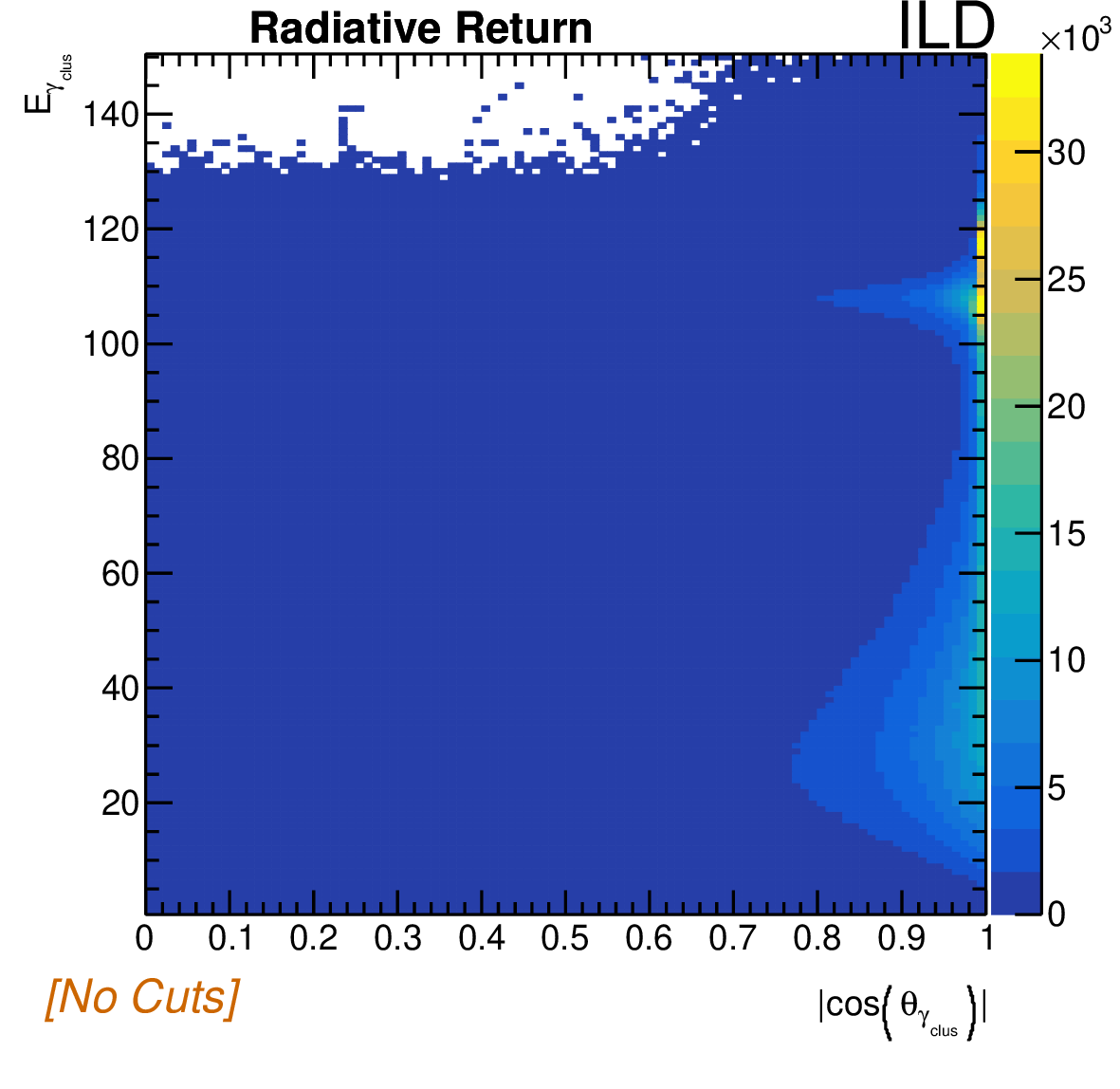} &
        \includegraphics[width=0.3\textwidth]{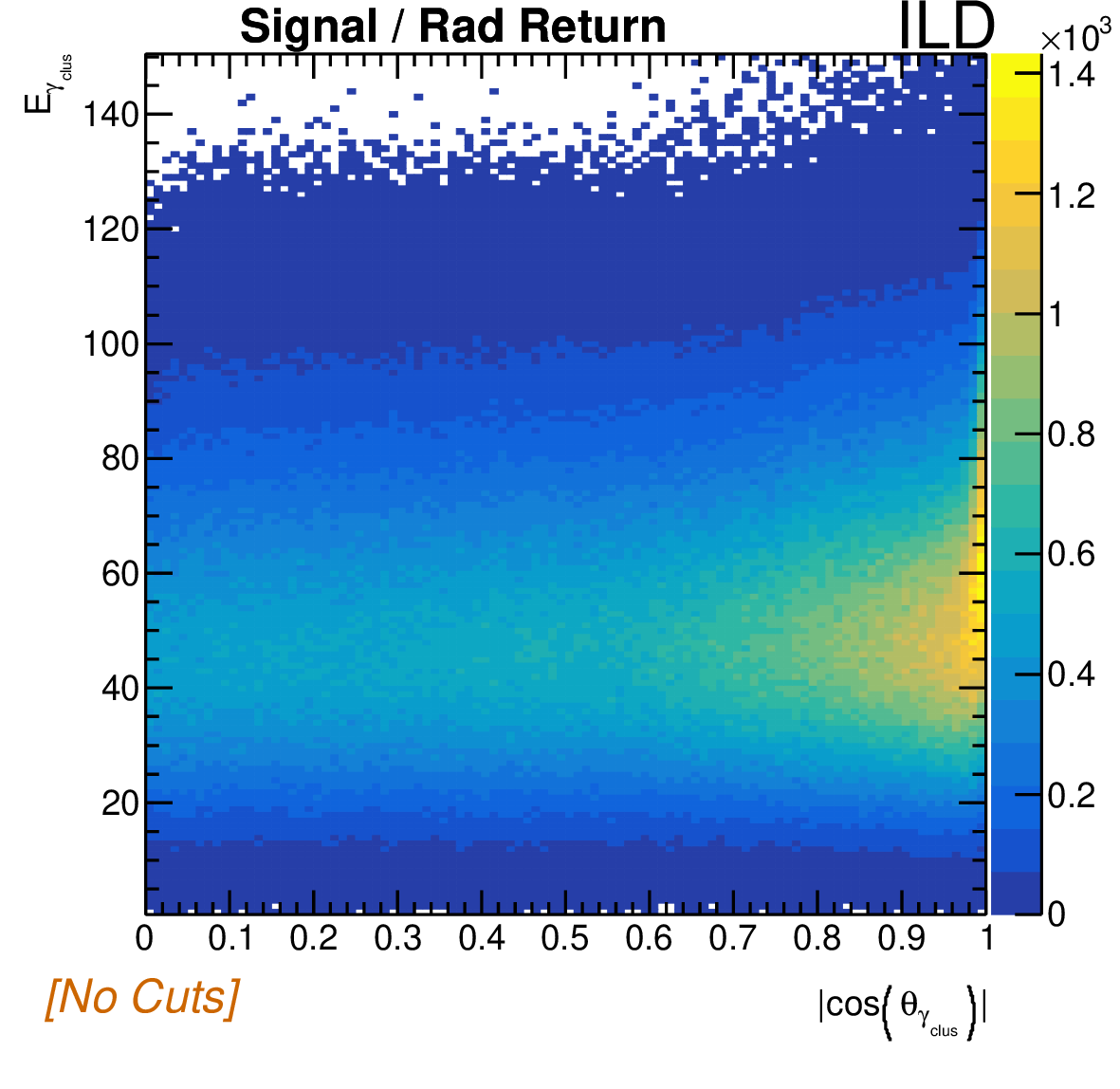} &
        \includegraphics[width=0.3\textwidth]{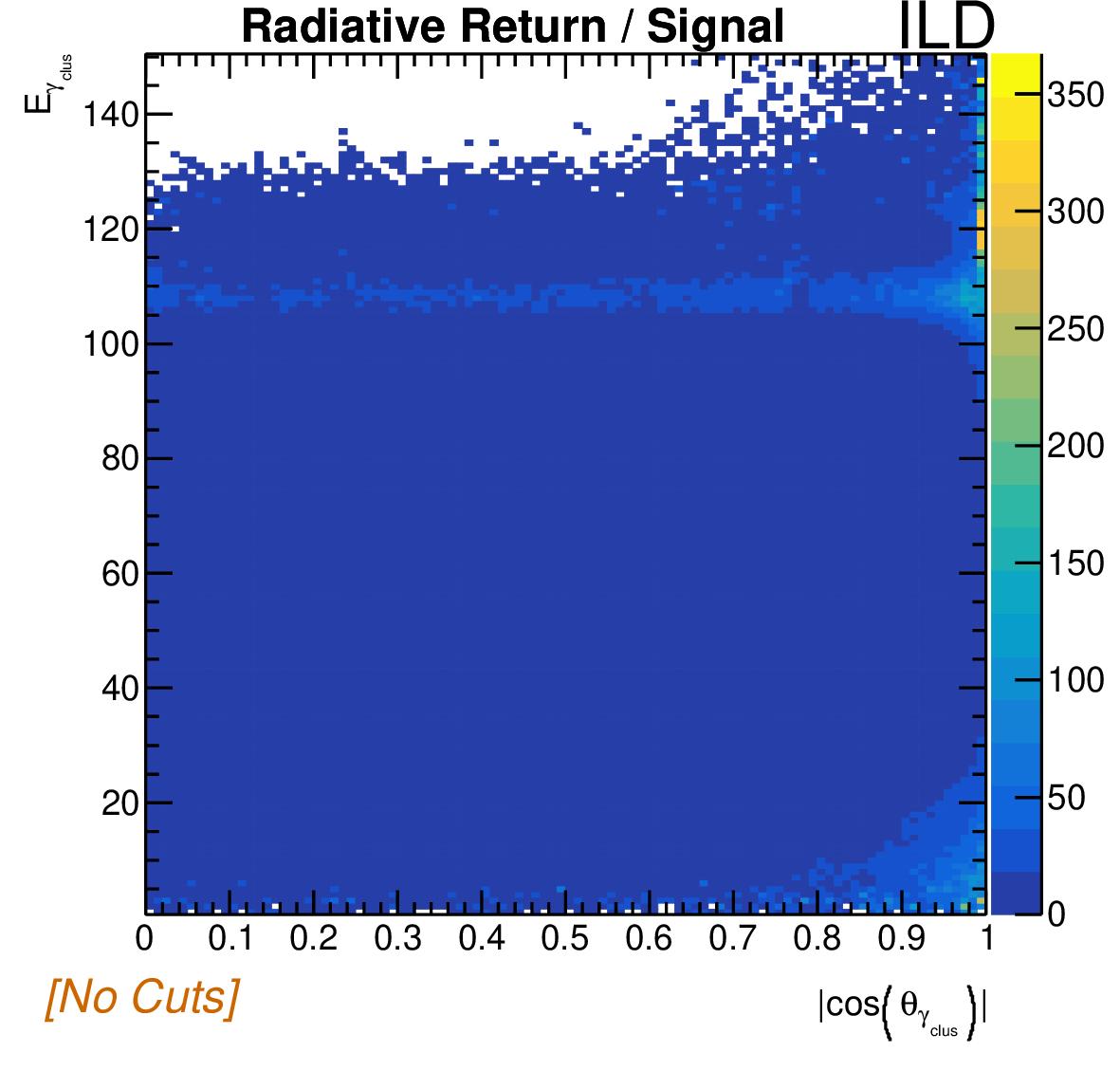}\\
        \end{tabular}
\caption{Two-dimensional maps used for vetoing events with reconstructed photons, as described in the text. The first column corresponds to the radiative return background; the second to the signal; and, the third, to the ratio between both.}
\label{fig:photon}
\end{figure}

The first kinematic variable that helps to distinguish between jets originating 
from a highly energetic quark from jets originating from highly energetic ISR is the number of PFOs in the jet.
For the case of ISR-jets, the number of PFOs will be equal to one in a significant fraction of cases.
Secondly, all the PFOs identified as neutral PFOs by the Pandora PFA inside each jet are added together.
The resulting object is defined as $\gamma_{clus}$.
The energy and angular distributions of each of these clusters (one per jet, as maximum)
are highly different for ISR and signal events.
Figure \ref{fig:photon} shows these two discrimination distributions.
The events with less than two PFO in any of the jets
and with  $\gamma_{clus}$ with energy larger than 115 GeV or 
reconstructed at $|\costheta|>0.97)$ are vetoed.
These two cuts improve the efficiency of background selection by a factor of two 
while not affecting the signal selection efficiency.

%\subsubsection*{Identification of events with non-detected ISR}
%\label{sec:photon}

\subsubsection*{Acollinearity cut}%Identification of high energy ISR photons}

The acollinearity of the two reconstructed jets is defined as in Eq. \ref{eq:acol}, but using the jet directions, not the quark directions. 
A cut on this kinematic variable is performed, likewise, as for the theoretical definition of the cross section. The acollinearity distribution and its cut value are shown in Figure \ref{fig:kinematics}, left plot.

\subsubsection*{Invariant mass cut}%Identification of high energy ISR photons}

\begin{figure}[!ht]
  \centering
      \begin{tabular}{ccc}
        \includegraphics[width=0.3\textwidth]{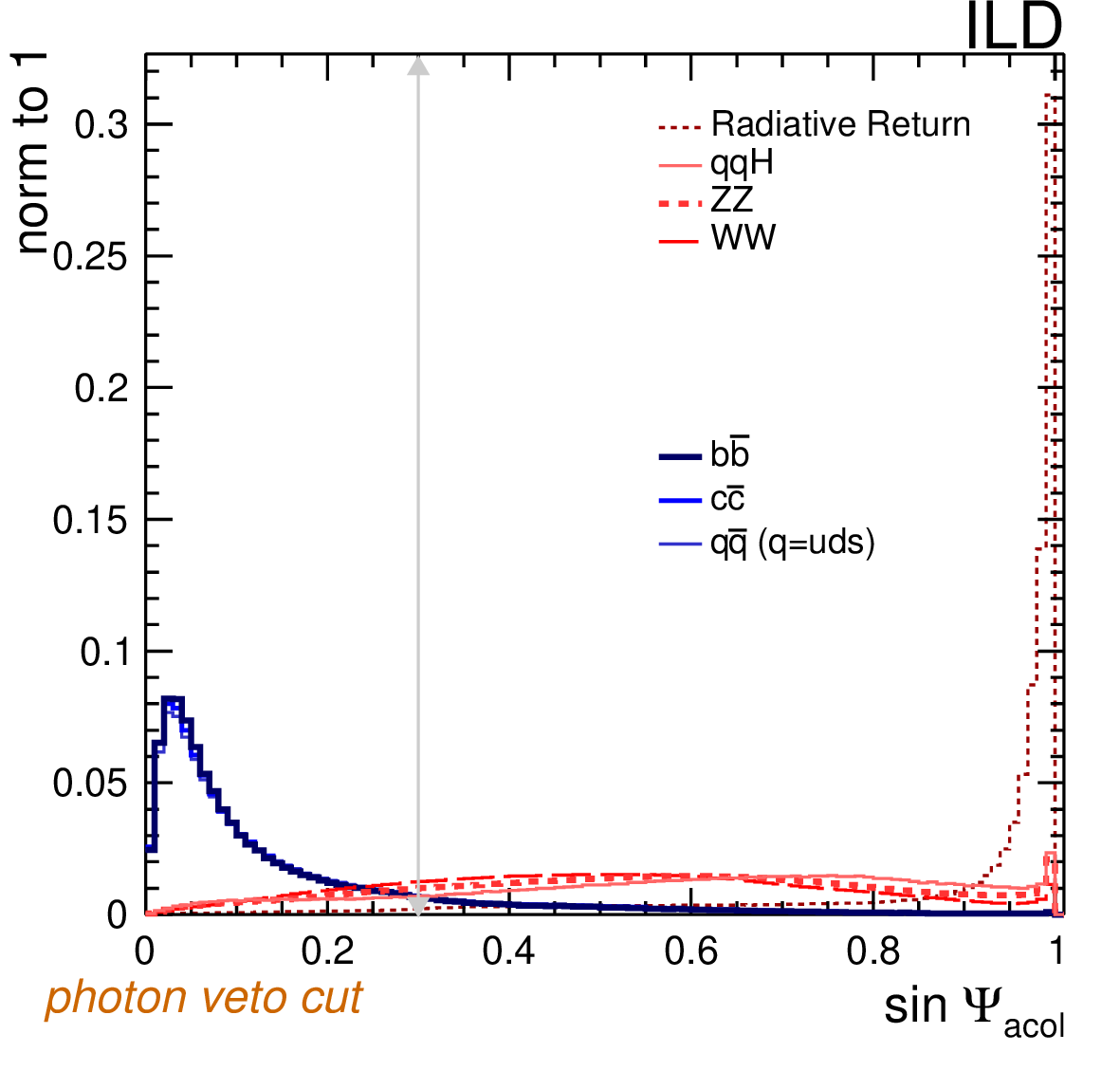} 
        \includegraphics[width=0.3\textwidth]{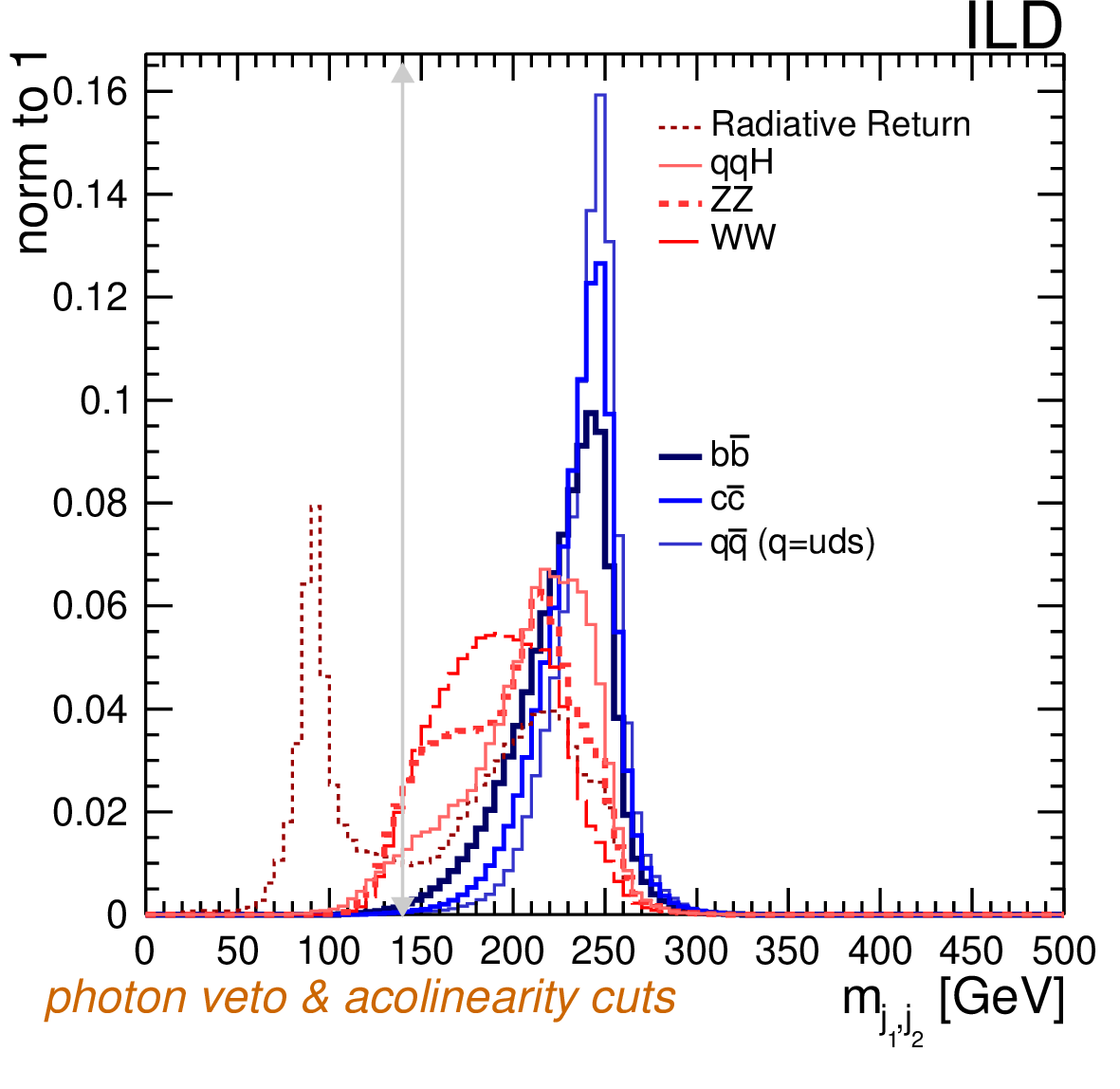} 
        \includegraphics[width=0.3\textwidth]{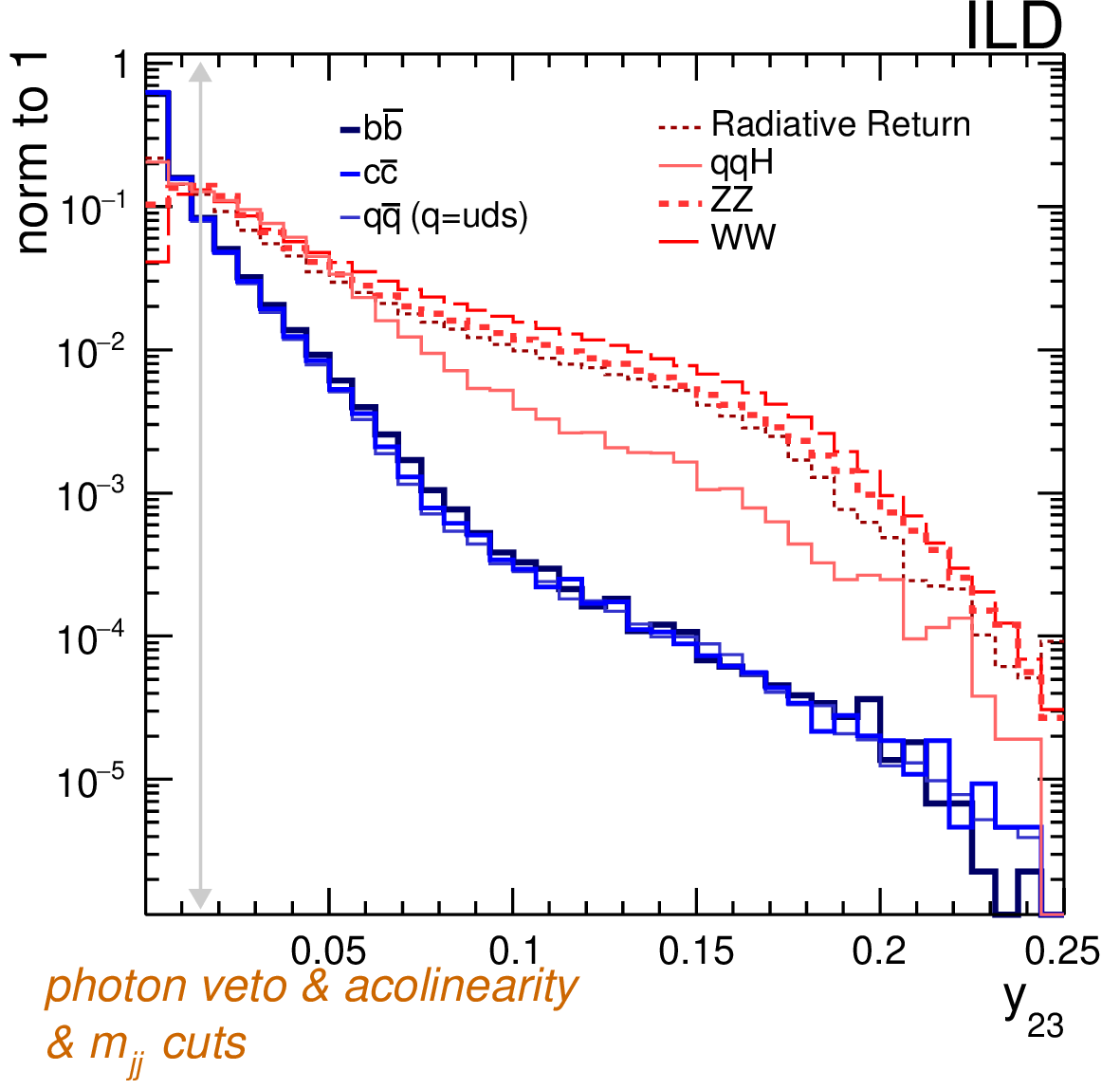}
      \end{tabular}
      \caption{\label{fig:kinematics} Kinematic variables used for the signal and background separation. The legends describe the type of process for each curve, and the labels in the bottom left part of the figure describe the cuts already performed for each case. For all plots, every distribution is normalised such that the integral of each curve is equal to 1. }
\end{figure}

A cut on the invariant mass of the two reconstructed jets is also applied. This distribution and its cut value are shown in the middle plot of Figure \ref{fig:kinematics}.
This cut improves the rejection of radiative return backgrounds.

\subsection{Cuts against pairs of heavy bosons backgrounds}
\label{sec:analysis_preselection_bkg}

After removing radiative return events, there is still some contamination from the heavy boson-induced background since only 90$\%$ of these events are filtered out. Although most of these processes have relatively small cross sections (see Table \ref{tab:crosssection_bkg}),
for some cases, the $WW$ contribution is sizeable, especially for the pure left electron beam polarisation.
A cut on the distance $y_{23}<0.02$ is defined to further suppress these backgrounds.
This variable $y_{23}$ refers to the jet distance (defined in the $VLC$ algorithm) at which
a two-jet system would be reconstructed as a three-jet system. 
This cut, shown in the last plot of Figure \ref{fig:kinematics}, also helps to reduce some remaining
radiative return events in which the QED ISR is not hard enough to 
be removed by the methods described before. 
Furthermore, $y_{23}$ is also sensitive to final state QED and QCD radiation. 
Therefore, it allows, in particular, for controlling the modelling of QCD radiation.

\subsection{Summary of the pre-selection procedure}

The cuts used to enrich the sample with signal events and remove all background contamination are:

\begin{itemize}
    \item[Cut 1:] Photon veto cuts. An event is rejected if at least one of the following conditions is fulfilled.  
    \begin{enumerate}
        \item at least one of the jets contains a reconstructed $\gamma_{cluster}$ with $E>115$ GeV or located in the forward region $|\costheta|>0.97$.
        \item at least one of the jets contains only one reconstructed PFO.
    \end{enumerate}
    \item[Cut 2:] events with $\sin{\Psi_{acol}}>0.3$ are rejected.
    \item[Cut 3:] events with $m_{jj}<140$ GeV are rejected.
    \item[Cut 4:] events with $y_{23}>0.02$ are rejected.
\end{itemize}

\begin{figure}[!h]
  \centering
      \begin{tabular}{cc}
        \includegraphics[width=0.45\textwidth]{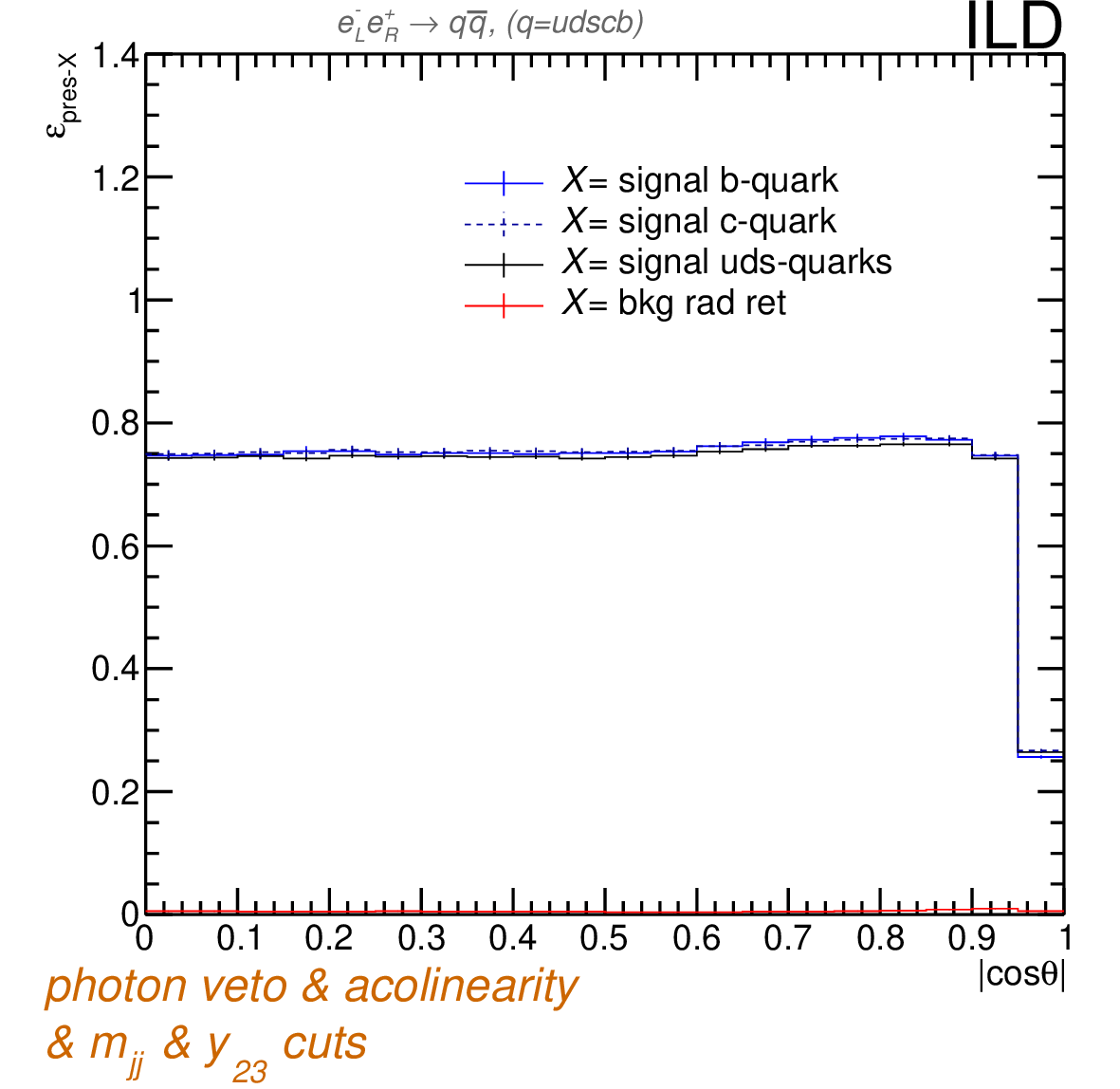}&
        \includegraphics[width=0.45\textwidth]{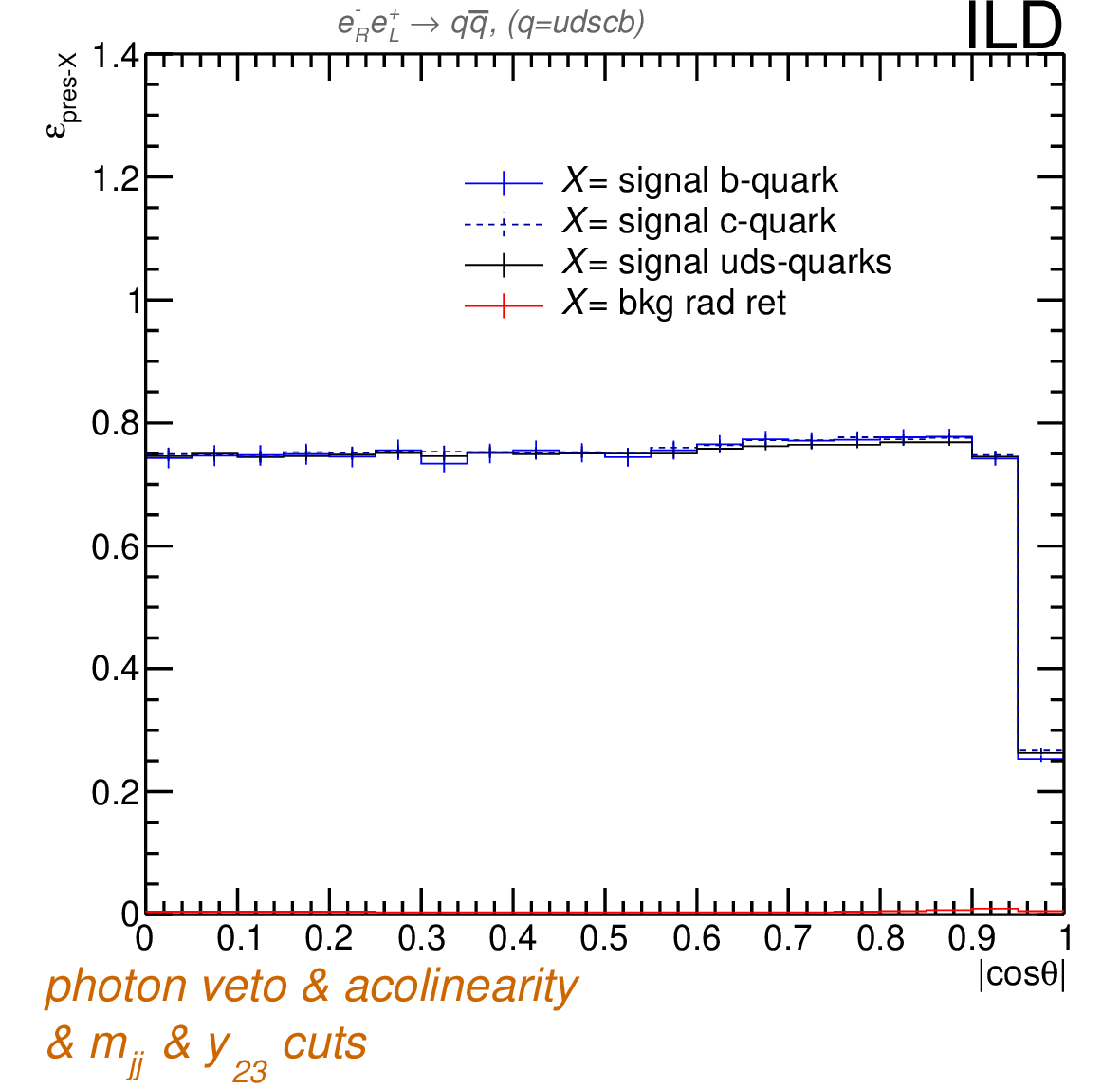} 
      \end{tabular}
      \caption{\label{fig:preselection} Efficiency of the preselection for the different quark flavours, $\varepsilon_{pres}$ , vs the angular distribution of the two jet system.}
      % \epsilonhad, vs the angular distribution of the two jet system.}
\end{figure}

The impact of each cut is shown in Table \ref{tab:cutflow}.
Figure \ref{fig:preselection} shows the selection efficiency for events \eeqq for the different quark flavours and two polarisation scenarios.
This pre-selection leaves $\sim70$\% of events in the barrel region ($|\costheta|<0.8$).
The angular distributions of the resulting pre-selection efficiencies 
are very similar for the five flavours at the 1\% or lower level.
The efficiency remains constant in most of the detector volume, except in the very forward region (defined as $\mbox{|\costheta|>0.9}$).

\begin{table}[!ht]
 %\renewcommand{\arraystretch}{1.6}
 % \scriptsize
 \begin{center}
  \begin{tabular}{c|ccc|cccc}
    \hline
    \multicolumn{8}{c}{{\textbf{ Efficiency of selection for $\eLpR \rightarrow X$ [\%]} }}\\
      \hline
    & \multicolumn{3}{c}{\textbf{ Signal}} & \multicolumn{4}{c}{ \textbf{ Background}}\\
      & \bbbar & \ccbar  & \qqbar($q=udscb$) & Rad. Ret. & $ZZ$ & $WW$ & $q\bar{q}H$ \\
    \hline
    Cut 1 & 93.4\% & 93.3\% & 92.9\% & 53.8\% & 89.9\% & 91.3\% & 93.1 \%\\
    + Cut 2 & 80.1\% & 79.4\% & 78.2\% & 1.7\% & 18.6\% & 16.0\% & 14.5 \%\\
    + Cut 3 & 79.7\% & 79.3\% & 78.1\% & 1.1\%  & 17.8\% & 15.1\% & 13.8 \%\\
    + Cut 4 & 71.6\% & 71.7\% & 71.0\% & 0.6\% & 5.9\% & 6.2\% & 6.1 \%\\
      \hline
      \hline
    \multicolumn{8}{c}{{\textbf{ Efficiency of selection for $\eRpL \rightarrow X$ [\%]}}}\\
      \hline
    & \multicolumn{3}{c}{\textbf{Signal}} & \multicolumn{4}{c}{ \textbf{Background}}\\
      & \bbbar & \ccbar  & \qqbar($q=udscb$) & Rad. Ret. & $ZZ$ & $WW$ & $q\bar{q}H$ \\
    \hline
    Cut 1 & 93.1\% & 93.4\% & 93.0\% & 51.5\% & 94.3\% & 89.6\% & 93.1 \%\\
    + Cut 2 & 79.8\% & 79.5\% & 78.4\% & 1.6\% & 14.9\% & 16.7\% & 14.6 \%\\
    + Cut 3 & 79.5\% & 79.4\% & 78.4\% & 1.0\%  & 13.3\% & 15.8\% & 13.9 \%\\
    + Cut 4 & 71.4\% & 71.8\% & 71.3\% & 0.6\% & 1.9\% & 6.8\% & 6.1 \%\\
      \hline
  \end{tabular}
  \caption{\label{tab:cutflow} Cut flow for signal and background events.}
 \end{center}
\end{table}

%% file: sections/6_analysis_R.tex
\section{Differential measurement of $R_{q}$}
\label{sec:analysis_R}

\subsection{Double Tag method (DT)}
\label{sec:analysis_doubletag}

%To reach the maximum of precision for the measurement of \Rb and \Rc, this ratio needs to be measured
%at the same time as the $b/c-$quark tagging efficiencies are measured.
%This is done by applying the Double Tag (DT) approach described in \cite{ALEPH:2005ab}, but in a fully differential way, profiting from the higher
%detector acceptance expected at ILD.
The Double Tag method (DT-method) has been introduced in \cite{ALEPH:2005ab}. For each event, the detector is subdivided into two hemispheres. For the present analysis, the separation of the hemispheres is defined by the $xy$-plane of the ILD coordinate system that crosses the origin of the ILD coordinate system. The ratio $f_1$ is the number of hemispheres in which a quark $q$ has been tagged divided by the number of available hemispheres, i.e. two per event. The ratio  $f_2$ is the fraction of events in which both hemispheres have been tagged. 
More precisely $f_1$ and $f_2$ are defined as:

%The Double Tag method (DT-method)\cite{ALEPH:2005ab} is based on the measurement of two ratios $f_{1}$ and
%$f_{2}$ for the simultaneous extraction of the tagging efficiency, $\epsilon_{q}$ and the \Rq. 
%The two ratios, for the \bquark and \cquark cases
%are defined as:
%\begin{equation}
%  \begin{aligned}
%f_{1b}(|\costheta|)=\frac{N_{b}(|\costheta|)-N^{bkg.}_{b}(|\costheta|)}{2\times (N_{0}(|\costheta|)-N^{bkg.}_{0}(|\costheta|))}=\\
%=\epsilon_{b}(|\costheta|)\Rb + \tilde{\varepsilon}_{c}(|\costheta|) \Rc + \tilde{\varepsilon}^{b}_{uds} (|\costheta|)(1-\Rb- \Rc)\\
%f_{2b}(\costheta)=\frac{N_{2b}(|\costheta|)-N^{bkg.}_{2b}(|\costheta|)}{N_{0}(|\costheta|)-N^{bkg.}_{0}(|\costheta|)}=\\
%=\epsilon_{b}^{2}(|\costheta|)(1+\rho_b)(|\costheta|)\Rb + \tilde{\varepsilon}_{c}^{2}(|\costheta|) \Rc + \tilde{(\varepsilon}^{b}_{uds})^{2}(|\costheta|) %(1-\Rb- \Rc)
%\label{eq:Rb}
%  \end{aligned}
%\end{equation}
%and, for the \cquark
%\begin{equation}
%  \begin{aligned}
%f_{1c}(|\costheta|)=\epsilon_{c}(|\costheta|)\Rc + \tilde{\varepsilon}_{b}(|\costheta|) \Rb + \tilde{\varepsilon}^{c}_{uds} (|\costheta|)(1-\Rb- \Rc)\\
%f_{2c}(\costheta)=\epsilon_{c}^{2}(|\costheta|)(1+\rho_c)(|\costheta|)\Rc + \tilde{\varepsilon}_{b}^{2}(|\costheta|) \Rb + %\tilde{(\varepsilon}^{c}_{uds})^{2}(|\costheta|) (1-\Rb- \Rc) 
%\label{eq:Rc}
%  \end{aligned}
%\end{equation}

\begin{equation}
 \begin{aligned} 
 f_{1q}(|\costheta|)=\frac{N_{q}(|\costheta|)-N^{bkg.}_{q}(|\costheta|)}{2\times (N_{0}(|\costheta|)-N^{bkg.}_{0}(|\costheta|))}\\
f_{2q}(\costheta)=\frac{N_{2q}(|\costheta|)-N^{bkg.}_{2q}(|\costheta|)}{N_{0}(|\costheta|)-N^{bkg.}_{0}(|\costheta|)}
\label{eq:Rq0}
  \end{aligned}
\end{equation}
which is equivalent to
\begin{small}
\begin{equation}
 \begin{aligned}
 f_{1q}(|\costheta|)=\\
 \varepsilon_{q}(|\costheta|)\Rq(|\costheta|) + \tilde{\varepsilon}_{q\prime}(|\costheta|) \Rqp(|\costheta|) + \tilde{\varepsilon}^{q}_{uds}(|\costheta|)(1-\Rq(|\costheta|)- \Rqp(|\costheta|))\\
f_{2q}(\costheta)=\\
\varepsilon_{q}^{2}(|\costheta|)(1+\rho_q(|\costheta|))\Rq(|\costheta|) + \tilde{\varepsilon}_{q\prime}^{2}(|\costheta|) \Rqp(|\costheta|) + (\tilde{\varepsilon}^{q}_{uds})^{2}(|\costheta|) (1-\Rq(|\costheta|)- \Rqp(|\costheta|))
\label{eq:Rq}
  \end{aligned}
\end{equation}
\end{small}
with $q,\,q\prime=b,\,c$ or $c,\,b$, respectively.
The variables $N_{0}$ and $N_{0}^{bkg.}$ are the total number of pre-selected di-jet events (see Section \ref{sec:analysis_preselection}) and
the estimated number of background events after the pre-selection, respectively.
Additionally, $N_{q}$ and $N^{bkg.}_{q}$ are the number of jets from signal and background that are tagged as quark flavour $q$. The tagging efficiency is given by $\epsilon_{q}$ and the mis-tagging probabilities as $\tilde{\varepsilon}_{q\prime}$ 
and $\tilde{\varepsilon}^{q}_{uds}$. 
%and $N^{bkg.}_{q}$ are the number of jets that are
%tagged as being originated by a quark type $q$ but that are generated via radiative return process or diboson production processes.
Furthermore, $N_{2q}$ and $N^{bkg.}_{2q}$ are signal and background events featuring a double tag for quark flavour $q$. The double-tag efficiency 
is given by the product $\epsilon_{q}^{2}\cdot(1+\rho_q)$.
%is the number of events with two jets tagged as originated by a \qqbar pair and $N^{bkg.}_{2q}$ the ones that come from background processes.
%Both ratios, for both flavours, depend on the mistagging efficiencies $\tilde{\varepsilon}_{uds}$ and $\tilde{\varepsilon}_{q^{\prime}}^{q}$ and 
%on the correlation variable 
The factor (1+$\rho_q$) is known in the literature as hemisphere correlation and parameterises the deviation of the double-tag efficiency from $\epsilon_{q}^{2}$. Correlations are introduced by a common primary vertex or hard QCD radiation. A further source of correlations is coherent noise in a detector, which however is not taken into account in this study. The parameter encompasses also asymmetries in detector efficiencies that are in the first approach uncorrelated.    
%due to displacements of primary vertex determination (common for both jets),
%purely geometric correlations associated to differences between the detector
%inhomogeneities and QCD related effects associated to hard gluon emission.
The hemisphere correlation for the quark flavour $q$ is derived from a simulated sub-sample consisting only of events of that flavour. In that case, it is simply:
\begin{align}
(1+\rho_q)=\frac{f_{2q}}{f^2_{1q}}
\end{align}    
%Ideally $(1+\rho_q)=1$.
%Using $(1+\rho_q)$ and the contributions from light quark flavours as input, Eq.~\ref{eq:Rq} leads to a system of four linearly independent equations to determine the observables ${\varepsilon}_{b}$, $\Rb$, ${\varepsilon}_{c}$ and $\Rc$.
%The hemisphere correlation is estimated by comparing the $q$-flavour tagging efficiency for single jets
%and two jet events in Monte Carlo samples made of \qqbar events of the flavour under study: 

%\begin{align}
%(1+\rho_Q)=\frac{4\cdot N_{\qqbar}^{sample}\codt N_{2q}^{sample}}{(N_{q}^{sample})^{2}}
%\end{align}    

%where, from a given sample, $N_{\qqbar}^{sample}$ reconstructed as two jets, the term $N_{2q}^{sample}$ represents
%the events with the two jets identified as $q$
%and $N_{q}^{sample}$ accounts for all single jets that are identified as $q$.

%The pre-selection has been designed to minimise these backgrounds and
%it is also expected that the modelling of backgrounds will
%be reliable at the percent level or better.
%In addition, to use this method it is assumed that the \Rb and \Rc measurements are performed
%in parallel. Further, it is assumed that the ILD will be able to estimate (or even measure) a
%t the level of the 1\% of relative accuracy the mis-tagging efficiencies 
%of wrongly tagging other quarks as the quark under study. 
%The $N_{0}$, $N_{1}$ and $N_{2}$ distributions as a function of $|\costheta|$
%are shown in Figure \ref{fig:N0N1N2} for the $b$ and \cquark cases.

\subsection{Results}
\label{sec:analysis_R_results}
\begin{figure}[!ht]
\begin{center}
    \begin{tabular}{ccc}
      \includegraphics[width=0.3\textwidth]{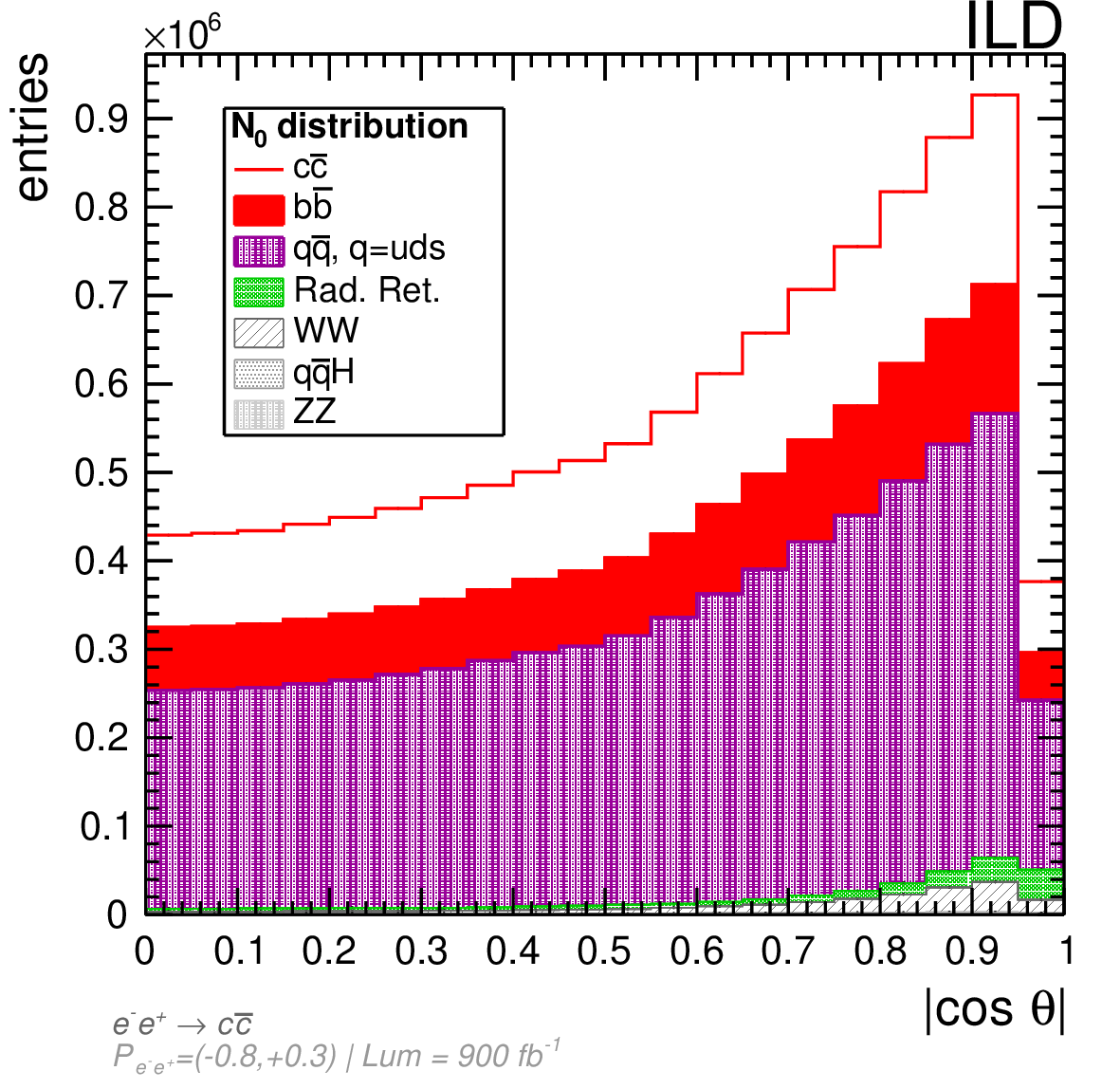} &
      \includegraphics[width=0.3\textwidth]{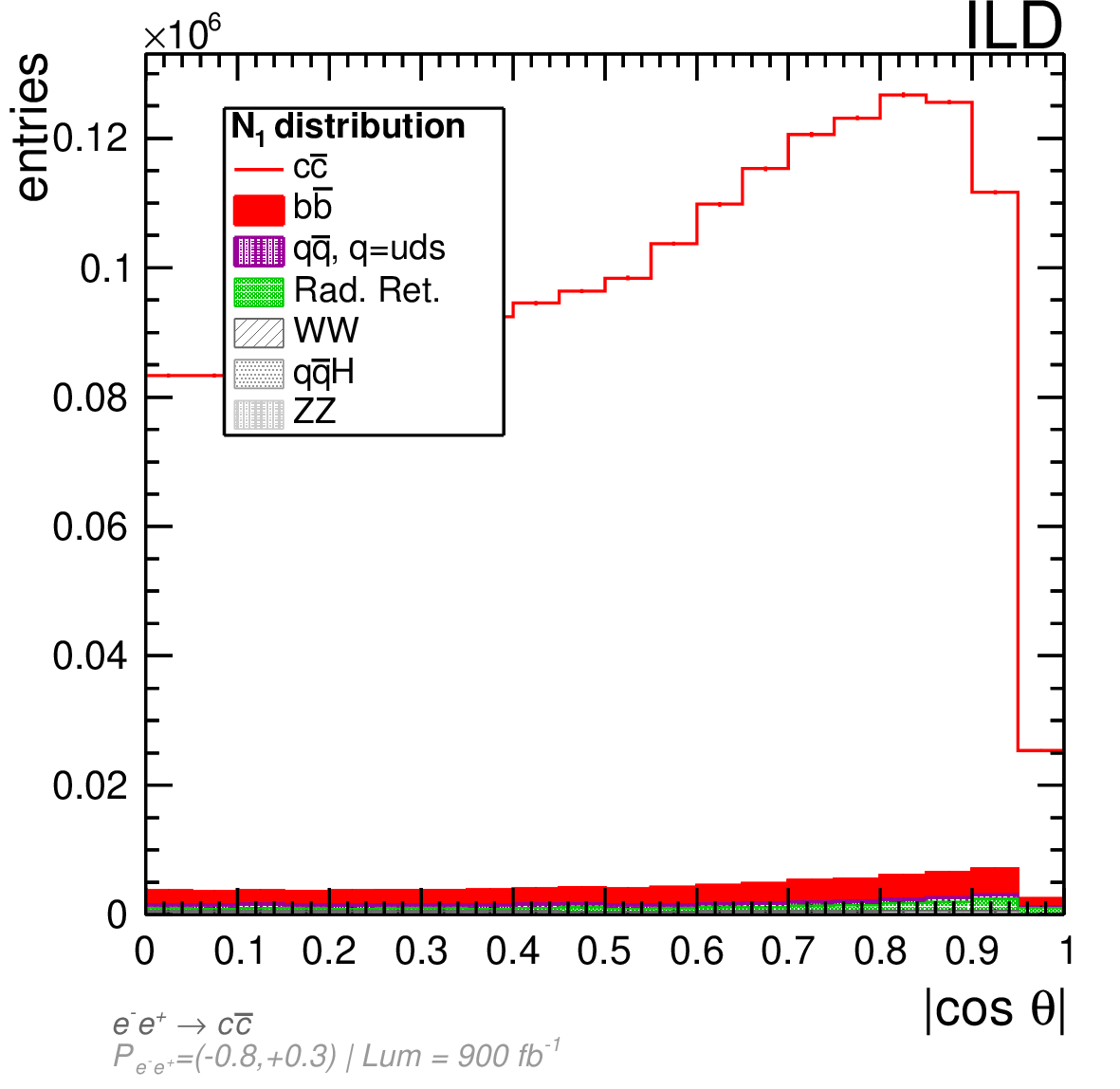} &
      \includegraphics[width=0.3\textwidth]{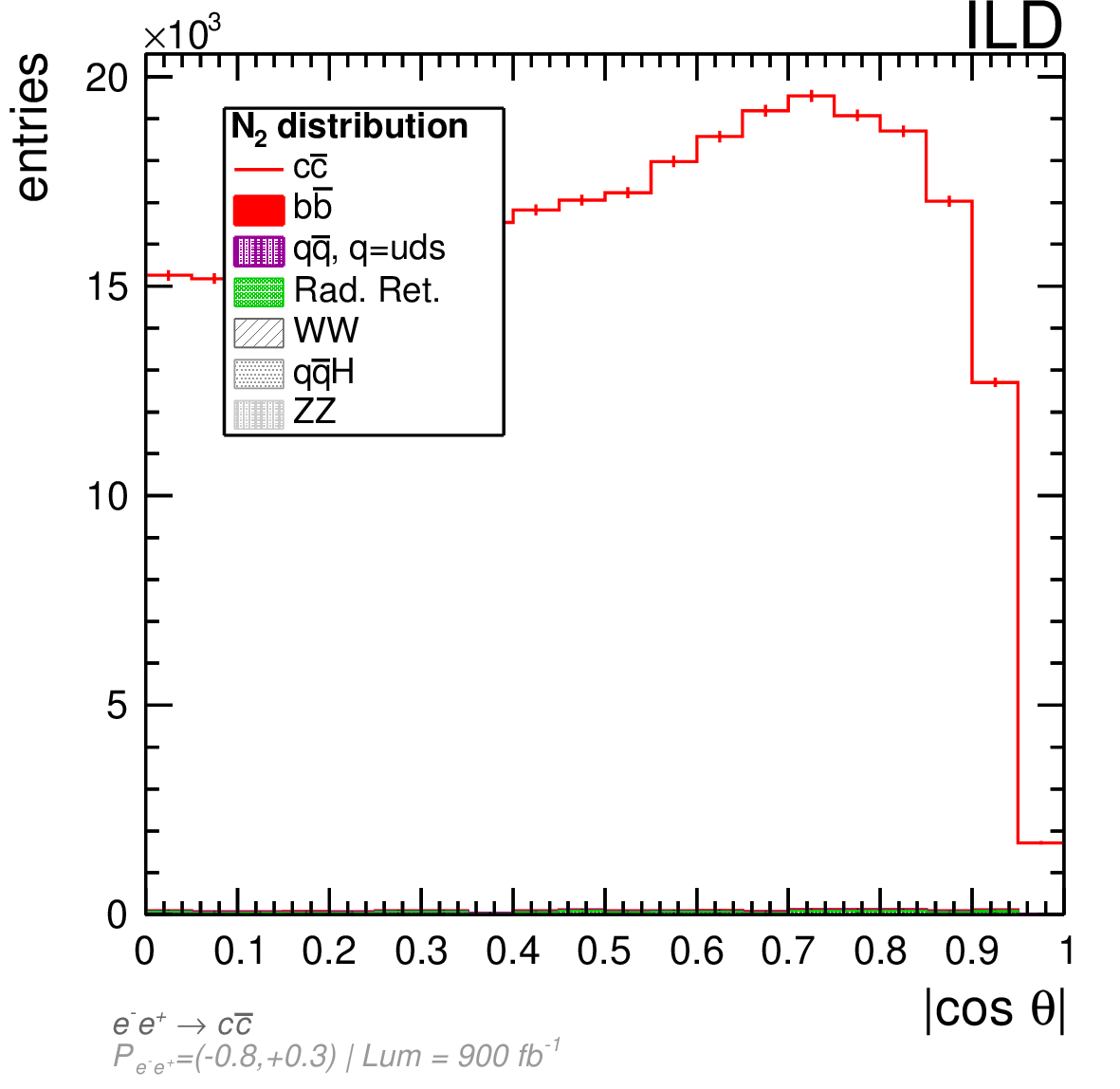} \\
      
      \includegraphics[width=0.3\textwidth]{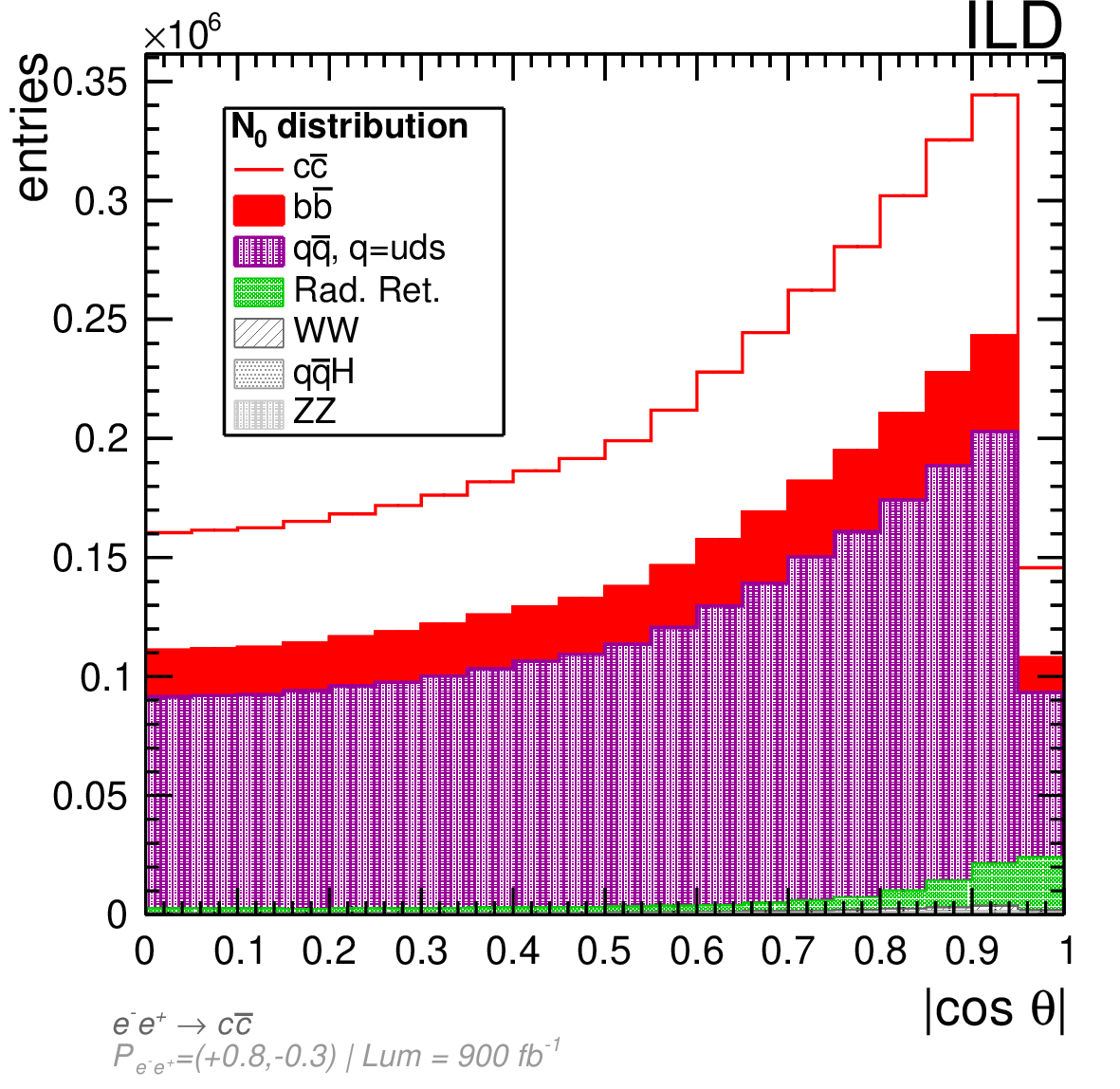} &
      \includegraphics[width=0.3\textwidth]{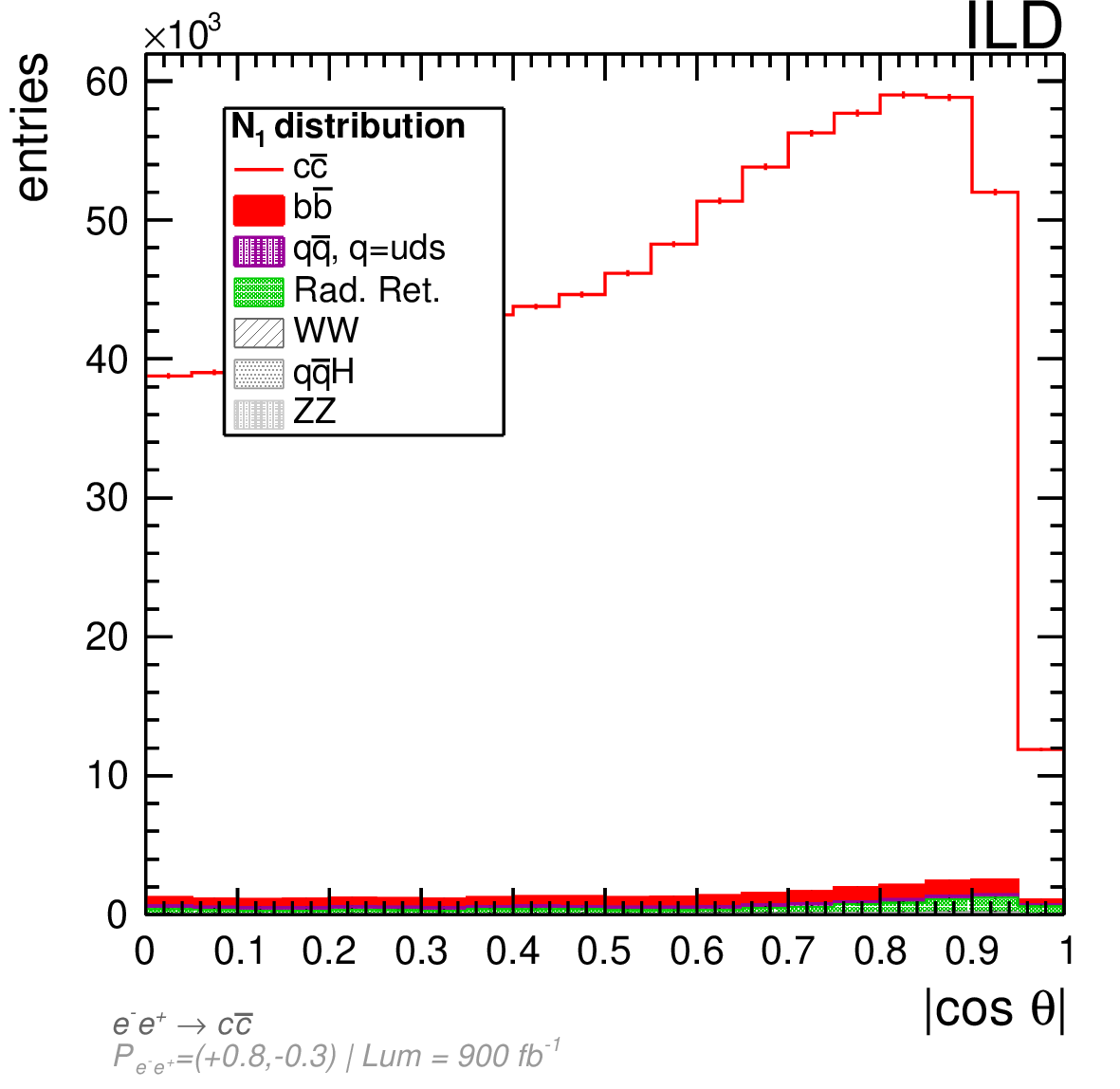} &
      \includegraphics[width=0.3\textwidth]{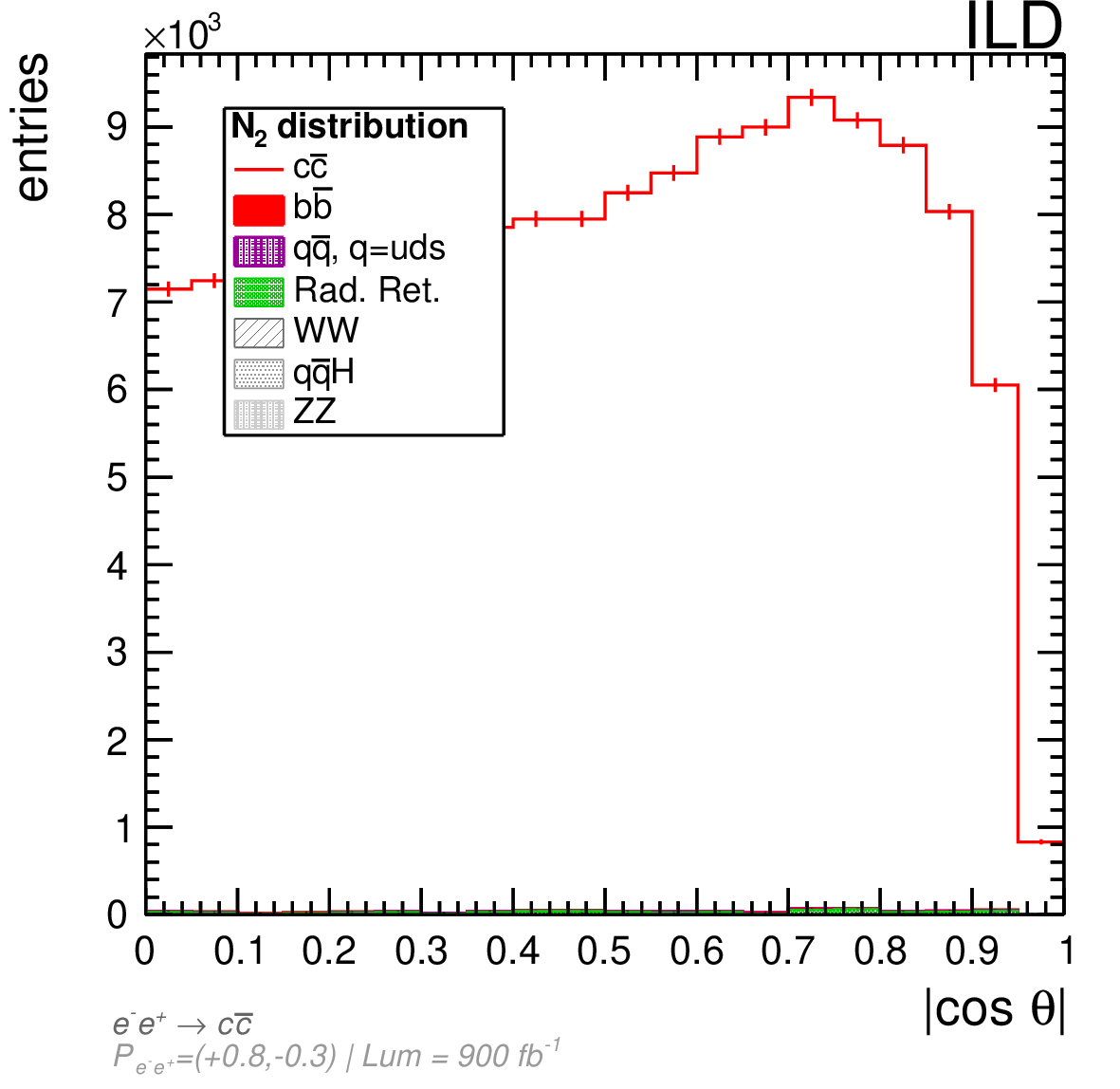} \\
      
      \includegraphics[width=0.3\textwidth]{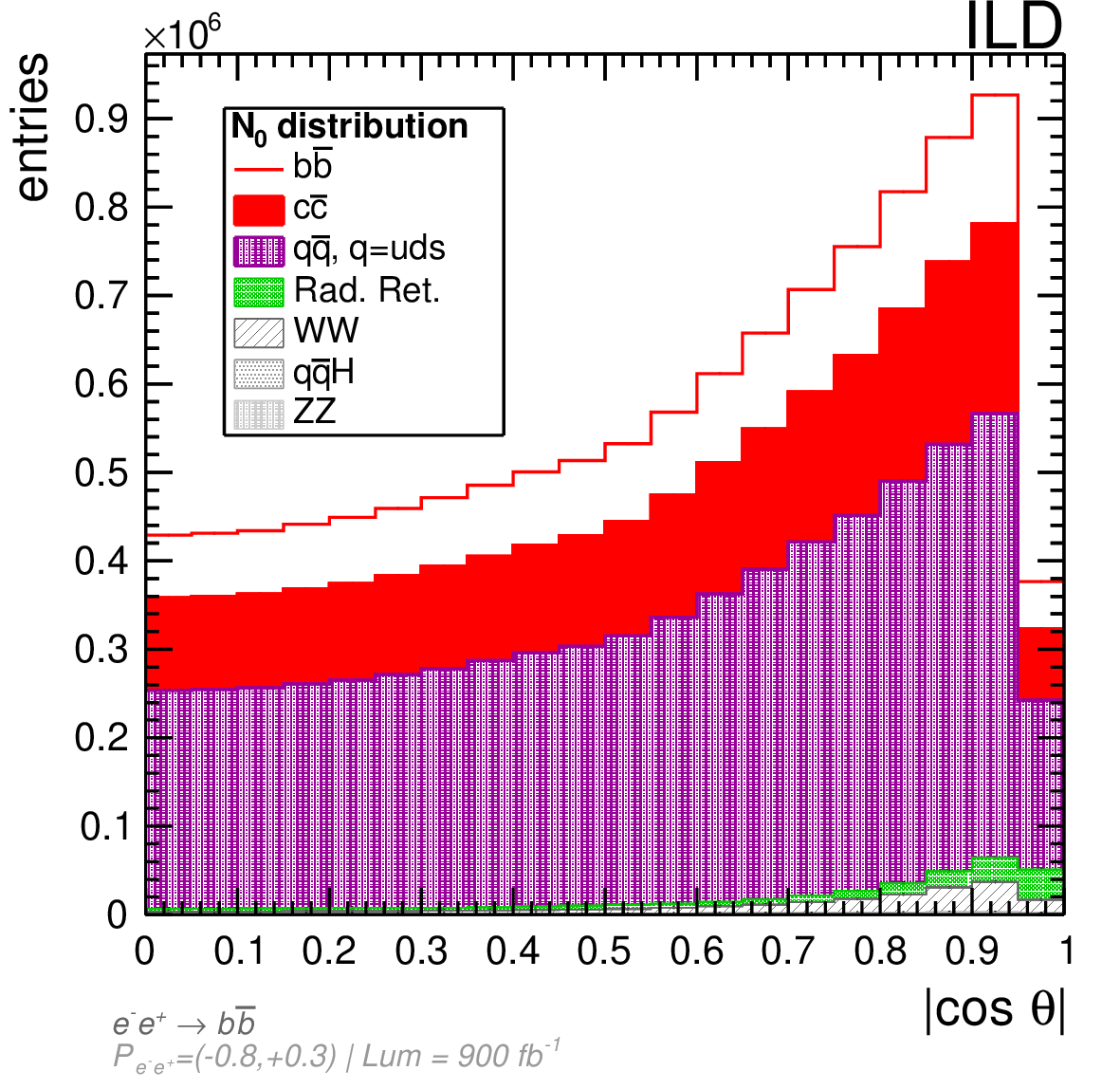} &
      \includegraphics[width=0.3\textwidth]{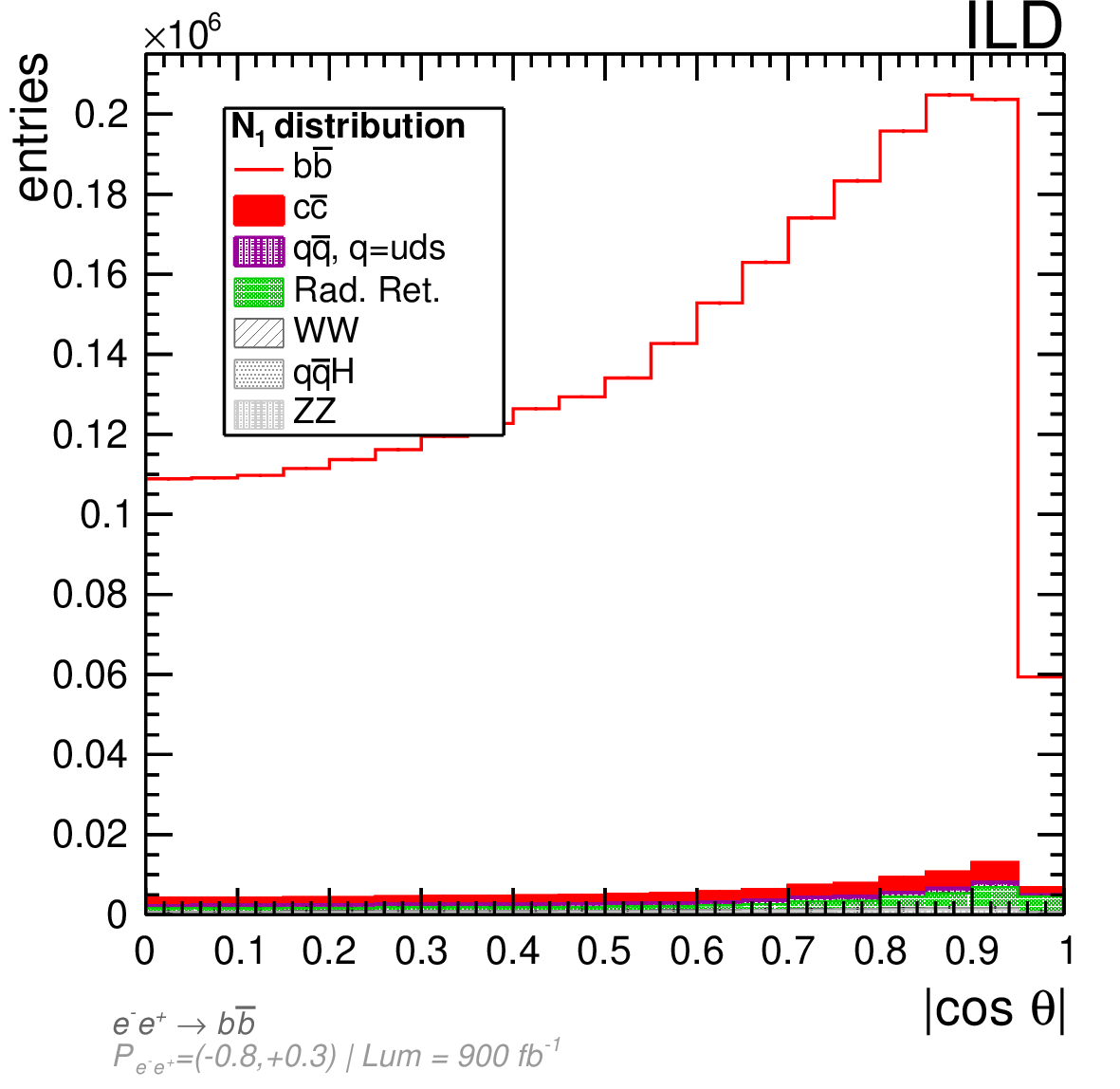} &
      \includegraphics[width=0.3\textwidth]{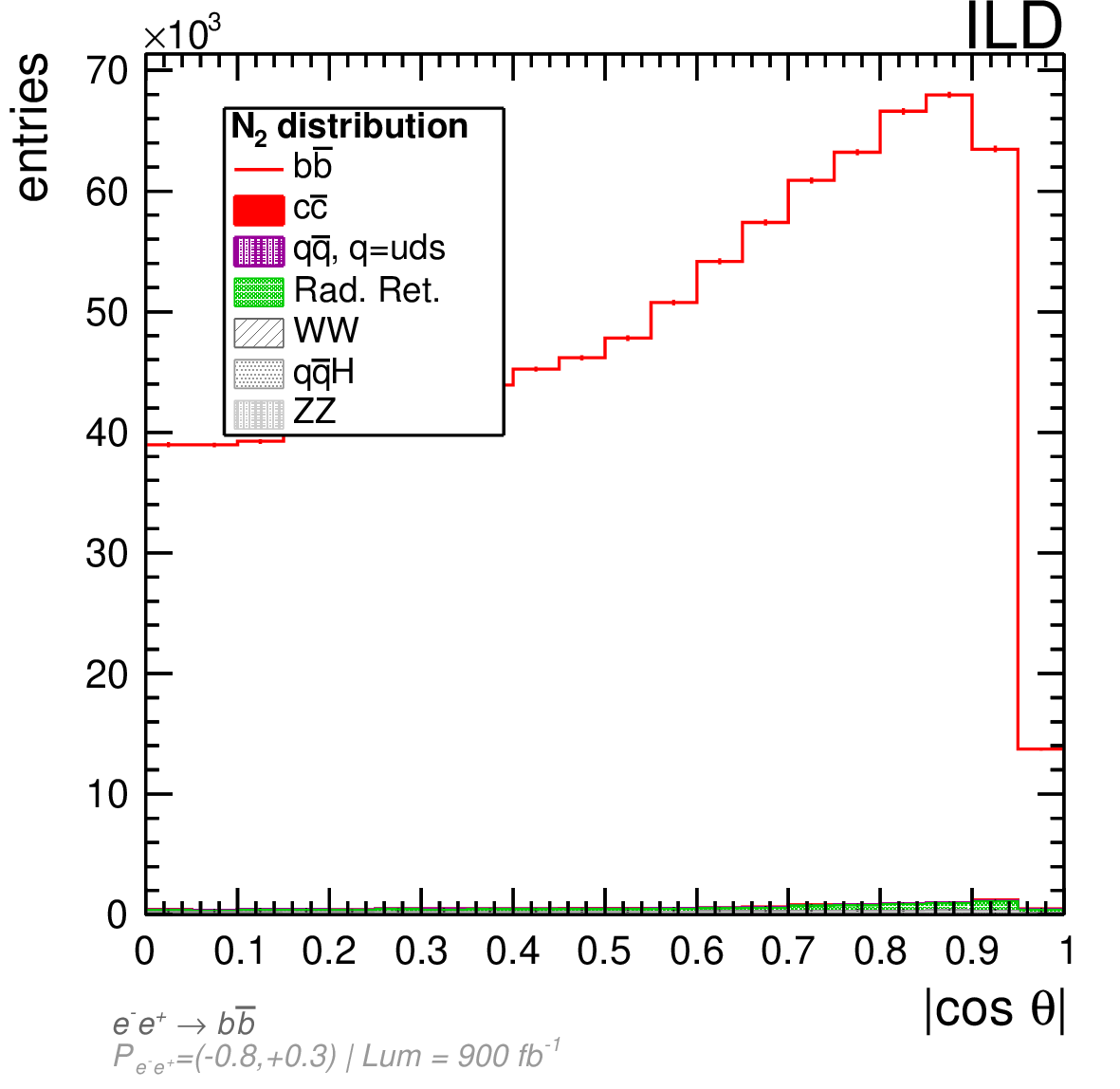} \\
      
      \includegraphics[width=0.3\textwidth]{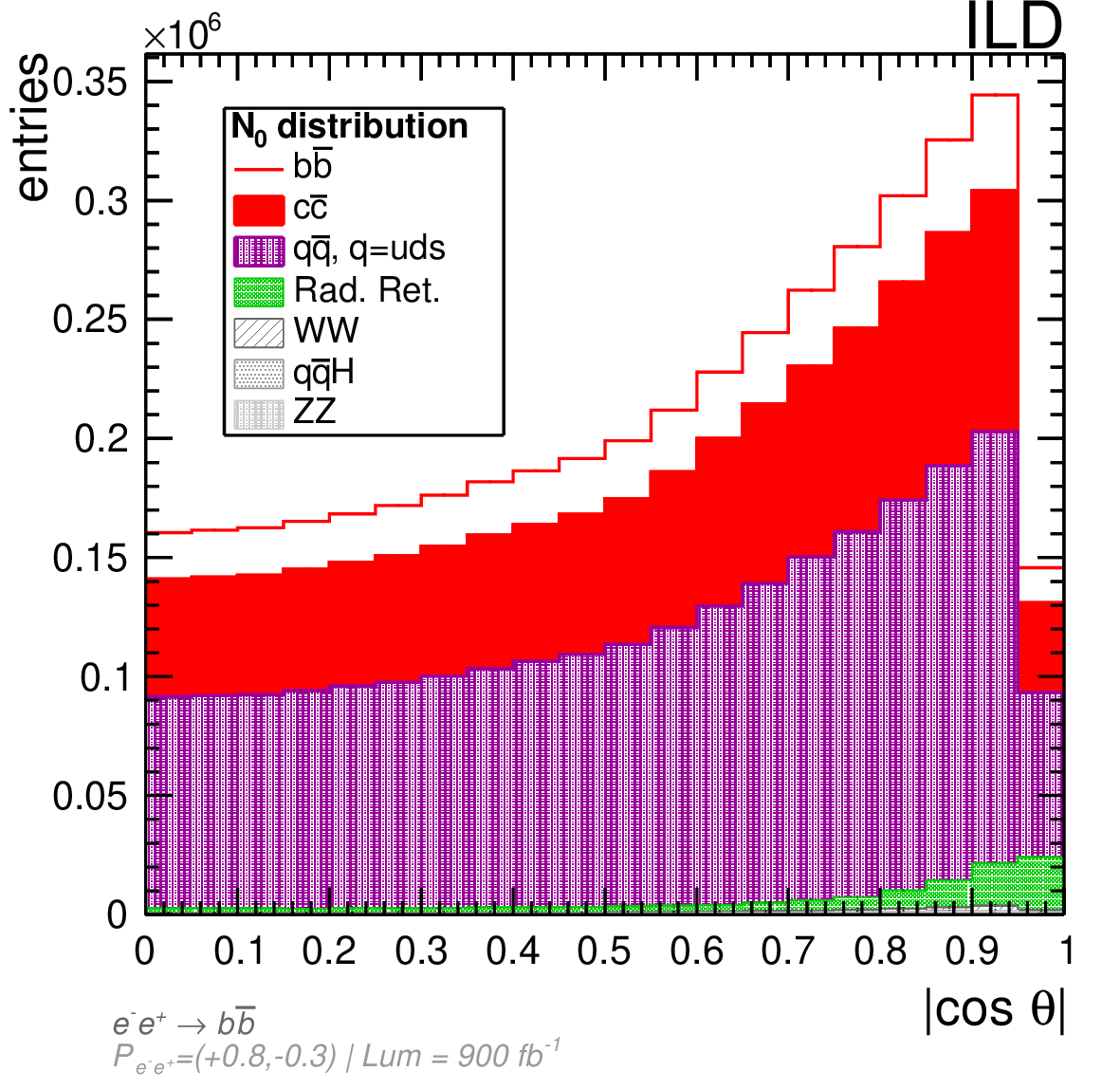} &
      \includegraphics[width=0.3\textwidth]{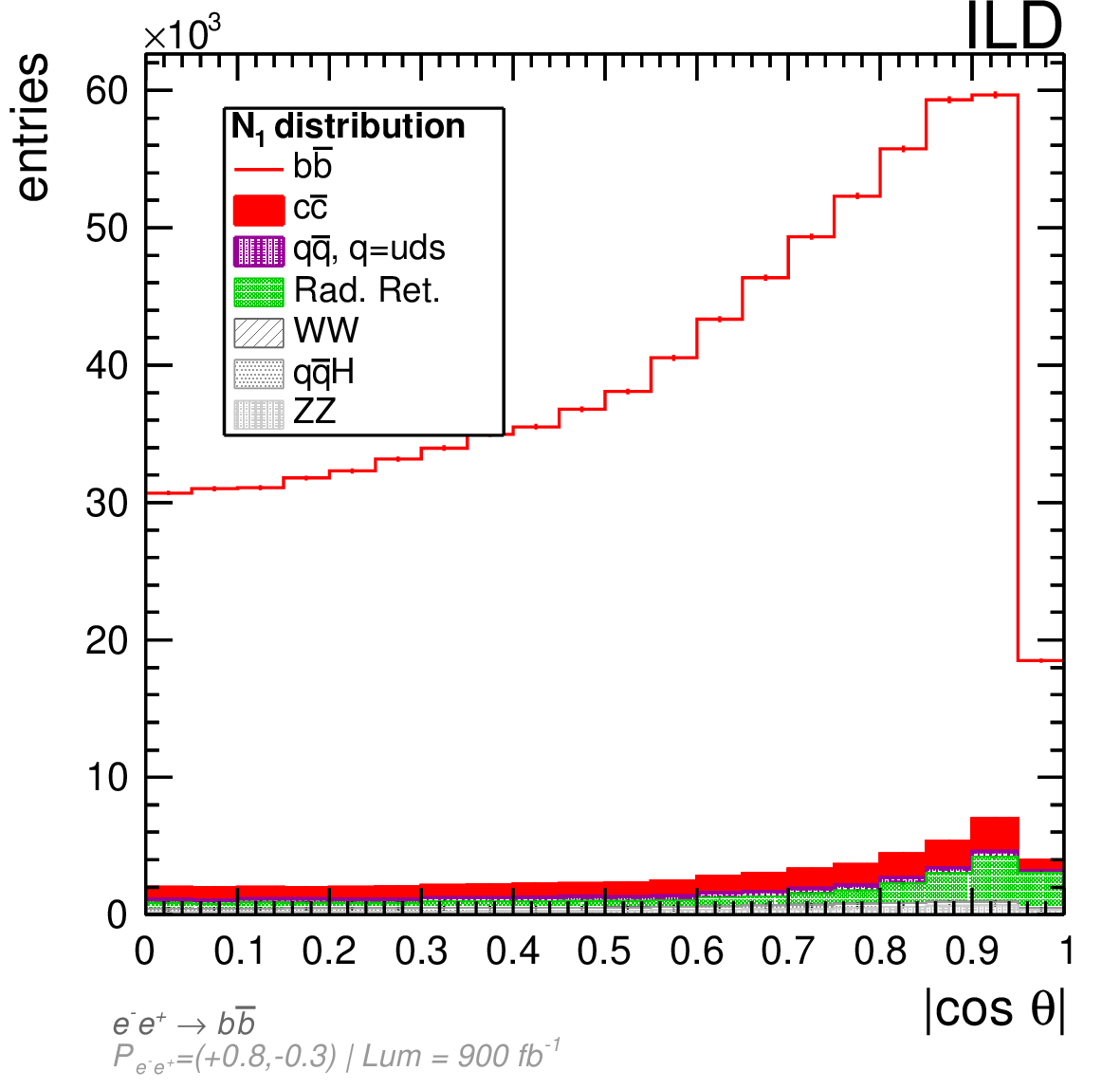} &
      \includegraphics[width=0.3\textwidth]{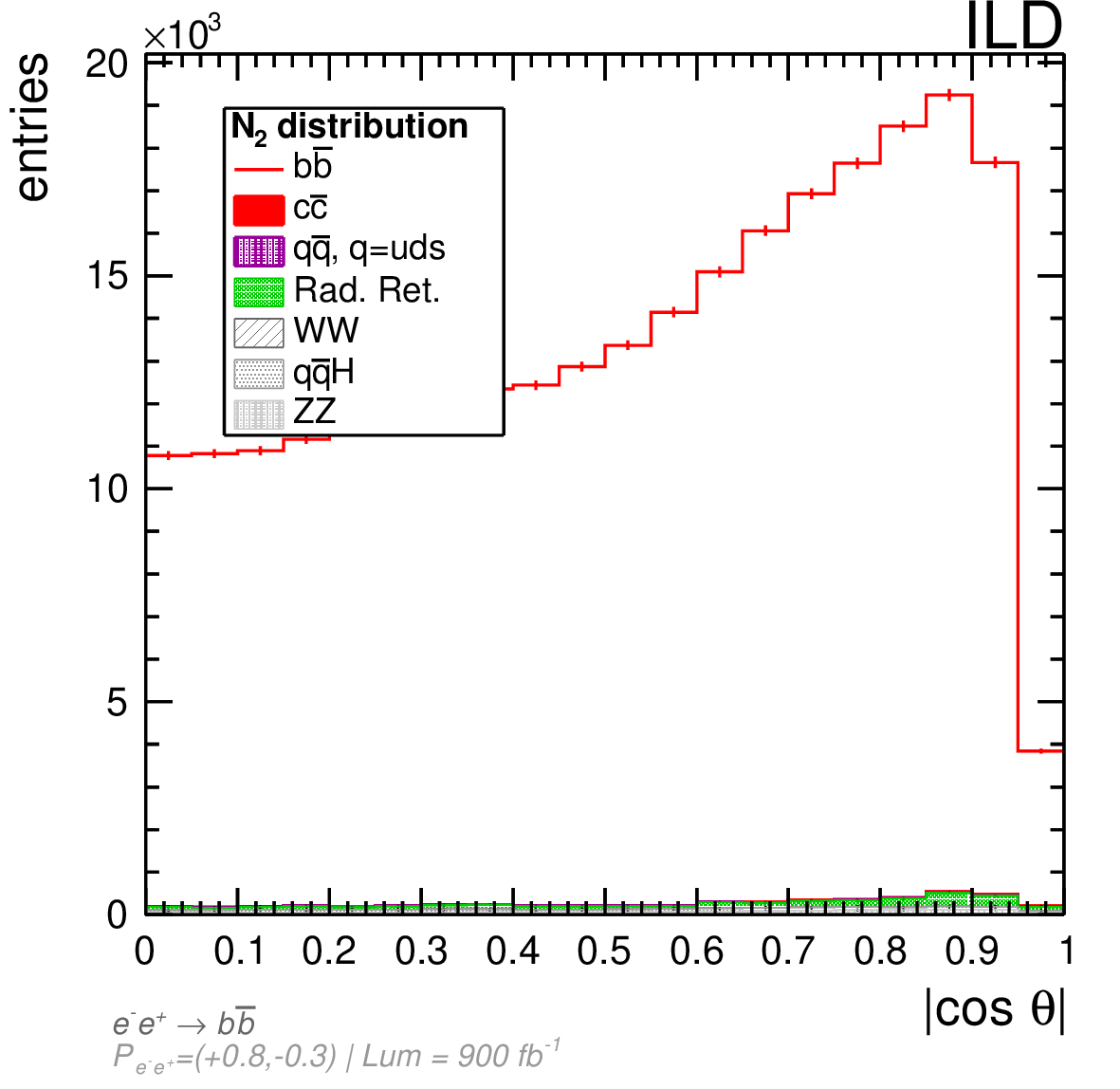} \\
    \end{tabular}
\caption{The $N_{0}$, $N_{1q}$ and $N_{2q}$ (signal and background events featuring none, at least one or two tags for the $q$ flavour respectively) stacked histograms are shown in three columns, before background rejection. Each row corresponds to a different process or beam polarisation condition. The ILC250 luminosity scenario is assumed.
\label{fig:N0N1N2}}
\end{center}
\end{figure}

\begin{figure}[!ht]
\begin{center}
    \begin{tabular}{cc}
      \includegraphics[width=0.45\textwidth]{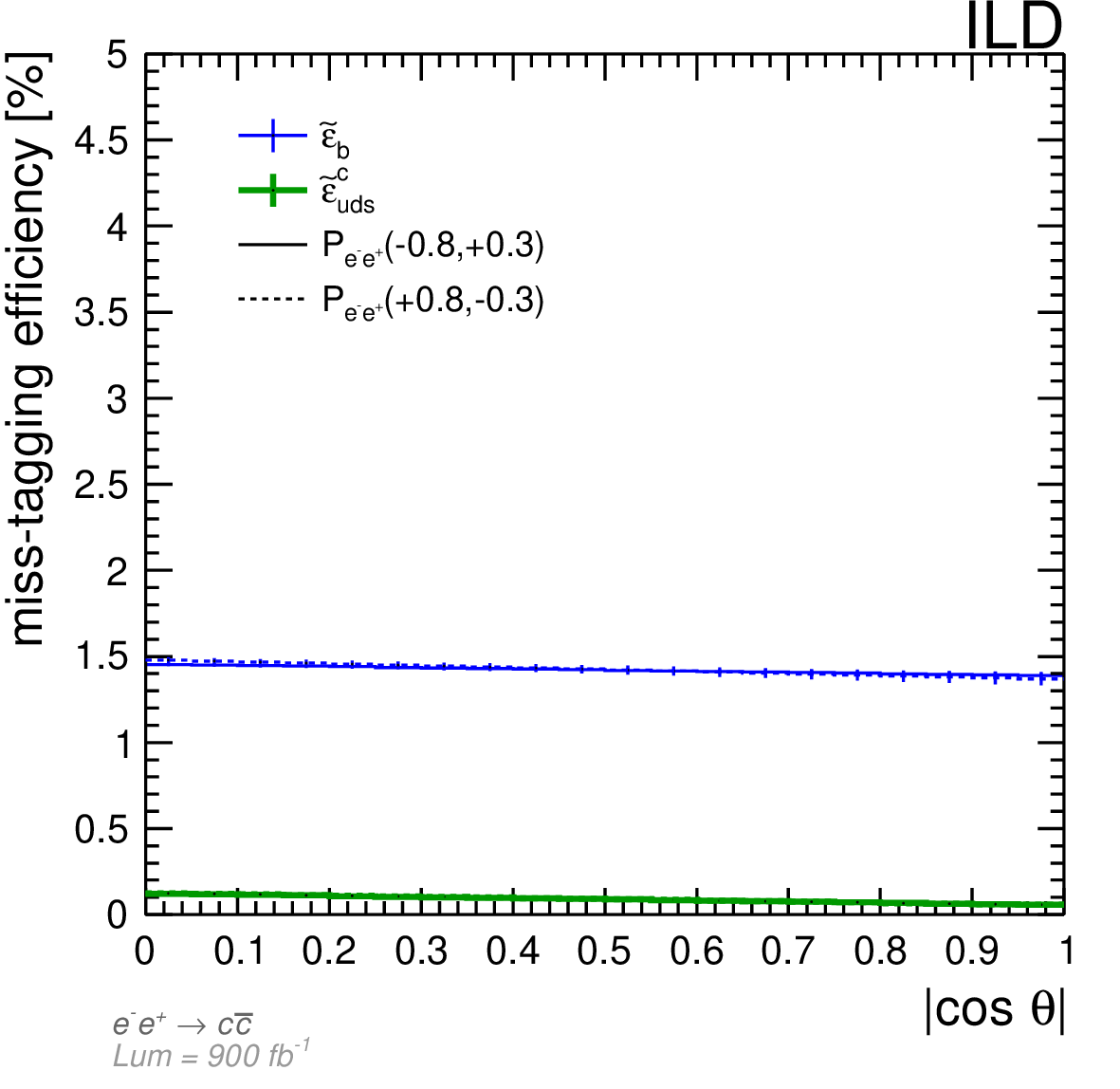} & \includegraphics[width=0.45\textwidth]{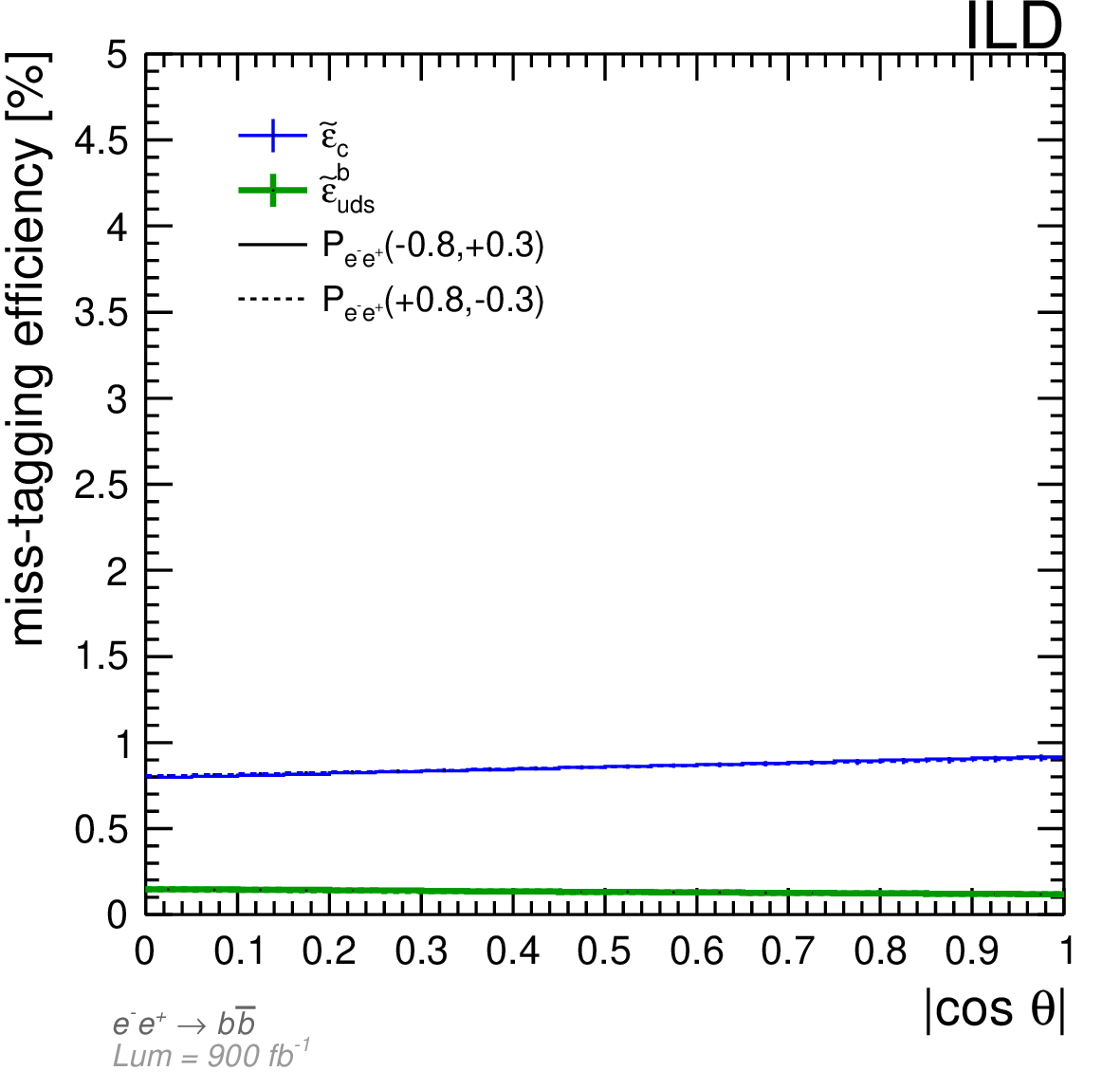} \\
      \includegraphics[width=0.45\textwidth]{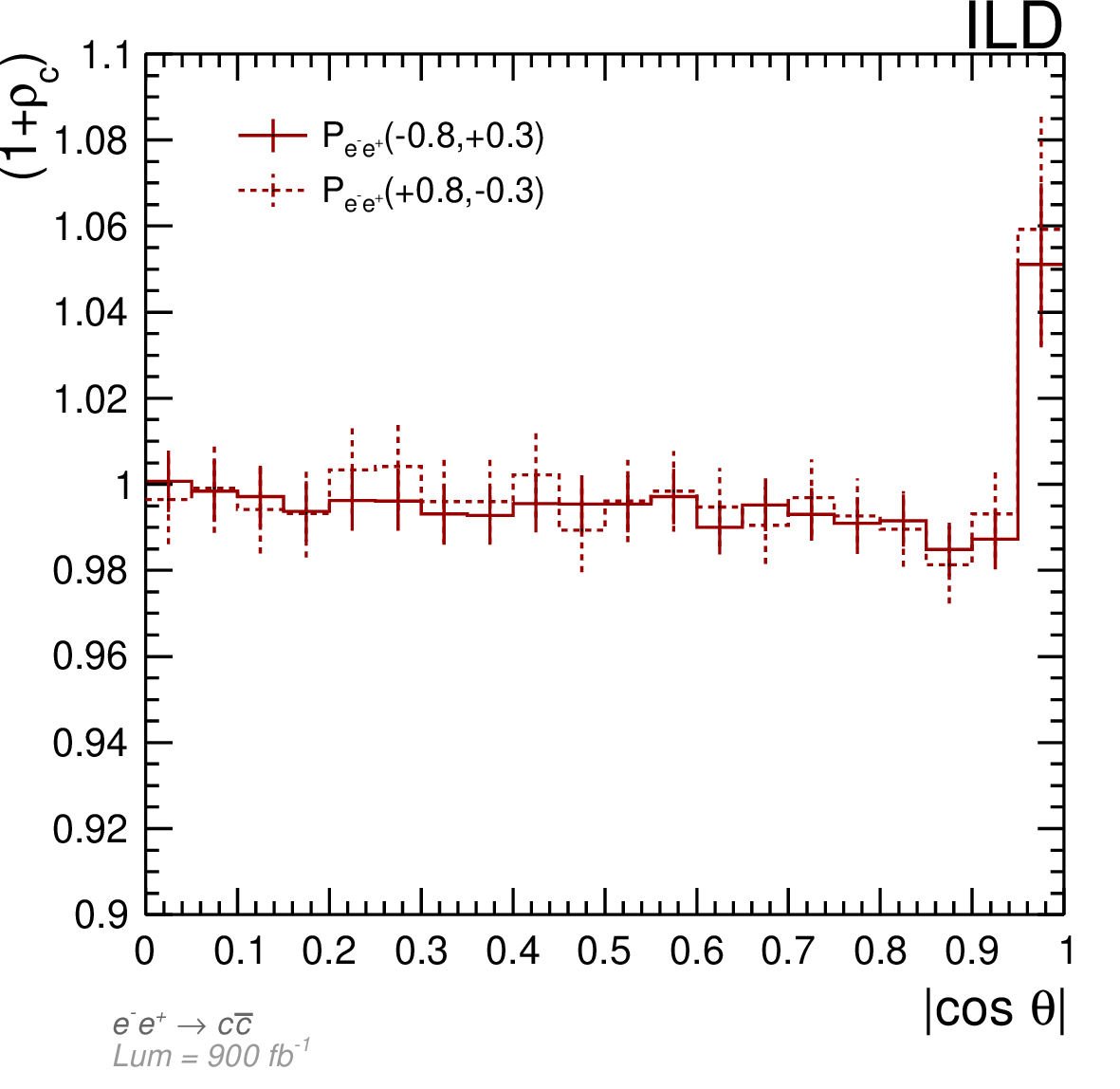} & \includegraphics[width=0.45\textwidth]{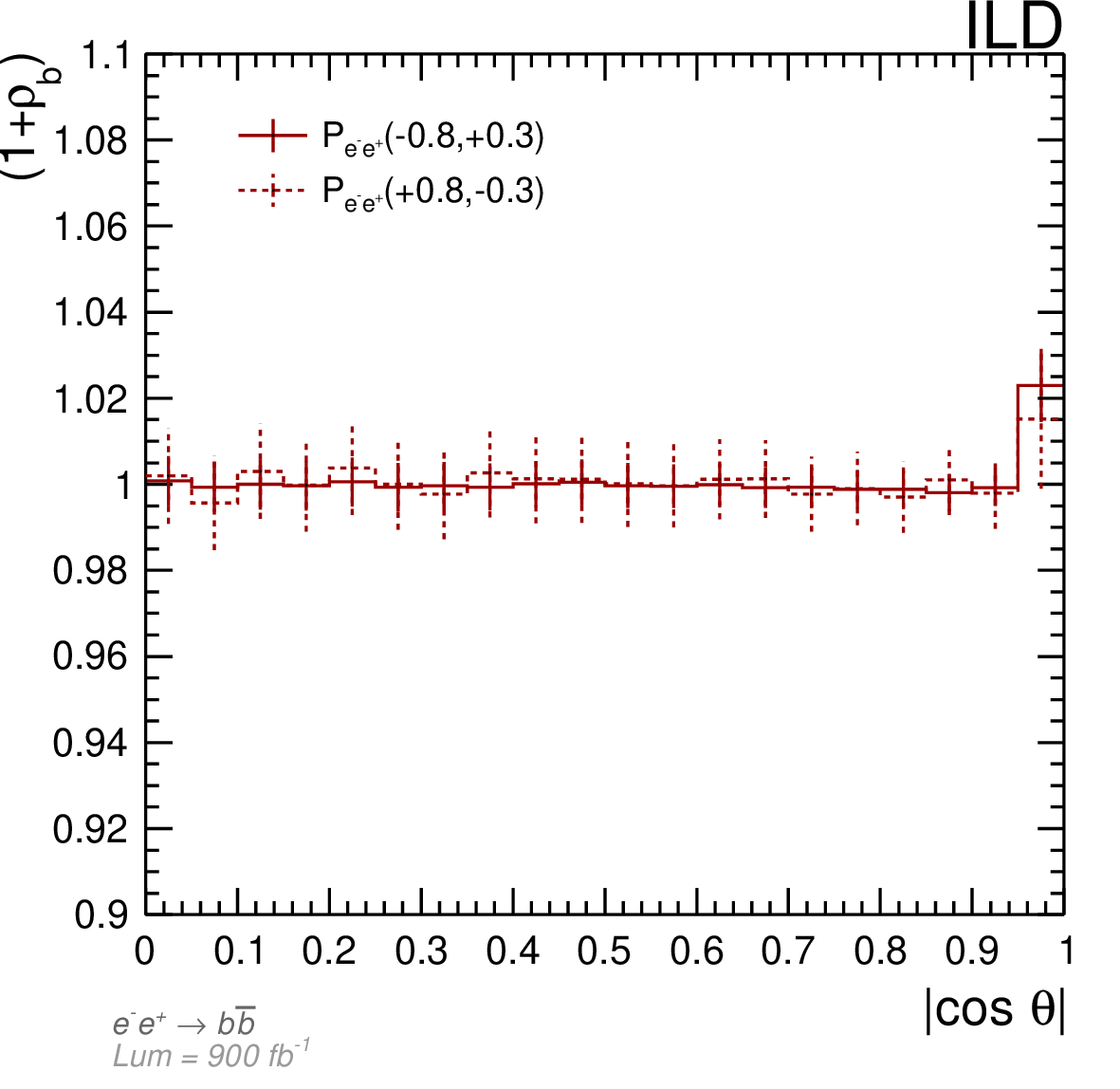}  
    \end{tabular}
\caption{Monte Carlo based estimations for the mis-tagging efficiencies (upper row) and geometrical angular correlation factors (lower row) for \eecc (left) and \eebb (right). }
\label{fig:mistag}
\end{center}
\end{figure}

\begin{figure}[!ht]
\begin{center}
    \begin{tabular}{cc}
      \includegraphics[width=0.45\textwidth]{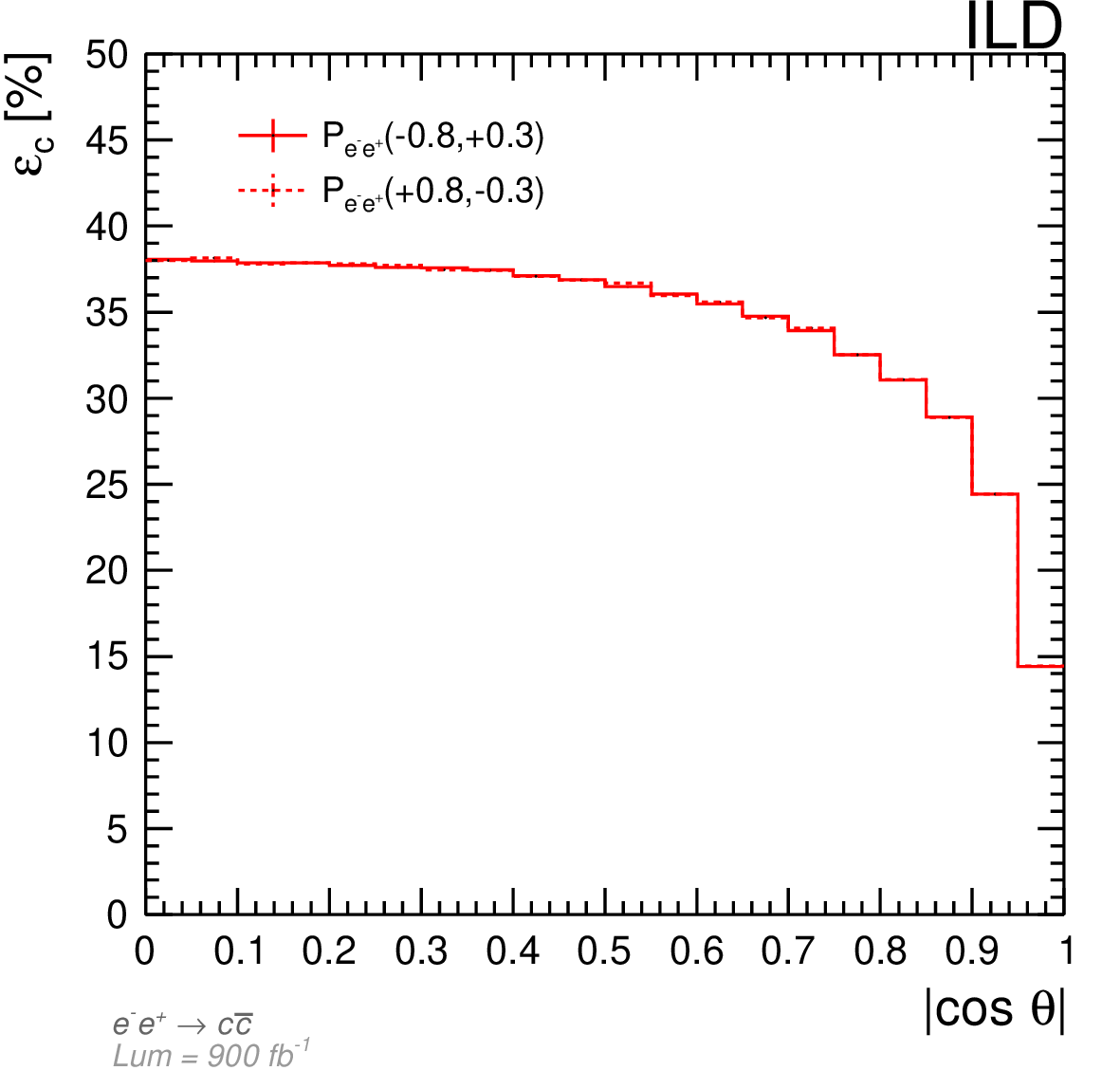} & \includegraphics[width=0.45\textwidth]{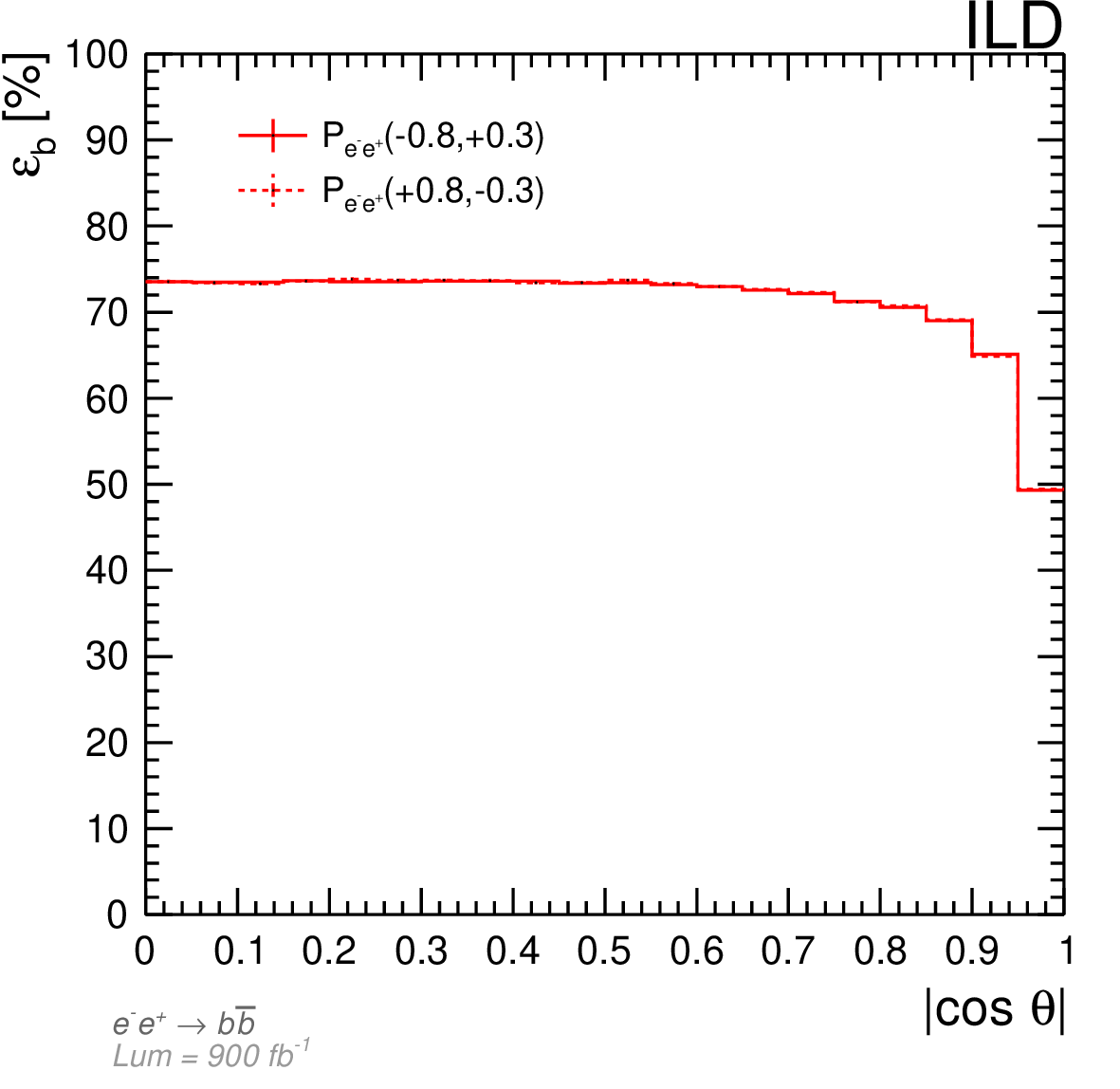} \\
      \includegraphics[width=0.45\textwidth]{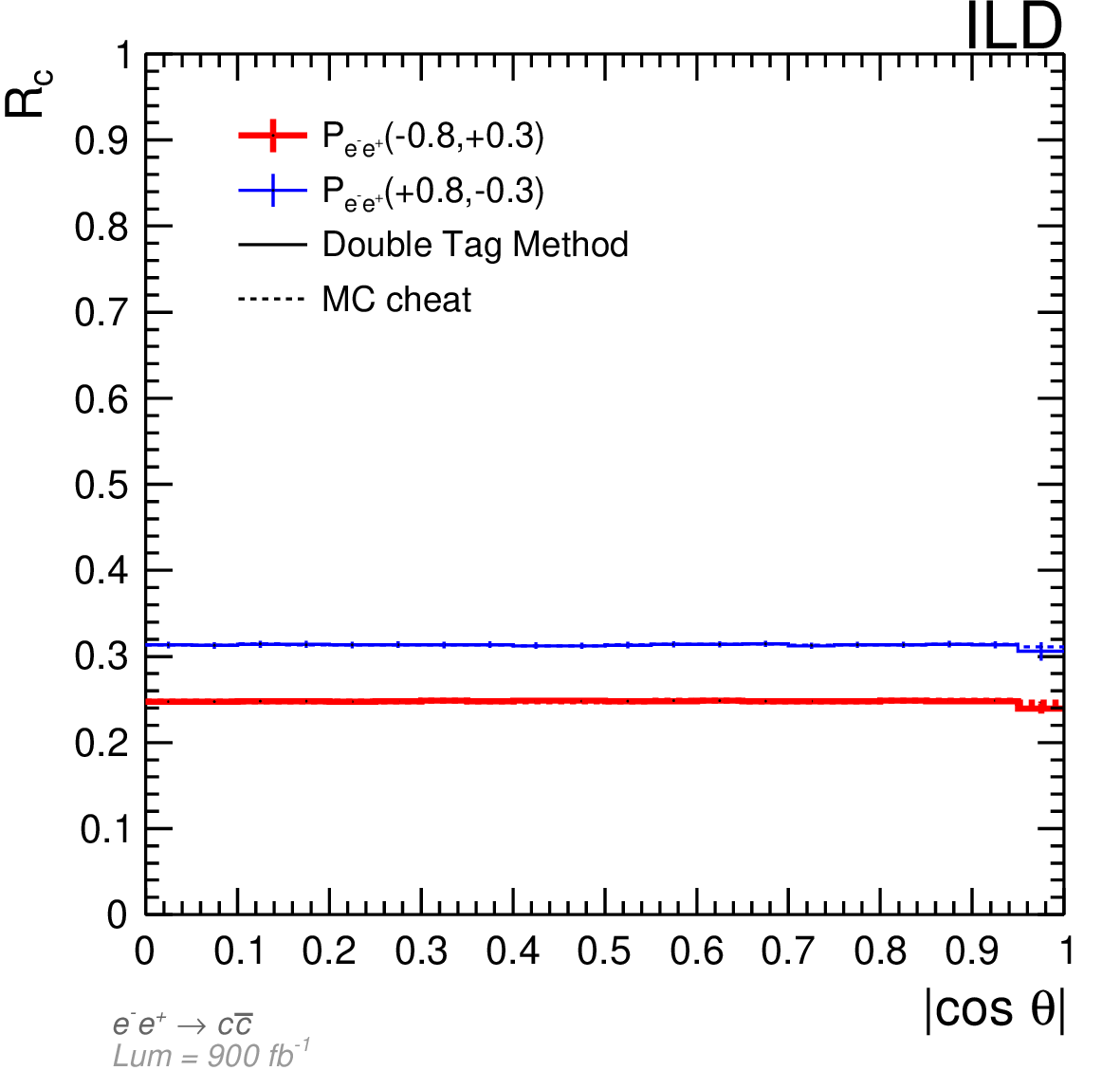} & \includegraphics[width=0.45\textwidth]{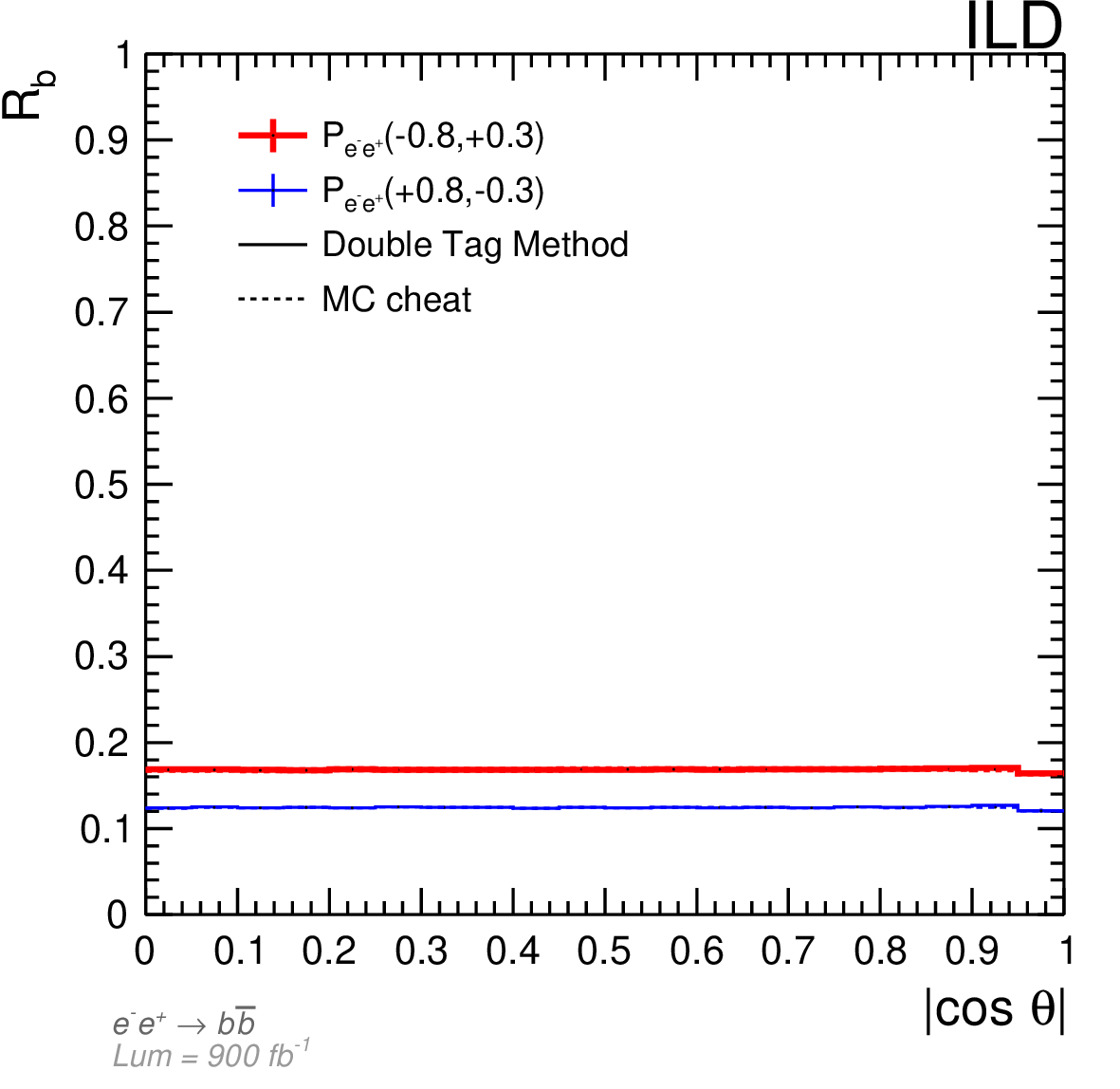}
    \end{tabular}
\caption{Extracted $\epsilon_{q}$ and \Rq using the DT method for \eecc (left) and \eebb (right). For the estimation of \Rq the comparison with the case in which the Monte Carlo is used to estimate the $\epsilon_{q}$ is included, showing no difference with the DT expectations.
\label{fig:RbRc}}
\end{center}
\end{figure}

The $N_{0}$, $N_{1q}$ and $N_{2q}$ distributions including background as a function of \costheta 
are shown in Figure \ref{fig:N0N1N2} for the $b$ and \cquark cases. 
All three columns demonstrate the suppression power of the pre-selection cuts 
introduced in Sec.~\ref{sec:analysis_preselection}. The background is at most 10\% in the case of single-tagged events and almost negligible for double-tag events. The \bquark and \cquark samples mutually contaminate each other and both are contaminated by light quarks, see Eq.~\ref{eq:Rq}. The mis-tagging rates have been determined using the Monte Carlo truth information. As shown in Fig.~\ref{fig:mistag} they are of the order of 1.5\% (0.8\%) for mis-tagging a \bquark (\cquark) as a \cquark (\bquark). The mis-tagging rate for light quarks is as small as 0.1\%. All mis-tagging rates are constant over the entire polar angle.  
The lower two panels of Fig.~\ref{fig:mistag} show the hemisphere correlations. The expected ($1+\rho_{q}$) for ILC250 
is constant in most of the detector volume and $\rho_{q}$ amounts to a negligible value smaller than $0.2\%$. We emphasized that the beam polarisation does not impact our estimations of the discussed quantities.
The visible deviation from one towards $|\costheta|=1$ can be qualitatively explained by the degradation of
the quality of the vertex reconstruction due to the acceptance limits of the vertex detector. 
%In this method, the statistical uncertainty is determined by the size of
%the DT sample, which is proportional to the square of the tagging efficiency. 
%The angular distribution of the estimated mis-tagging efficiencies and correlation factors 
%are shown in 
%Figure \ref{fig:mistag} for the \cquark and \bquark cases.
%For the estimation of these quantities, and the backgrounds, the truth information in the simulations is used.
%For the estimation of the tagging efficiencies and the \Rc and \Rb,
%the simulations are treated as data and the DT method is applied. 
Using mis-tagging rates, hemisphere correlations and the respectively other set \Rqp, $\tilde{\varepsilon}_{q\prime}$ as input, the equation system in Eq. \ref{eq:Rq} can be solved in each bin of $|\costheta|$ for \Rb (\Rc) and $\varepsilon_{b}$ ($\varepsilon_{c}$). The results are shown in Fig.~\ref{fig:RbRc}.
The maximum value 0.75 for $\varepsilon_{b}$ is reached for $|\costheta|=0$ and at around $|\costheta|=0.9$ it is still about 0.65 before a drop can be observed at acceptance limit.  For $\varepsilon_{c}$ the maximum value of 0.38 is reached for $|\costheta|=0$. The progressive degradation is stronger than in the case of $\varepsilon_{b}$. However, at $|\costheta|=0.9$ $\varepsilon_{c}$ is still around 0.25 before again a stronger drop can be observed at the acceptance limit. 

%By measuring the ratios
%defined in Eq. \ref{eq:Rb} and \ref{eq:Rc} the efficiencies can be measured
%as a function of the angle, without depending on Monte Carlo.
%The same is possible for \Rb and \Rc as it is shown in Figure \ref{fig:RbRc} 
%how these quantities are measured for different $\costheta$ angles.

The observable \Rq has been extracted in two ways. The first is given by the simple ratio of the reconstructed \bquark or \cquark di-jet events to the total number di-jet events. The second is done using differential observables:  the lower panel of Figure \ref{fig:RbRc} \Rb and \Rc have been determined bin-wise as a function of $|\costheta|$; then the constant term of a linear fit to this distribution is used to extract \Rb and \Rc. In the Standard Model the slope of the linear fit is expected to be zero as is the case in Fig.~\ref{fig:RbRc}. A deviation from zero would be a hint for either new physics or for an insufficient correction for detector effects. For this analysis, it is important to note that the two ways of extracting \Rb and \Rc lead to the same results, which is by itself an useful consistency check. As a further consistency check \Rq has also been extracted using Monte Carlo truth for the calculation of $\epsilon_{q}$. This is labelled as \textit{MC cheat} in Figure \ref{fig:RbRc}. 

%Two ways of extracting \Rq have been used, leading to the same results: perform a linear fit 
%of the extracted $\Rq(|\costheta|)$ and by integrating the distribution of reconstructed events (since we know the 
%differential form of the efficiencies).  The former method gives us extra information
%since allows for a fully differential measurement and could be used to isolate from the analysis 
%detector regions with non-modelled issues as could be dead or low efficiency sectors.

The study of uncertainties, including the most relevant systematic uncertainties, will be addressed in Sec.~\ref{sec:systematics}.

%% file: sections/7_analysis_AFB.tex
\section{Measurement of differential cross sections and determination of \Afb}
\label{sec:analysis_AFB}

The measurement starts with doubly tagged (DT) \bquark and \cquark samples, as described in the last section. There it is shown that these samples are almost background free.
The measurement of the differential cross-sections and hence of \Afb requires the measurement of the jet polar angles \emph{and} of the jet charges. 
For the latter two jet-charge methods will be introduced; the $Vtx$ and $K$-methods.

\subsection{Jet-charge measurement method 1: Vertex Charge, \Bc}
\label{sec:analysis_Bc}

The charge of the jet is estimated as the charge of the vertex, defined as the sum of the charges of all tracks in the secondary vertices in the jet. The quarks fragment into charged and neutral hadrons. The charged hadrons are those that can be used for this method. A \bquark fragments with a probability of around 60\% into charged hadrons. 
The method requires that all charged tracks of the $b$ or $c$-hadron decays are correctly measured and are associated with the secondary vertex within the jet. The probability of losing a track is very small, as discussed in Section 
\ref{sec:simulation}, but its effect is enhanced in the case of the \bquark due to the large number of secondary tracks per jet.

\subsection{Jet-charge measurement method 2: Kaon Charge, \Kc}
\label{sec:analysis_Kc}

In this case, the charge of the jet is reconstructed as the sum of all the identified kaons reconstructed in secondary vertices inside the jet.

\begin{itemize}
    \item \textbf{Charm quark:} $c$-quarks mostly fragment into $D^{0}/D^{\pm}/D_{s}$-mesons. The decay branching ratio of $D^0$ to charged kaons is $\sim50\%$. For the $D_{s}$ this number is somewhat lower: $\sim33\%$. The $D^{\pm}$ produce one and three prongs in their decays, with only $\sim30\%$ of the cases having a charged kaon in the final state. In all cases, identifying a kaon in a secondary vertex gives direct information on the charge of the original $c$-quark.
    \item \textbf{Bottom quark:} The CKM matrix elements $|V_{cb}|$ and $|V_{us}|$ are significantly different from 0, and $|V_{cs}|\approx 1$. Therefore, $B$-hadron decays yield a sizeable fraction of charged Kaons in the final state. 
    It is expected to have $\sim$0.8 charged kaons and $\sim$3.6 charged $\pi$ per $B$-hadron decay while the multiplicity of protons is of $\sim$0.13 \cite{ParticleDataGroup:2020ssz}. 
\end{itemize}

For the kaon identification the TPC \dEdx is used to identify charged kaons
in the secondary tracks as described in Section \ref{sec:simulation}.
For the charge measurement, it is allowed to use more than one charged kaon:
$K^{-}K^{-}$, $K^{-}K^{-}K^{+}$ combinations (and inverted signs) are accepted while the
$K^{-}K^{+}$ combination are not used.

\subsection{Double Charge method (DC)}
\label{sec:analysis_migrations}

The Double Charge method (DC) requires two opposite-charged jets.
It starts with a selection of $N$ events containing two jets with measured charges. Jets with opposite charge are accepted. Those with the same charge are rejected. Let \Pb be the probability that the jet charge reproduces the sign of the charge of the quark of the hard scattering. Hence, $\Qb=1-\Pb$ is the probability that this is not the case. Supposing that the jet-charge measurements are independent and symmetric between two hemispheres it is simply:  

\begin{equation}
    \begin{aligned}
    %N^{M}_a(\costheta>0)=\\ =\Pb^2(|\costheta|)\cdot f^{2}_{M}(|\costheta|)N_{sample}(\costheta>0) + \Qb^2(|\costheta|)\cdot f^{2}_{M}(|\costheta|)N_{sample}(\costheta<0)\\
    %N^{M}_a(\costheta<0)=\\ =\Pb^2(|\costheta|)\cdot f^{2}_{M}(|\costheta|)N_{sample}(\costheta<0) + \Qb^2(|\costheta|)\cdot f^{2}_{M}(|\costheta|)N_{sample}(\costheta>0)\\
    N^{M}_{acc.}(|\costheta|)=\\ =\Pb^2(|\costheta|)\cdot N(|\costheta|) + \Qb^2(|\costheta|)\cdot N(|\costheta|)
%    N^{M}_{rej.}(|\costheta|)=2\cdot\Pb(|\costheta|)\cdot \Qb(|\costheta|)\cdot N(|\costheta|)
    \end{aligned}
    \label{eq:purity}
\end{equation}
where $N^{M}_{acc.}$ is the number of events with compatible charges in both jets (opposite sign) 
measured with the method $M$. 

The obtained value of \Pb, after solving Eq. \ref{eq:purity}, allows us to calculate how the events found in a bin of $|\costheta|$ should be distributed between the bins at 
either $+|\costheta|$ or $-|\costheta|$ in the following way:
\begin{equation}
\begin{aligned}
   N^{corr.}(|\costheta|)=\frac{\Pb^{2} N_{acc.}(\costheta>0) - \Qb^{2} N_{acc.}(\costheta<0)}{\Pb^{4}- \Qb^{4}} \\
   N^{corr.}(-|\costheta|)=\frac{\Pb^{2} N_{acc.}(\costheta<0) - \Qb^{2} N_{acc.}(\costheta>0)}{\Pb^{4}- \Qb^{4}} 
\end{aligned}
\label{eq:corr}
\end{equation}
For simplicity, in this equation, only the case of using the same charge measurement method in both jets is shown.
The generalisation is straightforward.

\begin{figure}[!ht]
\begin{center}
    \begin{tabular}{cc}
      \includegraphics[width=0.45\textwidth]{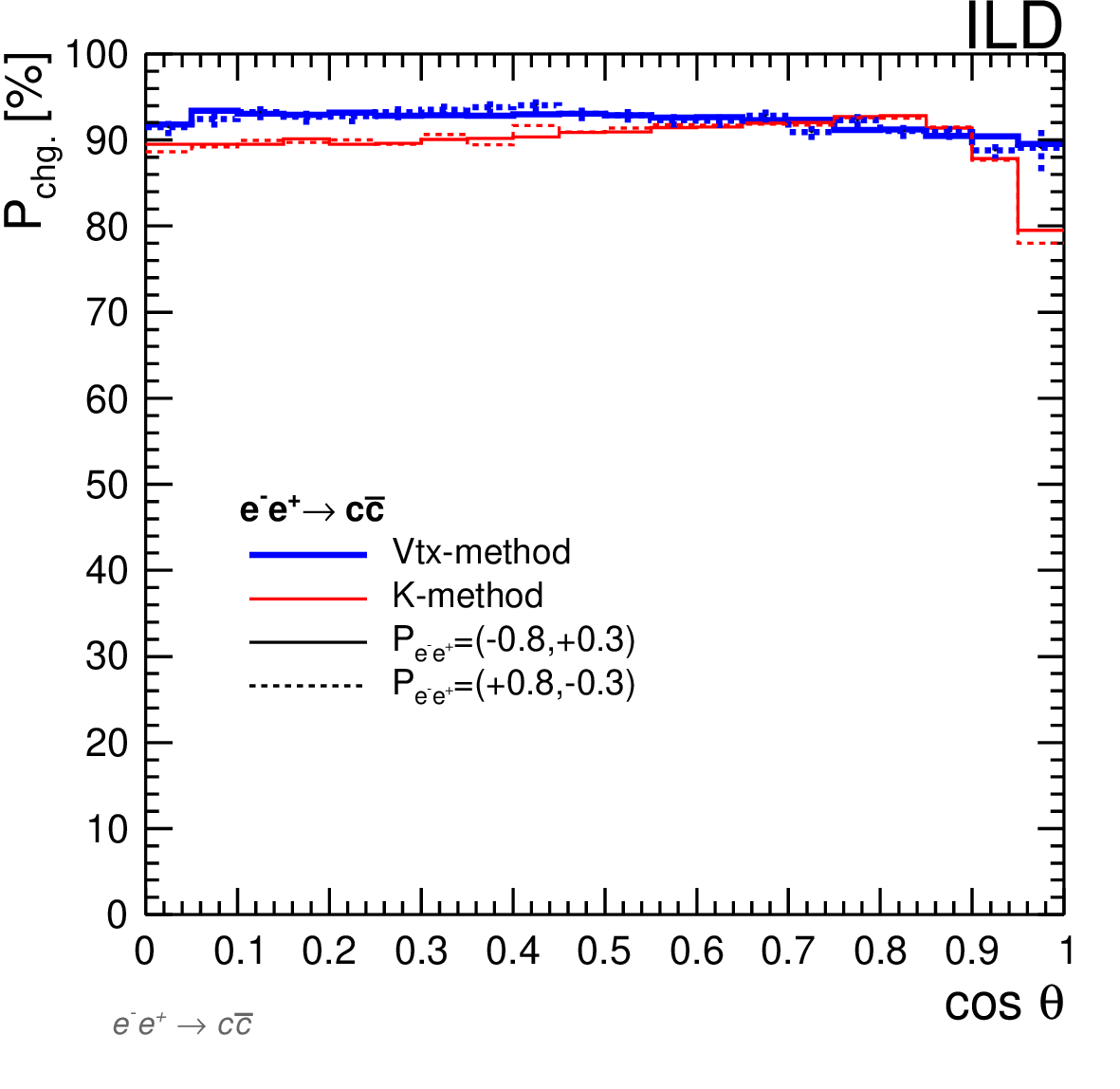} &
      \includegraphics[width=0.45\textwidth]{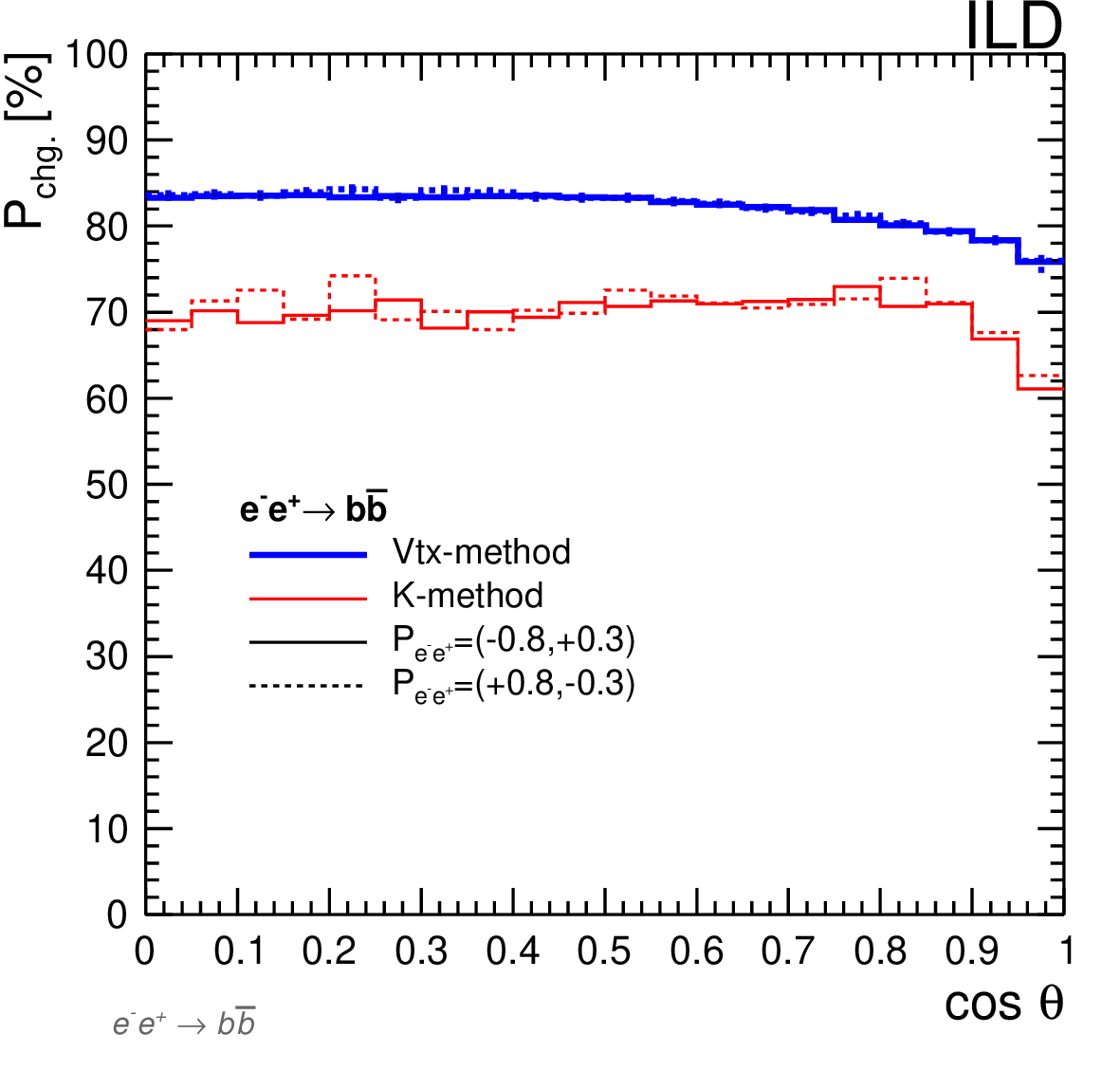} 
    \end{tabular}
\caption{Distributions for \Pb for the \Bc (blue) and the \Kc (red) for \ccbar (left) and \bbbar (right). The results are shown for different beam polarisation scenarios, using different types of lines. \label{fig:purity}}
\end{center}
\end{figure}

The resulting \Pb for the two different methods are shown in Figure \ref{fig:purity}. 
%for the \cquark and \bquark cases. 
One observes an approximately constant $\Pb(\costheta)$  except for the very forward region $|\costheta|>0.9$.
A slight slope is observed when using the \Kc. This effect compensates with a slightly smaller 
efficiency of the $K$-method for larger polar angles, as observed in Figure \ref{fig:purity}.

\begin{figure}[!ht]
\begin{center}
    \begin{tabular}{cc}
      \includegraphics[width=0.45\textwidth]{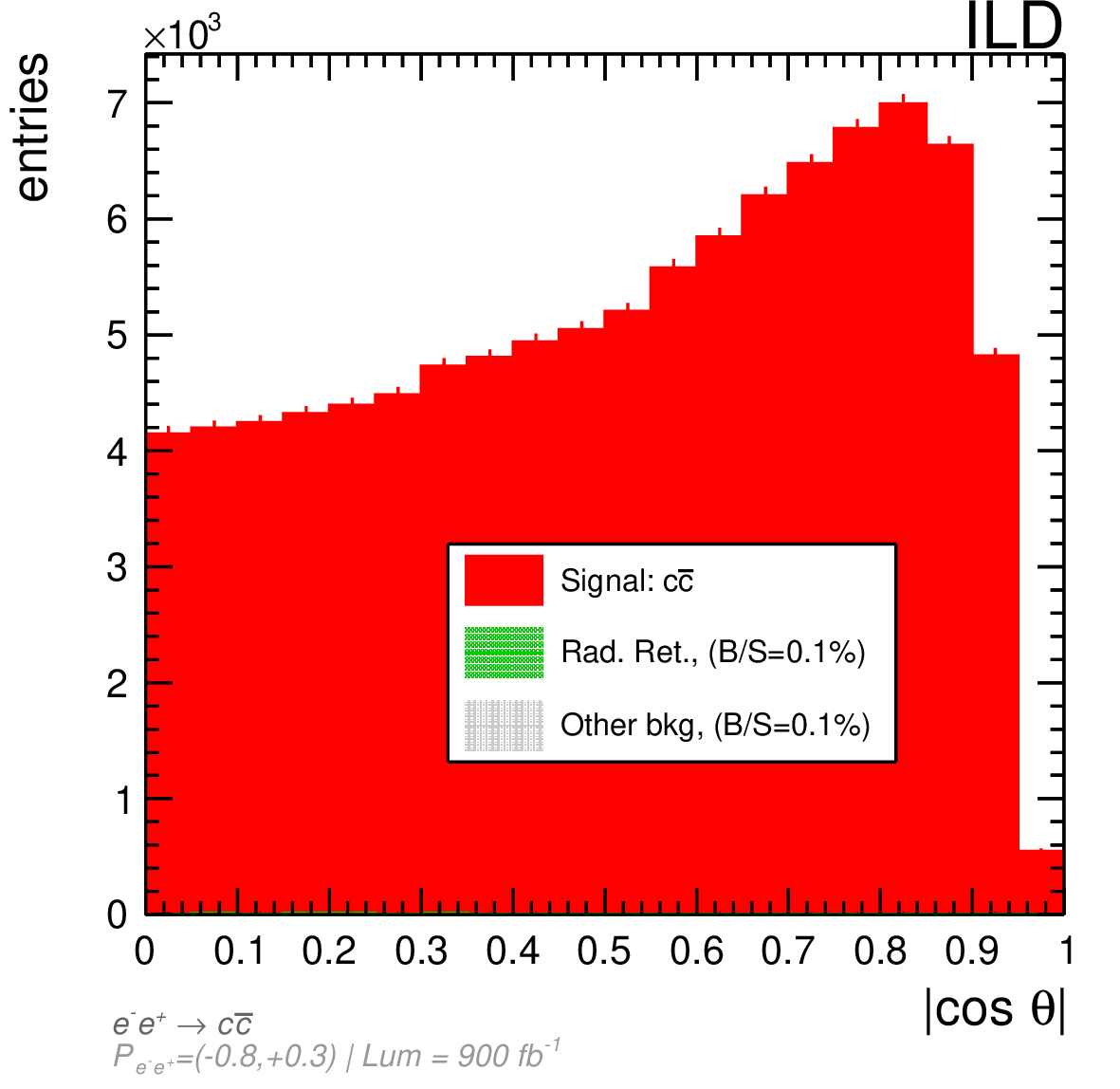} &
      \includegraphics[width=0.45\textwidth]{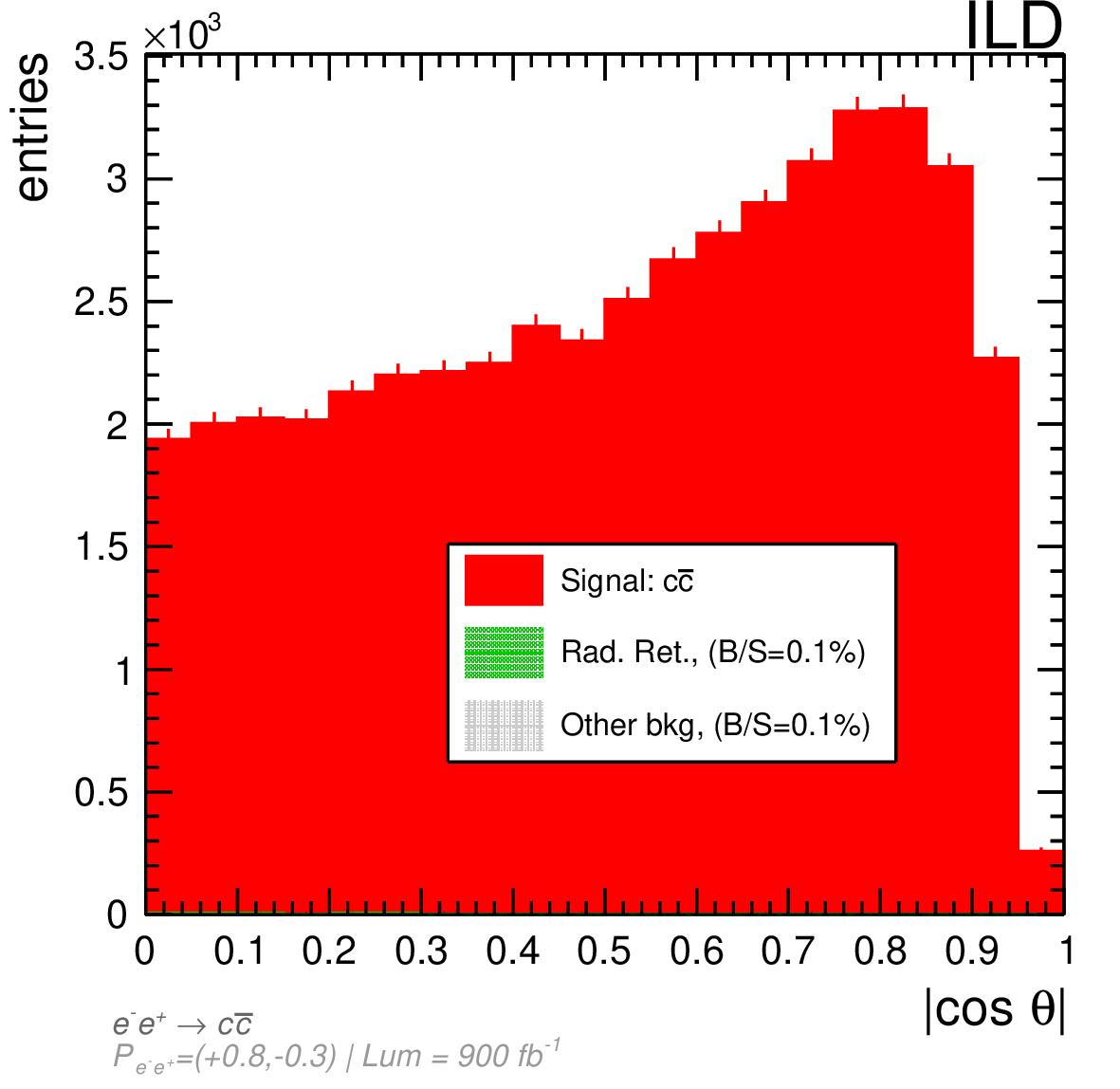} \\ 
      \includegraphics[width=0.45\textwidth]{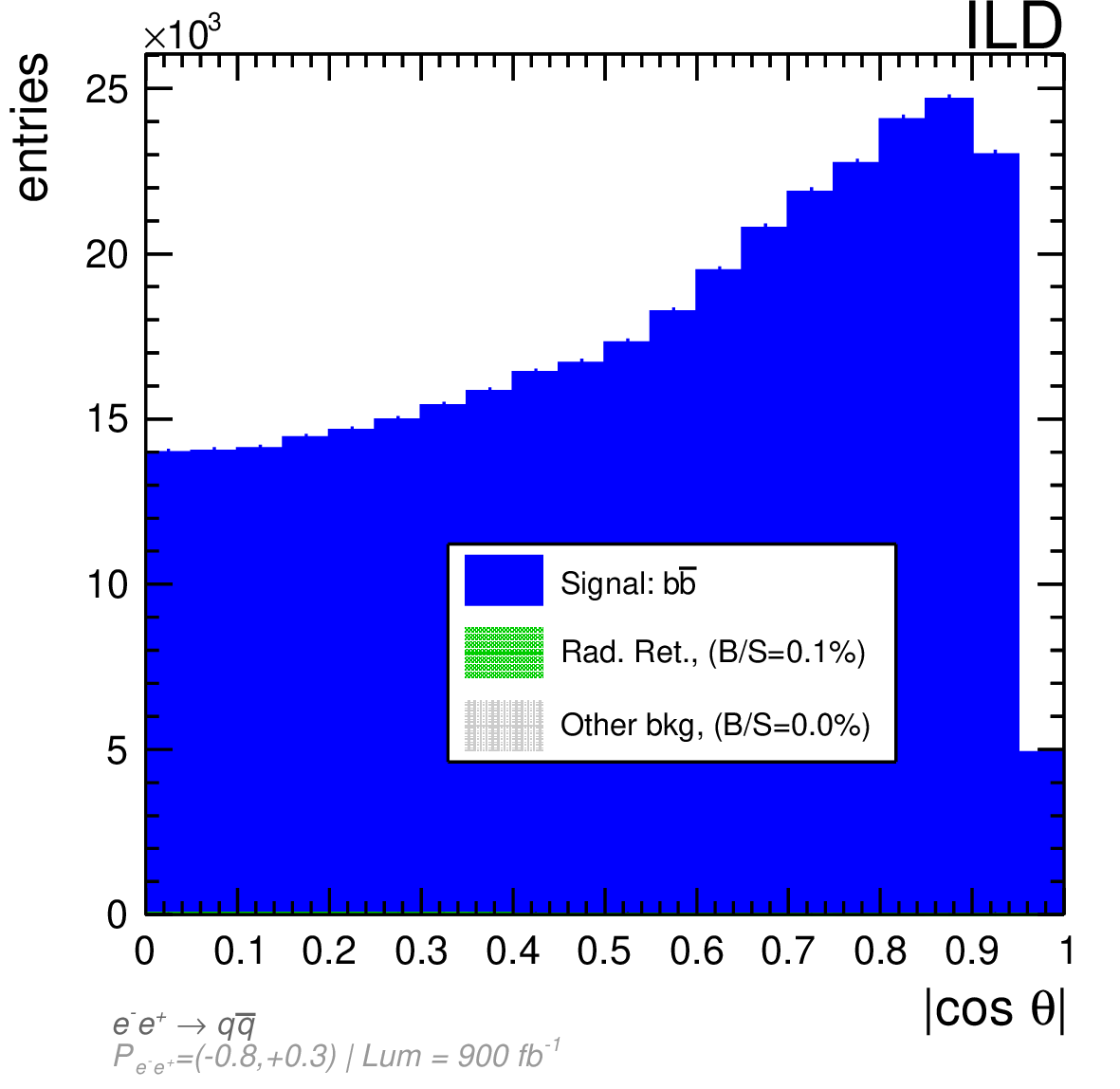} &
      \includegraphics[width=0.45\textwidth]{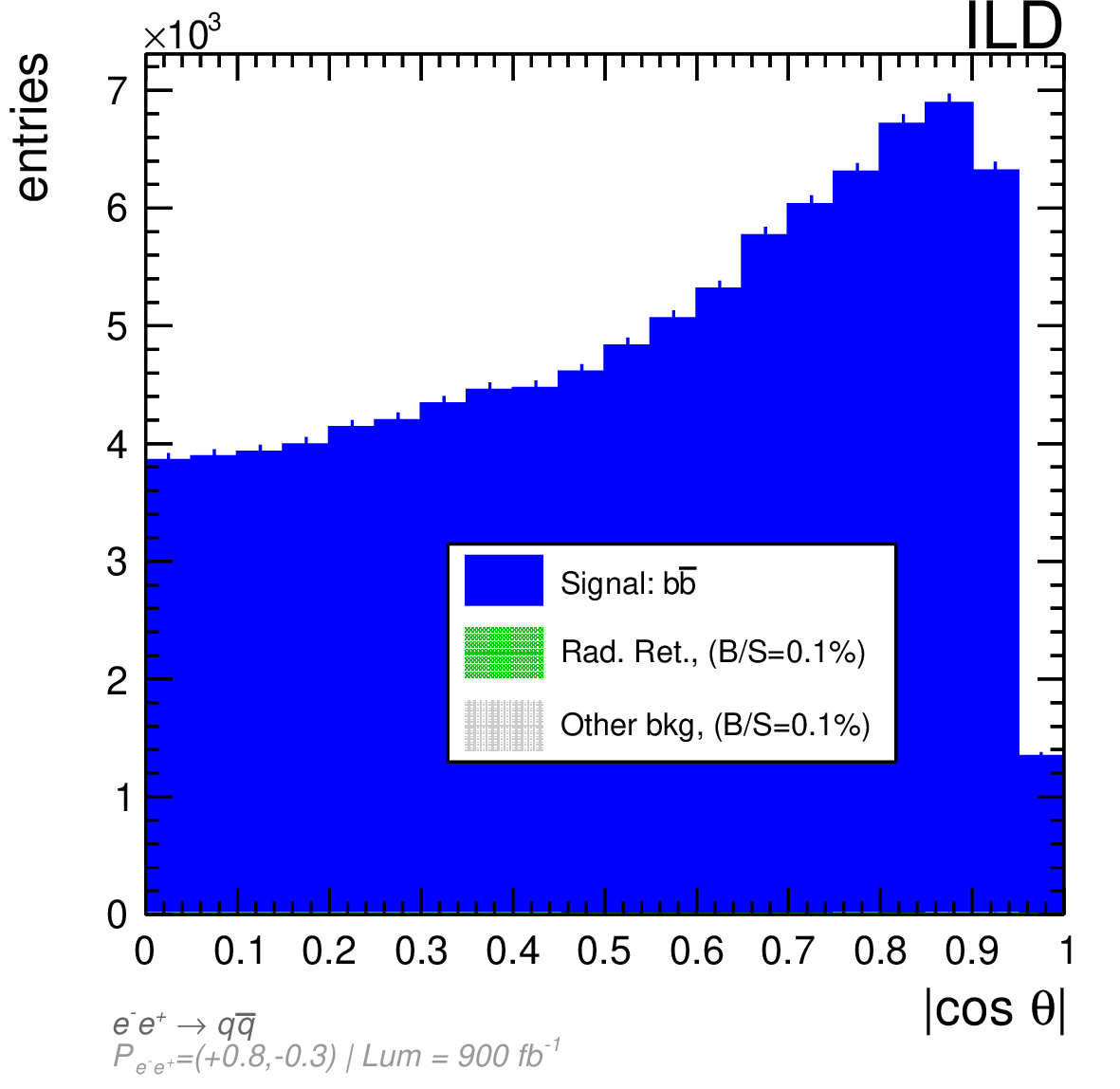} 
    \end{tabular}
      \caption{Signal and background sharing on the selected events for the \AFB measurement, before the determination of the sign of \costheta. The first row shows the distributions for the \cquark case and the second row for the \bquark case, while the left column shows the distributions for the $P_{\ee}=(-0.8,+0.3)$ and the right column for the $P_{\ee}=(+0.8,-0.3)$ cases. The backgrounds contribute with $\sim 0.1\%$ of the events in each histograms and, therefore, are barely visible in the figures. \label{fig:bkg}}
\end{center}
\end{figure}

Furthermore, since the \Bc and the \Kc have similar values of \Pb, it is also possible
to use mixed cases in cases where opposite jets do not use the same method.
In order to define the different categories, the leading method is defined as the one with higher efficiency. 
This allows to define a set of categories, $Cat._{i}$, for our double charge measurements: 
\begin{itemize}
    \item[$Cat._{1}$:] $M_{1}/M_{1}-$method in which both jets charge has been measured with the method $M_{1}$.
    \item[$Cat._{2}$:] $M_{1}/M_{2}-$method in which one of the jets had no measurement of the charge using the method $M_{1}$ but had it with method $M_{2}$.
    \item[$Cat._{3}$:] $M_{2}/M_{2}-$method in which none of the jets had a $M_{1}$ charge measurement but both had their charge measured by the method $M_{2}$
\end{itemize}
with the $M_{1}$-method being the \Kc and the $M_{2}$-method being the \Bc for the \cquark case and opposite for the \bquark case.
The distributions after the full reconstruction procedure, including the three categories, are shown in Figure \ref{fig:bkg}, before the calculation of the direction of the quark or anti-quark.  

\begin{figure}[!ht]
\begin{center}
    \begin{tabular}{cc}
      \includegraphics[width=0.45\textwidth]{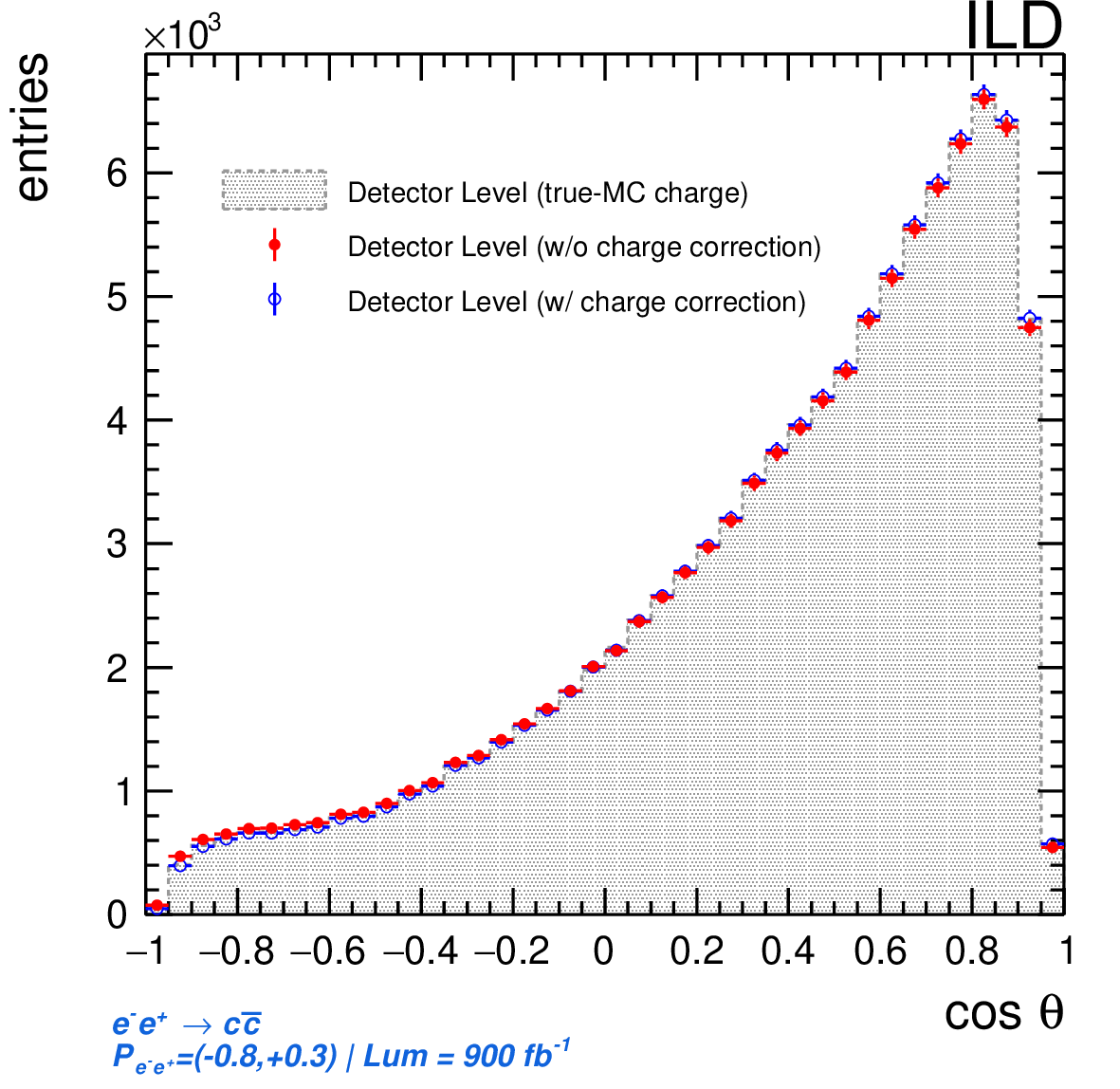} &
       \includegraphics[width=0.45\textwidth]{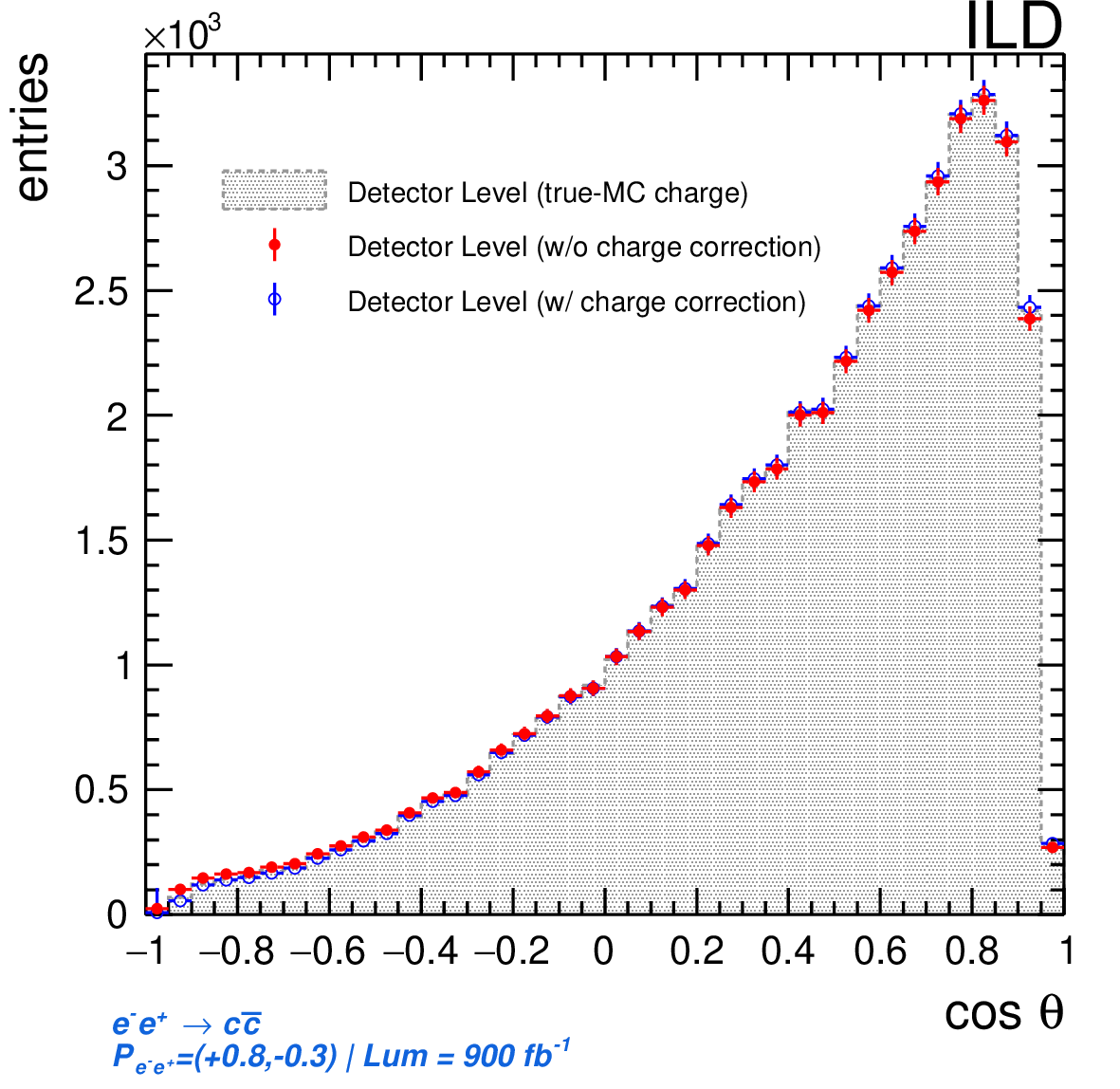} \\
       \includegraphics[width=0.45\textwidth]{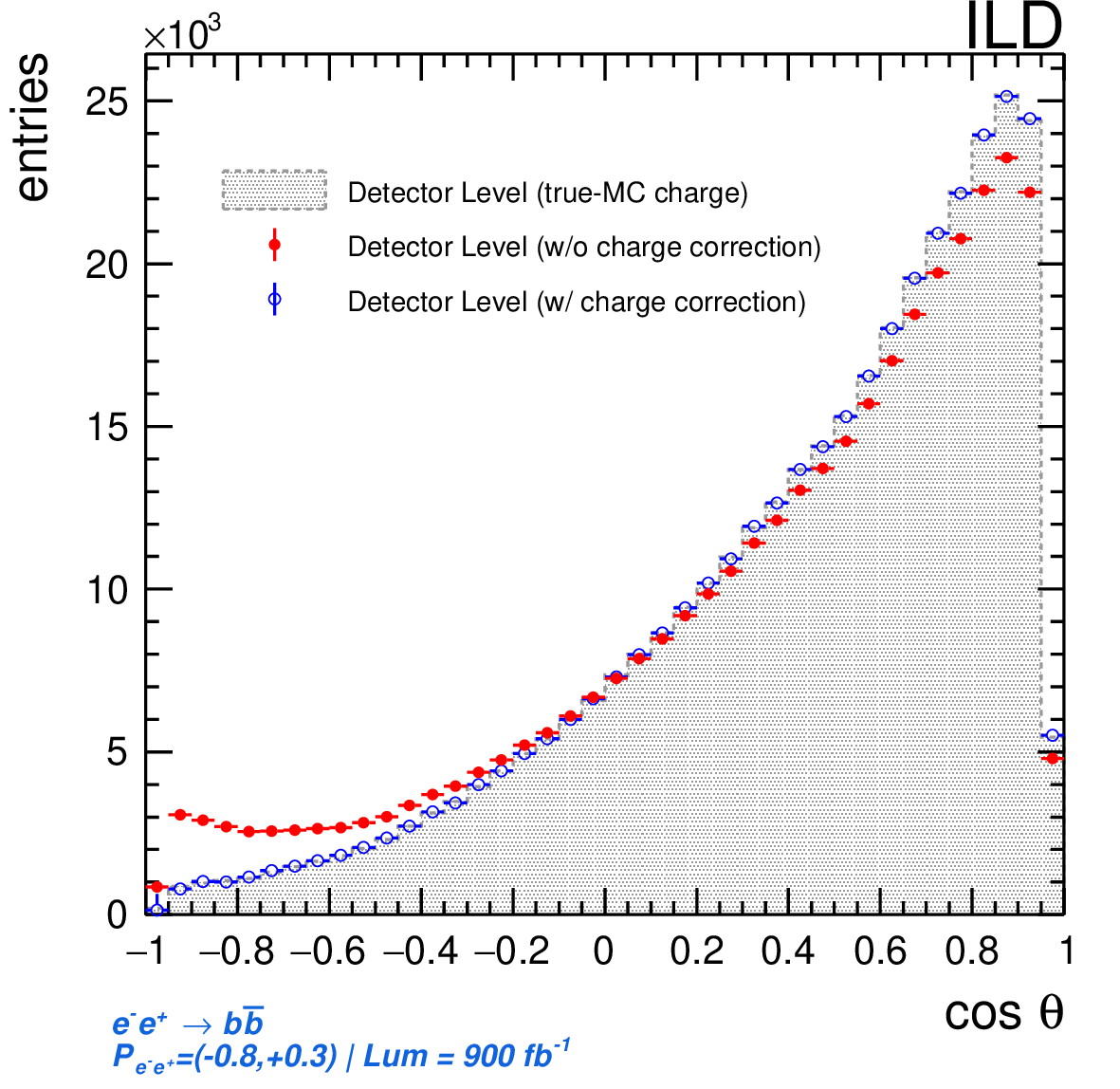} &
       \includegraphics[width=0.45\textwidth]{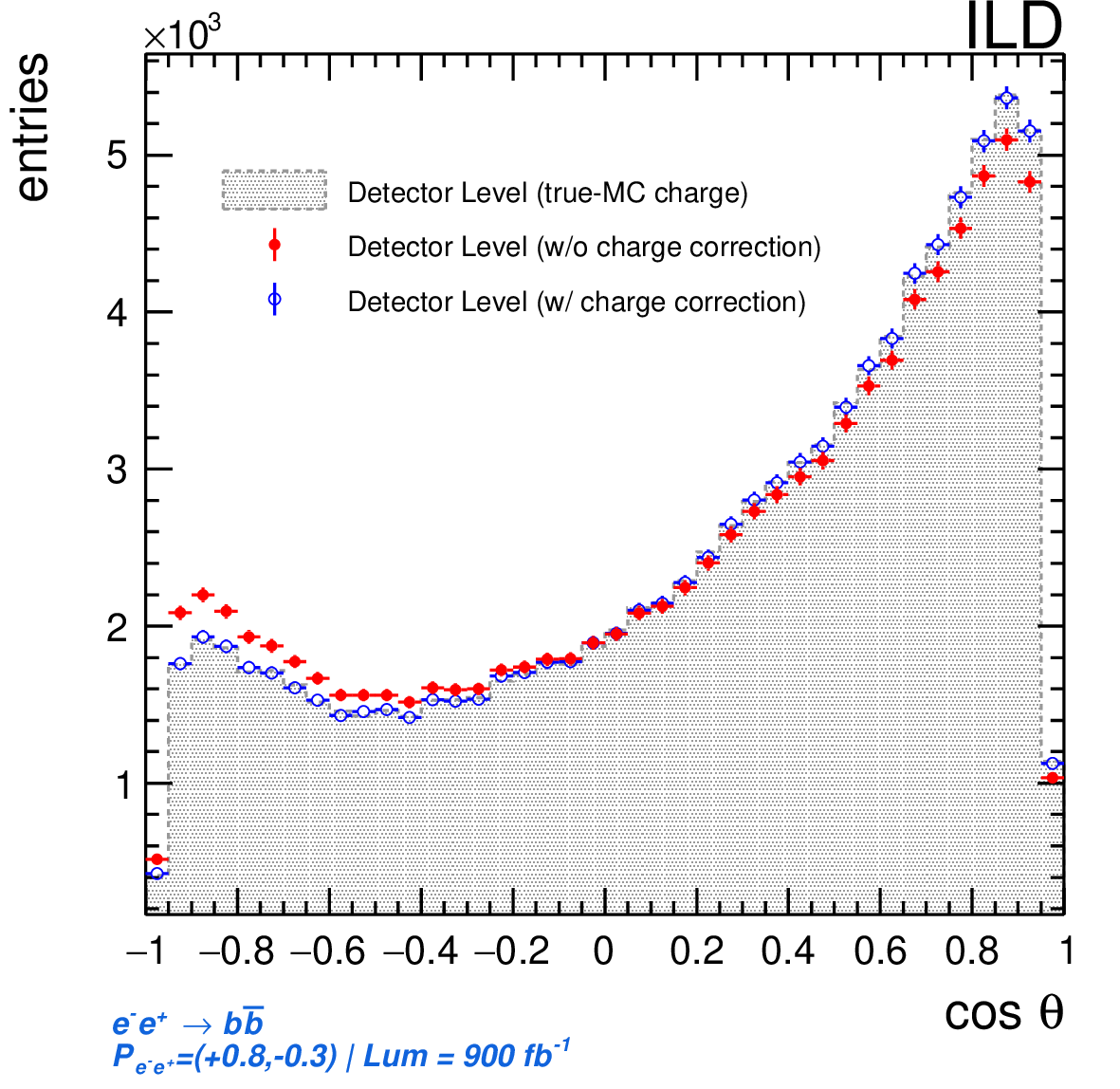} \\
      \end{tabular}
    \caption{Distribution of the detector level (\textit{i.e.} without efficiency/acceptance corrections) of the number of \eeqq events reconstructed after identification of the \costheta sign. For the filling of the shaded area the truth Monte Carlo is used to calculate the charge of the jet and the sign of \costheta. The red dots correspond to the DC measurement but before the charge correction. Finally, the blue dots include the charge correction, which is the last step of the DC method.
     \label{fig:migration}}
\end{center}
\end{figure}

The result of applying Eqs.~\ref{eq:purity} and~\ref{eq:corr} is shown in Fig.~\ref{fig:migration}.
The agreement between the Monte Carlo truth charge and the data-driven corrected distribution is excellent. It is important to remark that this correction is sizeable in the case of the \bquark and especially for the \eLpR beam polarisation scenarios. It is almost negligible for the \ccbar case. The qualitative explanation is that the number of prongs is small for the \cquark and secondary vertexes are mostly made of two tracks (Fig. \ref{fig:sectracks}). Hence if a track is lost in the association to the vertex, in most cases the full vertex is not reconstructed resulting in a decrease of the efficiency but not of \Pb. 

\subsection{Efficiency corrections}
\label{sec:analysis_charge_correction}

The goal is to measure $\frac{d\sigma}{d\costheta}$ at the parton level and extract the forward-backward asymmetry.
However, for each category, $Cat.i$, the measured distribution is:
\begin{equation}
    N_{Cat.i}(\costheta) = \mathcal{L} \left[\epsilon_{pres.}(|\costheta|) \epsilon_{Cat.i}(|\costheta|) \frac{d\sigma}{d\costheta} + \epsilon_{bkg} \frac{d\sigma_{bkg}}{d\costheta}\right] 
    \label{eq:N_AFB}
\end{equation}
with $\mathcal{L}$ being the total collected luminosity.

\begin{figure}[!ht]
\begin{center}
    \begin{tabular}{cc}
      \includegraphics[width=0.45\textwidth]{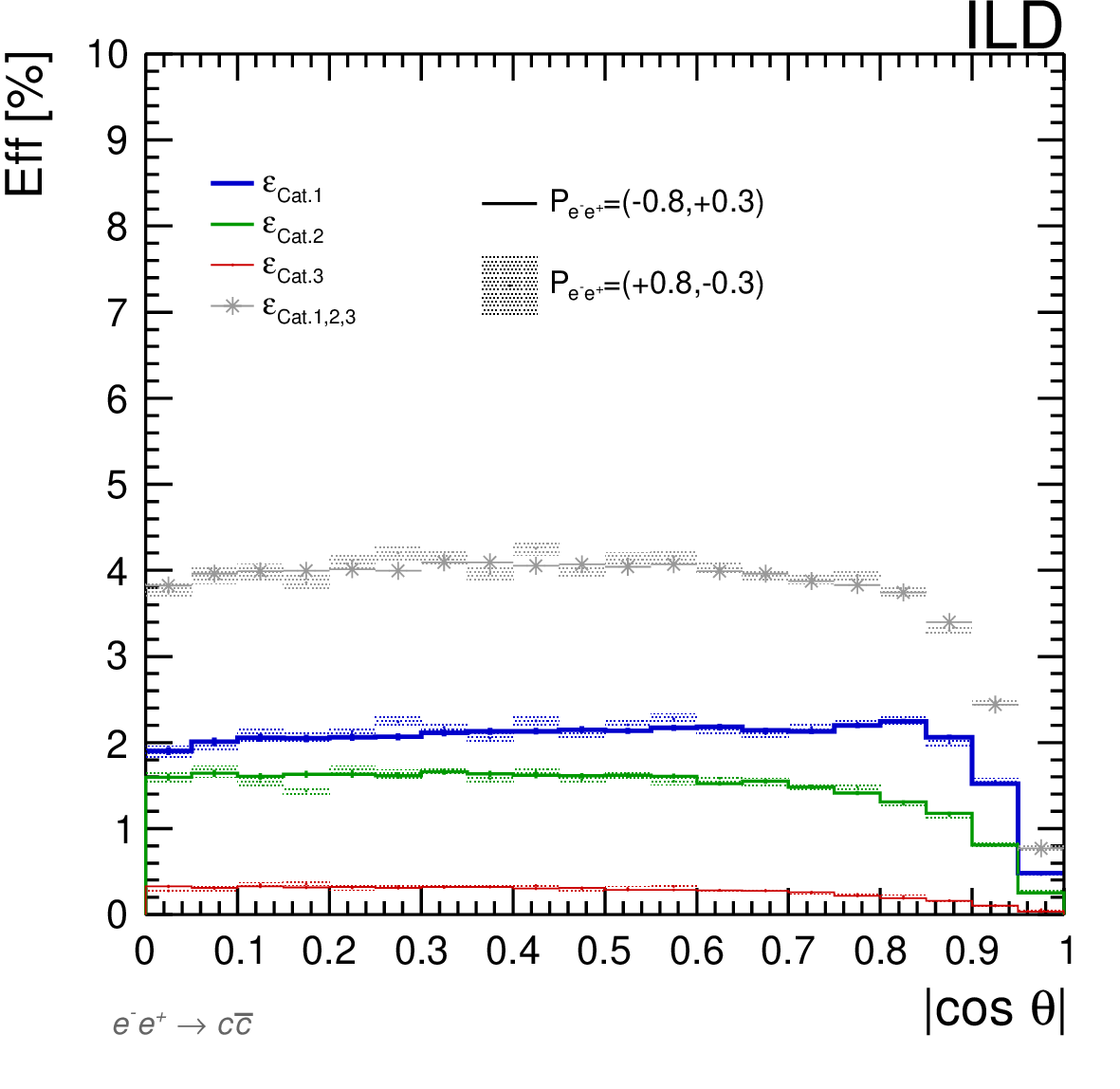} &
      \includegraphics[width=0.45\textwidth]{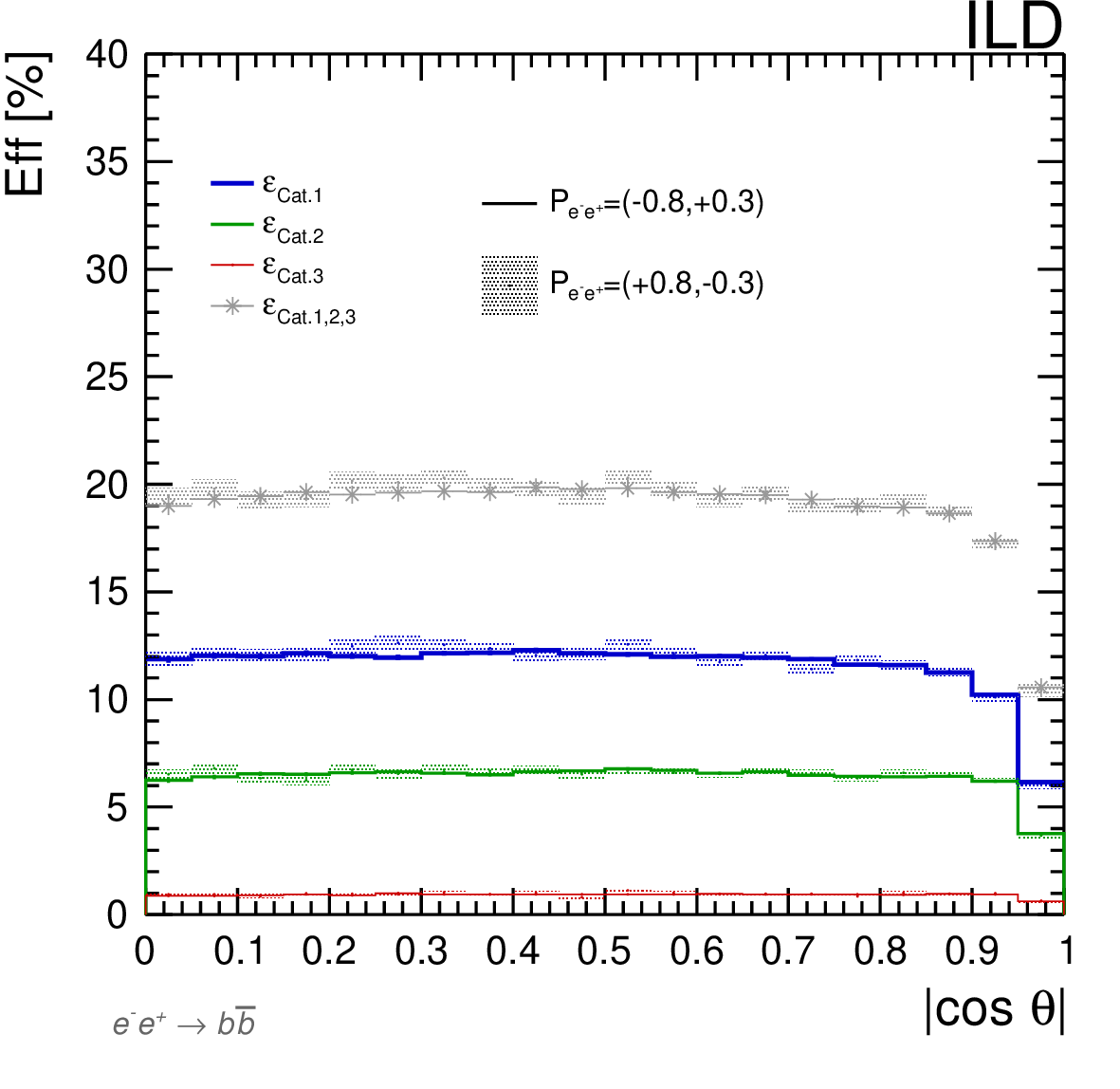} 
    \end{tabular}
      \caption{Distribution of the different selection efficiencies for \ccbar (left) and \bbbar (right) for the \AFB measurement, as described in Eq. \ref{eq:eff_cat}. For the \ccbar (\bbbar) case, the $Cat.1$ corresponds the case in which both jets had at least a \Kc (\Bc) measurement, while in the $Cat.3$ only the \Bc (\Kc) is available for both jets.\label{fig:eff_cat}}
\end{center}
\end{figure}

The different $\epsilon_{Cat.i}$
can be be expressed in terms of the different, $(1+\rho_{q})$, $\epsilon_{q}$ and \Pb described in this section and the previous one:
\begin{flalign}
    &\epsilon_{Cat.1}=(1+\rho_{q}) \cdot (\epsilon_{q} f_{M_{1}})^{2} (\PbB^{2} + \QbB^{2} ) \\
    &\epsilon_{Cat.2}=2 (1+\rho_{q}) \cdot (\epsilon_{q} f_{M_{1}}) \cdot (\epsilon_{q} f_{M_{2}}) \cdot (\PbB\cdot\PbK+\QbB\cdot\QbK)  \\
    &\epsilon_{Cat.3}=(1+\rho_{q}) \cdot (\epsilon_{q} f_{M_{2}})^{2} (\PbK^{2} + \QbK^{2}) 
    \label{eq:eff_cat}
\end{flalign}
The $f_{M_{j}}$ is defined as the fraction of jets
for which at least one jet would have a charge measurement using the method $M_{j}$ .
The only three quantities 
that are not 
estimated using the data as input are the correlation factors,
$(1+\rho_{q})$, the pre-selection efficiency  
$\epsilon_{pres}$ and the efficiency of background rejection $\epsilon_{bkg}$. The different $\epsilon_{Cat.X}$ distributions are shown in Figure \ref{fig:eff_cat}.

\subsection{Results}
\label{sec:analysis_AFB_results}

\begin{figure}[!ht]
\begin{center}
    \begin{tabular}{cc}
      \includegraphics[width=0.45\textwidth]{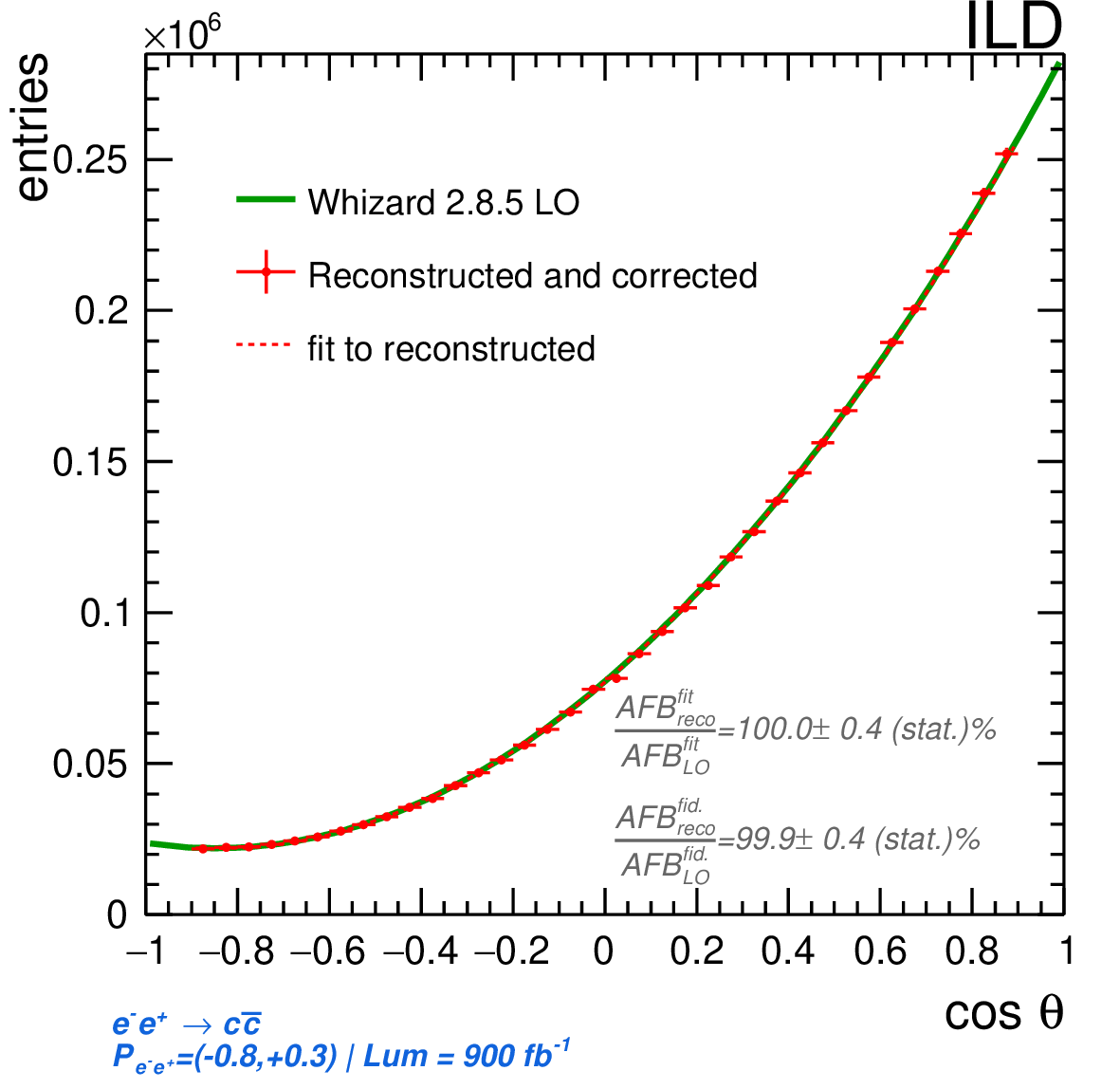} &
      \includegraphics[width=0.45\textwidth]{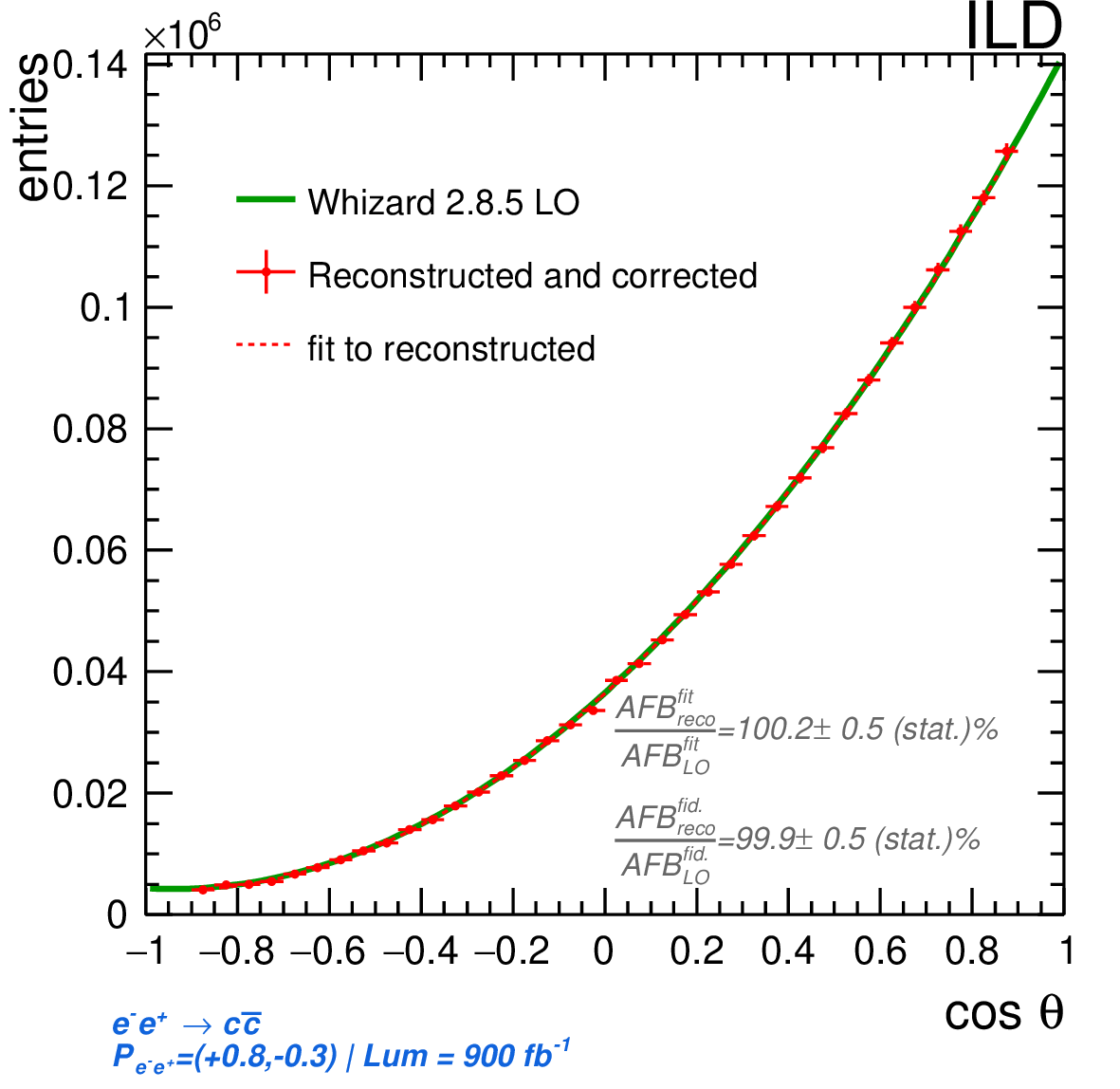} \\
      \includegraphics[width=0.45\textwidth]{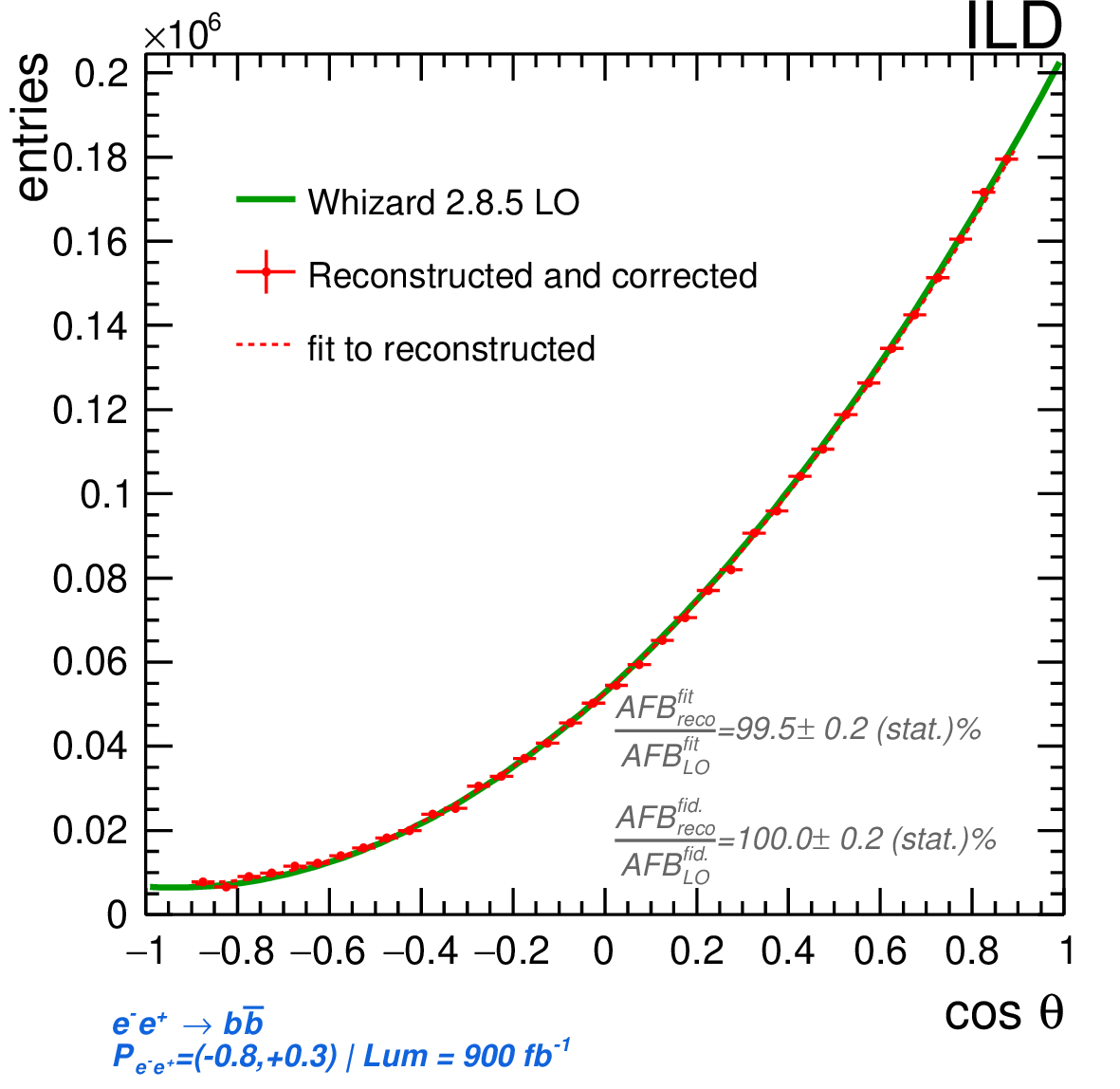} &
      \includegraphics[width=0.45\textwidth]{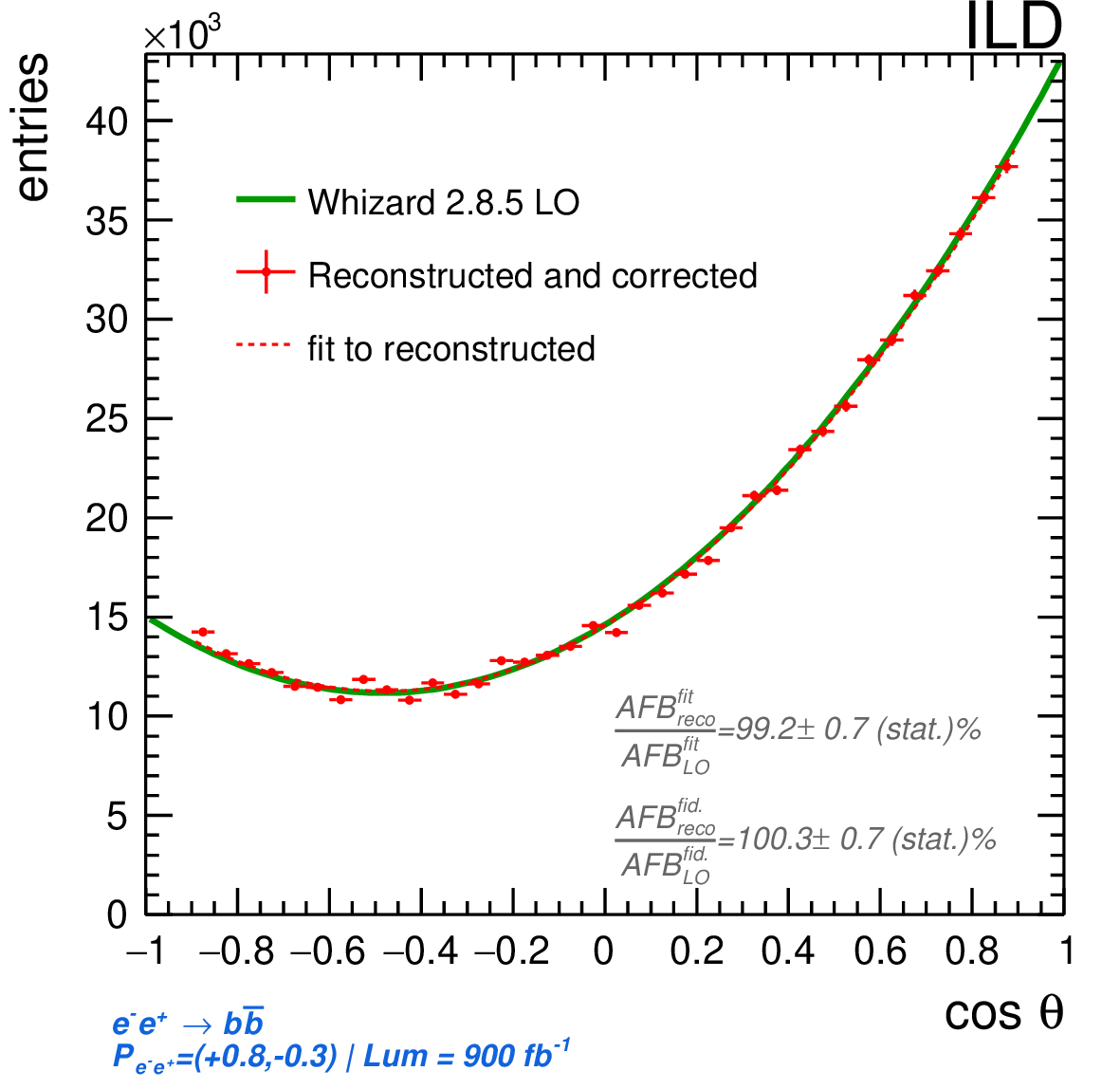} 
    \end{tabular}
      \caption{Fit (red graph) of the final distributions (red points) to the function described in Equation \ref{eq:fiteq}.
      The fit is performed between $-0.9<\costheta<0.9$ to avoid the regions with a large correction due to efficiency and acceptance losses.
      The result of the fit is extended to $[-1,1]$ for the \AFB estimation.
      The LO calculations are represented in the green graphs. The forward-backward asymmetry after the full reconstruction
      and the LO prediction are compared and are well in agreement with statistical uncertainties.\label{fig:AFB_fit}}
\end{center}
\end{figure}

The result of the full correction procedure is shown in Figure \ref{fig:AFB_fit}.
For the estimation of \Afb a fit to the distributions at the parton level
using the following function\footnote{This shape neglects the SM tensorial contribution $T\times \text{sin}^{2}\theta$
which is very small given the large $\gamma$ factor for $b$ and $c$-quarks. If this term is included in the fit, the same
results are found on the \Afb determination.} is performed:
\begin{flalign}
    \frac{d\sigma}{d\costheta} = S \left(1+\text{cos}^{2}\theta\right) + A  \costheta.
    \label{eq:fiteq}
\end{flalign}
The fit is performed in the range of \costheta with high reconstruction efficiency $-0.9\leq\costheta\leq 0.9$ 
thus avoiding the very forward regions where the efficiency of vertex reconstruction and/or particle identification drops. 
The \AFB is calculated by extrapolating the fitted function to the full range of \costheta.
As a crosscheck, the \AFB is also calculated with a simple count method and compared with the fit to the 
calculation, restricted to the fiducial acceptance region $-0.9\leq\costheta\leq 0.9$ (quoted as \textit{fid.} in the figures).

Considering the size of the samples for the assumed luminosity scenario,
the expected statistical uncertainty
on the determination of \AFBc and \AFBb at ILC250 is given in
Table \ref{tab:uncertainties}. The quoted uncertainty also accounts for the statistical uncertainties
on the efficiency and charge corrections.

The study of the systematic uncertainties is addressed in Section \ref{sec:systematics}.

%% file: sections/8_systematics.tex
\section{Systematic uncertainties}
\label{sec:systematics}

At ILC250, with 2000 \fb of integrated luminosity, the 
statistical errors are at the level of a few per mil. 
Therefore, a comprehensive assessment of the experimental
systematic uncertainties is required. The size of the main systematic uncertainties is described in the following and summarised in Table \ref{tab:uncertainties}.

\subsection{Preselection efficiency and background rejection}
\label{sec:syst_pres}

\subsubsection{Background removal}
\label{sec:syst_pres_bkg}

To use the DT and the DC methods, we are required to subtract the backgrounds previously.
The largest backgrounds come from ISR and $WW$ pairs. 
Mutual contamination between \bbbar or \ccbar (and light quarks)  are treated in Section \ref{sec:syst_flav}.
For the case of \AFB, these backgrounds are reduced to zero thanks to the simultaneous application of the DT and DC and only play a second-order 
role in the determination of flavour tagging efficiencies. Furthermore, it is expected that thanks to the large size of the expected 
data samples, the modelling of the backgrounds would be possible, at least at the per cent level.

Conservatively a $10\%$ uncertainty is assumed for every bin
of the $\epsilon_{bkg}$ distribution from Eq. \ref{eq:N_AFB} and on the number of events expected per bin in the $N_{0}$, $N_{1}$ and $N_{2}$ from Eq. \ref{eq:Rq}.
These factors are varied by $\pm 10\%$, and the $R$ and \AFB analysis are repeated for both variations. The total uncertainty is the difference between both
divided by two.

\subsubsection{Estimation of preselection efficiencies and the difference between flavours}
\label{sec:syst_pres_epsilon}

The measurement of \Rb and \Rc is not affected by preselections
if the same efficiency for each flavour is kept.
Figure \ref{fig:preselection} shows a MC comparison between 
the selection efficiency curves for \bbbar, \ccbar and light 
quarks, showing that universality stays within a \% level. 
Possible fluctuations due to QCD radiation and mass effects
are expected to be well understood and modelled with NLO QCD Monte Carlo generators \cite{Abreu:1997ey,rodrigomb,Fuster:2021ekh}.
The MC information allows us to correct for this small disagreement for the various flavours, and the smallness of this correction guarantees that 
the corresponding systematic error when measuring \Rb or \Rc can be kept well below 0.1\%, which 
translates into a negligible impact on the determination of $R$.

However, for the differential cross section measurement,
the preselection efficiency cannot be neglected since it affects the shape of the cross section, 
although part of the uncertainty is cancelled through the normalisation
in \Afb. Again, this is evaluated by producing pseudo-data distributions
assuming a $10\%$ accuracy, bin-wise, on the preselection efficiency factor.
This uncertainty is propagated to the \AFB estimation in the same way as explained in Section \ref{sec:syst_pres_bkg}.

This uncertainty is one of the leading systematic uncertainties in the case of the \Afb measurement
but still not competing with the statistical uncertainties.
It is important to remark that the $10\%$ accuracy assumption is to be confirmed with data and dedicated analysis, and we
expect that this assumption would turn out to be a conservative one.
It is hoped that this  could be further reduced profiting from more advanced techniques for event reconstruction and
background modelling (a maximum profile-likelihood approach \cite{CMS:2019jul}, for example or even newer machine learning techniques).

\subsection{Flavour tagging}
\label{sec:syst_flav}

Flavour tagging uncertainties are from the following three sources:

\begin{itemize}
\item Tagging efficiency, which depends on the details of the fragmentation modelling and $b/c$-hadrons decays, $\epsilon_{q}$.
\item Contamination from the other heavy quark type, $\tilde{\epsilon}_{q^{\prime}}$ .
\item Contamination from the lighter quark types ($uds$) and the process $g\rightarrow q\bar{q}$ which are poorly known and which, for single quark tagging, can contaminate all two-fermion processes, $\tilde{\epsilon}^{q}_{uds}$.
\end{itemize}

Using the DT method, the efficiency is extracted from data and not from the MC, which is plagued by all sorts of uncertainties ($b$-fragmentation function, $B$ and $D$-hadrons branching ratios, etc.). For a given selection with at least one jet tagged as originated from a quark of flavour $q$
with an efficiency $\epsilon_{q}$, one compares the number of events with double tagging, which is proportional to $\epsilon^{2}_{q}(1+\rho_{q})$, see
Eq.~\ref{eq:Rq}. 

The system of equations described in Equation \ref{eq:Rq} can be solved simultaneously for $R_{q}$ and $\epsilon^{2}_{q}$ 
provided $(1+\rho_{q})$, $\tilde{\epsilon}_{q^{\prime}}$ and $\tilde{\epsilon}^{q}_{uds}$ are known from either simulations or alternative methods.

For this study, we take $\tilde{\epsilon}_{q^{\prime}}$ and $\tilde{\epsilon}^{q}_{uds}$ as given by the simulation and perform pseudo experiments varying 
their values by $\pm 10\%$ in each bin and propagating the uncertainty as explained in previous sections. The hemisphere correlation $(1+\rho_{q})$ is kept constant. It will be discussed in Sec.~\ref{sec:syst_angular}.
Except for the \AFBb with the case of right-handed electron beam polarisation, the contribution by $\epsilon_{uds}$ to the uncertainty is negligible.
However, the $\tilde{\epsilon}_{q^{\prime}}$ is one of the dominant sources of uncertainties for all observables.
In all these cases, the uncertainty on the measured observables is of the order or smaller than one per mil. However, the assumed uncertainty on the different mistagging efficiencies is similar to what was estimated in past experiments, potentially improvable with more data and sophisticated detectors and algorithms.

\subsection{Hemisphere Correlations and Detector Asymmetries}
\label{sec:syst_angular}

The estimated ($1+\rho_{q}$) for ILC250 is constant in most of the detector volume, with $\rho_{q}$ being smaller than $0.2\%$. Since the implementation of the tracking system in the ILD simulation is symmetric and no coherent noise is simulated, a $\rho_{q}$ value different from zero can only result from occasional mis-measurements of the primary vertex or hard QCD radiation diluting the back-to-back configuration of the di-jet system. 
The small value indicates that both effects can be controlled to a high level. Hemisphere correlations due to a common vertex are suppressed if, in addition to an excellent tracking system, the actual beam size is very small. This is the case for ILC; see Sec.~\ref{sec:ILCILD}. Furthermore, hemisphere correlations can be eliminated by a high tagging efficiency since, by definition; it is $\rho_{q}=0$ if $\epsilon^2_{q}=1$.
To estimate the impact of the uncertainties on ($1+\rho_{q}$), we calculated the results for $\Rq$ and \AFB with and without applying the hemisphere correlation factor and found negligible dependence.
For the estimation of QCD effects, it would be required to have an NLO-QCD simulation with full detector effects.
However, following \cite{AlcarazMaestre:2020fmp}, we assume a $0.1\%$ uncertainty for all observables.
In the formalism followed in this article, detector asymmetries would also modify the hemisphere correlations. A simple way to control detector asymmetries is to count the number of charged tracks in all di-jet events. An asymmetry of 1\% has to be controlled at the 10\% level to yield an uncertainty of 0.1\% on the hemisphere correlation. Given the expected large number of charge tracks, this seems feasible.  
Detector asymmetries are also relevant for the studies in Sec.~\ref{sec:analysis_AFB}, notably for the actual value \Pb, i.e. the probability of correct measurement of the jet-charge. Strictly speaking, the values of \Pb determined in Sec.~\ref{sec:analysis_AFB} are an average over the corresponding forward and backward regions. In the first approach, the correction for detector asymmetries is linear for \Afb. Therefore, the detector asymmetries have to be controlled to a value better than the expected statistical precision of \Afb as given in Table~\ref{tab:uncertainties}, thus better than 0.24\% - 0.7\% depending on the beam polarisation. As before, counting the number of tracks will allow for estimating the detector asymmetry. A second way to control the detector asymmetry, closer related to the differential cross section measurement, is to measure the amount of ``unphysical" secondary vertices featuring double charge values (for example, a vertex with only two positive tracks).
A double charge vertex implies that a track has been lost. This could happen for low-resolution offset measurements (affecting both hemispheres equally) or because of inefficiencies in reconstructing a vertex inside the detector volume. The latter would be affected by any possible asymmetry of the detector.     
For both described methods, the large number of available tracks for these studies should allow us to achieve the required precision.
As before, the impact of detector asymmetries decreases with increasing \Pb since the asymmetry is zero if $\Pb^2=1$.

\subsection{Charge Measurements}
\label{sec:syst_charge}

The DC method described in Section \ref{sec:analysis_migrations} is entirely based on data-driven strategies.
We not only estimate the efficiency of each method but the quality of the technique itself, which can be affected by detector effects (lack of acceptance, low-resolution effects),
reconstruction features (missasignement between tracks and clusters or between track segments) and physical effects ($B^{0}$ oscillations).
All these effects are taken into account by the DC method.
To evaluate the uncertainty and avoid biases, we applied the purity and efficiency values extracted from a sample with a given beam polarisation 
to the sample with the other beam polarisation, finding no difference between the results.
Of course, the method has a statistical uncertainty due to the finite size of the $data$ sample. This uncertainty is taken into account 
and included in the total statistical uncertainty.

\subsection{Polarisation and luminosity}

At ILC250, the \eebb observables are quite different for the \eLpR and \eRpL cases; however, this is not the case
for \eecc, which shows a differential cross section with a similar shape for both polarisations.
Therefore, polarisation errors 
influence the two flavours very differently and only have a sizeable impact on the right polarisation 
scenario for the \bquark.
To estimate the uncertainties due to the measurement of the beam polarisation, we take
the numbers from \cite{Karl:424633}. These are quoted in Table \ref{tab:polerror}.
These uncertainties are propagated to the measured observable, and the associated uncertainty is estimated
by dividing the maximum difference between predictions (with $\pm1\sigma$) by two.

The luminosity uncertainties will cancel in all of our observables, turning out to be negligible.

\begin{table}[!ht]
 \begin{center}%\renewcommand{\arraystretch}{1.8}
%  \scriptsize
  \begin{tabular}{cccc}
    \hline
    \multicolumn{4}{c}{{\textbf{Beam polarisation uncertainty}}}\\
    \hline
    $\Delta P^{-}_{e^{-}}~ [\%]$ & $\Delta P^{+}_{e^{-}}~ [\%]$ &  $\Delta P^{+}_{e^{+}}~ [\%]$ &  $\Delta P^{-}_{e^{+}}~ [\%]$ \\
    \hline
    0.1 & 0.04 & 0.1 & 0.14 \\
    \hline
    \end{tabular}
  \end{center}
 \caption{\label{tab:polerror} Uncertainty on the beam polarisation. Numbers extracted from \cite{Karl:424633}.}
\end{table}

%% file: sections/9_results.tex
\section{Results and prospects}
\label{sec:results}

\begin{figure}[!ht]
  \centering
    \begin{tabular}{c}
        \includegraphics[width=0.9\textwidth]{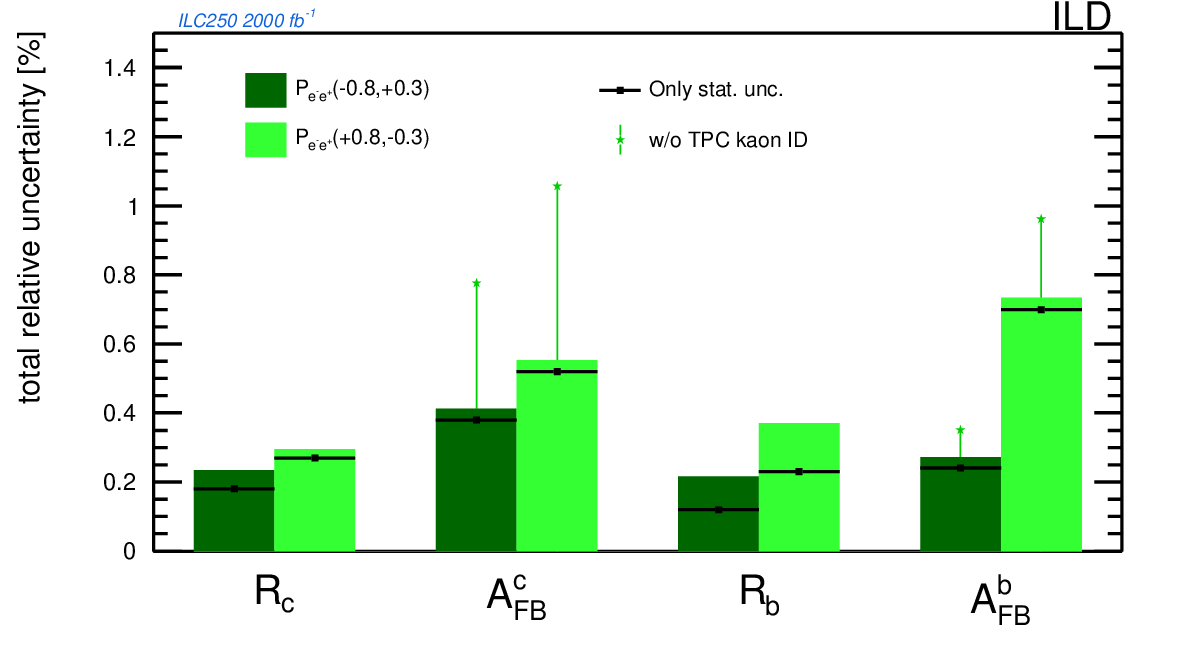}
     \end{tabular}	
  \caption{\label{fig:results} Expected achievable precision for the measurement of electroweak observables \Rq and \Afbq at ILC250 with the ILD. We include the scenario in which no charged hadron identification capabilities are provided by the TPC, hence having only the $Vtx$-method for the charge measurement used for \AFB.}
\end{figure}

\begin{figure}[!ht]
  \centering
    \begin{tabular}{c}
        \includegraphics[width=0.6\textwidth]{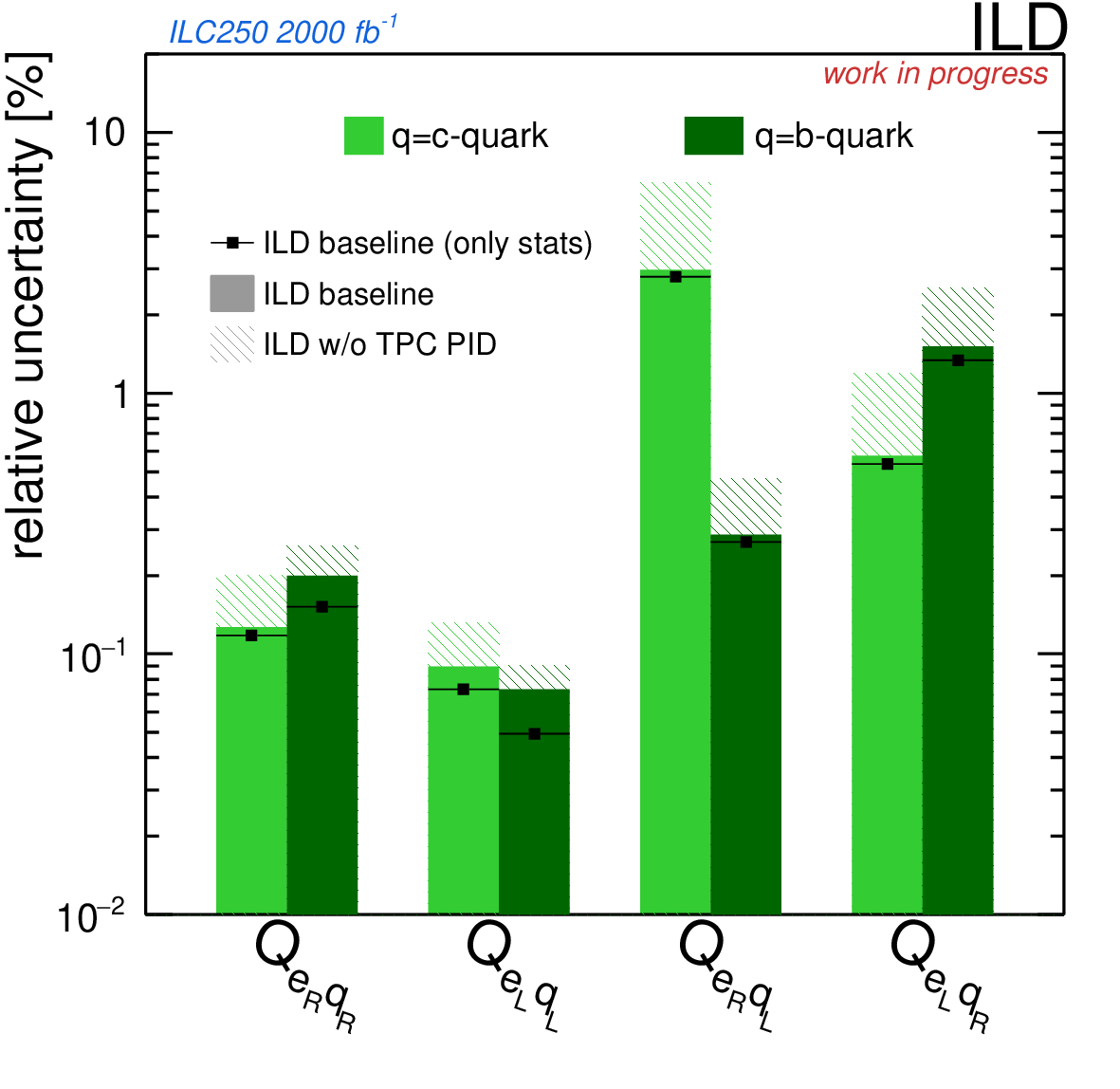}
     \end{tabular}	
  \caption{\label{fig:results2} Expected achievable precision for the extraction of the helicity amplitudes $Q_{e_{X}q_{Y}}$ as defined in Section \ref{sec:observable}. As in Figure \ref{fig:results}, we include the scenario where the TPC provides no charged hadron identification capabilities, hence having only the $Vtx$-method for the charge measurement used for \AFB.
  Further studies, including the prospects for ILC-\Zpole, ILC500 and ILC1000, will allow the inspection of the SM and eventual BSM terms inside the helicity amplitudes.}
\end{figure}

The results on the expected experimental precision foreseen for
the measurement of electroweak observables \Rq and \Afbq at the
ILC running at 250 GeV are summarised in Table \ref{tab:uncertainties} and
Figure \ref{fig:results}. 
For both observables and both polarisation scenarios, total experimental
uncertainties of few per mil are expected for the full 2000 \fb ILC250 program. 
This includes a comprehensive assessment of the systematic uncertainties.
Such accuracy poses a challenge
to theoretical higher-order corrections, particularly for what concerns electroweak corrections. It is
out of the scope of this document to discuss this issue.
In Figure \ref{fig:results2} we show the expected uncertainties on the extraction of the 
helicity amplitudes $Q_{e_{X}q_{Y}}$ as defined in Section \ref{sec:observable}. 
For this plot, SM values for such helicity amplitudes are assumed.

\begin{table}[!ht]
 %\renewcommand{\arraystretch}{1.6}
 % \scriptsize
 \begin{center}
  \begin{tabular}{c|cc|cc|cc|cc}
    Source & \multicolumn{4}{c|}{\eecc} & \multicolumn{4}{c}{\eebb} \\
    \hline
     & \multicolumn{2}{c|}{$P_{e^{-}e^{+}}(-0.8,+0.3)$} & \multicolumn{2}{c|}{$P_{e^{-}e^{+}}(+0.8,-0.3)$}  & \multicolumn{2}{c|}{$P_{e^{-}e^{+}}(-0.8,+0.3)$} & \multicolumn{2}{c}{$P_{e^{-}e^{+}}(+0.8,-0.3)$}  \\
     & \multicolumn{1}{c}{\Rc} & \AFBc &  \Rc & \AFBc &  \multicolumn{1}{c}{\Rb} & \AFBb &  \Rb & \Afbb \\
    \hline
    \textbf{Statistics} & \textbf{0.18\%} & \textbf{0.38\%} &  \textbf{0.27\%} & \textbf{0.52\%} &  \textbf{0.12\%} & \textbf{0.24\%} &  \textbf{0.23\%} & \textbf{0.70\%}\\
    \hline
    Preselection eff. & <0.01\% & 0.12\% &  0.02\% & 0.16\% &  <0.01\% & 0.08\% &  0.06\% & 0.12\% \\
    Background & 0.01\% & 0.01\% &  0.02\% & 0.02\% &  0.01\% & 0.01\% &  0.06\% & <0.01\% \\
    heavy quark mistag & 0.11\% & <0.01\% &  0.06\% & <0.01\% &  0.12\% & <0.01\% &  0.22\% & <0.01\% \\
    $uds$ mistag & 0.03\% & <0.01\% &  0.02\% & <0.01\% &  0.08\% & <0.01\% &  0.14\% & <0.01\% \\
    Angular correlations & 0.10\% & 0.10\% &  0.10\% & 0.10\% &  0.10\% & 0.10\% &  0.10\% & 0.10\% \\
    Beam Polarisation & <0.01\% & <0.01\% &  0.02\% & 0.01\% &  <0.01\% & 0.01\% &  0.03\% & 0.15\% \\
    \textbf{Systematics} & \textbf{0.15\%} & \textbf{0.16\%} &  \textbf{0.12\%} & \textbf{0.19\%} &  \textbf{0.18\%} & \textbf{0.13\%} &  \textbf{0.29\%} & \textbf{0.22\%}\\
    \hline
    \textbf{Total} & \textbf{0.24\%} & \textbf{0.41\%} &  \textbf{0.30\%} & \textbf{0.55\%} &  \textbf{0.21\%} & \textbf{0.27\%} &  \textbf{0.37\%} & \textbf{0.73\%}\\

  \end{tabular}
  \caption{\label{tab:uncertainties} Breakdown of statistical and systematic uncertainties for the different experimental observables.}
 \end{center}
\end{table}

The work presented here is focused on the identification of the significant advantages of ILD and ILC250 for this type
of studies but also on the identification of primary sources of systematic uncertainties.
We took a conservative approach, assuming today's state of the art on background knowledge
and reconstruction tools performance. Moreover, the event selection is based on a simple "cut-based" approach
which is not optimised to enhance the statistics. We believe that statistical 
uncertainties could be further reduced with more sophisticated techniques. 
However, the goal of the study is to estimate the overall potential of ILC250 and ILD and
identify any eventual showstopper due to the ILD design.
We identify that the highest quality tracking and vertexing systems and reconstruction tools are
mandatory to meet the high precision requirements
for the measurements described in this document. In particular, 
maximising the angular coverage of the tracking systems as close as possible to the beam axis is very important.
This work proves the forward region's relevance and encourages further optimising its design. Moreover, further development and optimisation of the reconstruction tools
are expected at the moment of the ILC realisation, further reducing the systematic uncertainties associated 
with the reconstruction.
The second key feature of ILD for this type of measurements 
is the charged hadron identification capability for relatively high momentum tracks (above 3-5 GeV).
This is possible with a TPC and \dEdx discrimination. We inspected the prospects of the achievable 
precision on the \AFB measurement assuming no particle identification performed by using the TPC in the ILD 
(seen as green stars in Figure \ref{fig:results} or dashed histograms in Figure \ref{fig:results2}).
The study shows a clear gain by the usage of a TPC in ILD, especially for the \cquark case 
\footnote{The charged kaon identification is even more crucial for $s$-quark studies, preliminary work is presented in \cite{Okugawa:2022zmt}.}. 
The impact of not using a TPC on particle flow is not covered in this study.
The ILD has also considered the possibility of using time-of-flight measurement systems to perform the charged hadron identification
. However, this type of technique would only increase the particle identification capabilities 
for relatively low momentum tracks (see \cite{ILD:2020qve}, Figure 8.6)
which are of little interest to our studies, at least at energies of ILC250 or larger.
The time-of-flight measurement for charged kaon identification could be an asset for low-energy runs at the \Zpole.

The unprecedented precision expected for \eecc and \eebb observables would provide an unambiguous
resolution of the SLC/LEP1 anomaly in the $\sin^{2} \theta_{W}$ determination.
Moreover, models with extended gauge structures \cite{Yoon:2018xud,Funatsu:2017nfm,Funatsu:2020haj} predict large deviations from
the standard model electroweak couplings. These models predict large modifications
on the electroweak observables studied in this document, and some of these models
predict such kinds of effects for all fermions (not only for the heaviest). 
Of particular importance is the fact that - thanks to the beam polarisation at the ILC - we could inspect the different
helicity amplitudes to distinguish between different models. 
In addition, the ILC will also offer the possibility to scrutinise the electroweak observables with high precision 
at the \Zpole (providing an improvement of at least one order of magnitude for the \ccbar and \bbbar $Z$-couplings measured at SLC and LEP) 
and at the energy frontier, 500 GeV and even 1 TeV, further enhancing the sensitivity to BSM.

